\documentclass[11pt]{aastex62}          

\usepackage{aas_macros}
\usepackage{graphicx}
\usepackage{amssymb}
\usepackage{amsmath}
\usepackage{ulem}
\usepackage{epstopdf}
\usepackage{color}
\usepackage{cancel}

             \font\sevenrm=cmr7

          \font\sixrm=cmr6       
               \font\sixbf=cmbx6      

\def\erg{\varepsilon}

\def\lambar{\lambda\llap {--}}
\def\omegaB{\omega_{\hbox{\sixbf B}}}
\def\rns{R_{\hbox{\sixrm NS}}}
\def\rlc{R_{\hbox{\sixrm LC}}}
\def\rmax{r_{\rm max}}
 
\def\ergfmax{\erg_f^{\rm max}}
\def\tauEM{\tau_{\hbox{\sixrm EM}}}

\def\Edotrot{{\dot E}_{\hbox{\sixrm ROT}}}
\def\sigt{\sigma_{\hbox{\sixrm T}}}

\def\thetaBr{\theta_{\hbox{\sevenrm B}r}}
\def\ThetaBn{\Theta_{\hbox{\sevenrm B}n}}
\def\thetaB{\theta_{\hbox{\sixbf B}}}
\def\muB{\mu_{\hbox{\sixbf B}}}
\def\phiB{\phi_{\hbox{\sixbf B}}}

\def\fsc{\alpha_{\hbox{\sevenrm f}}}

\def\muBhat{\hat{\boldsymbol{\mu}}_{\hbox{\sixrm B}}}
\def\thetamin{\theta_{\hbox{\sevenrm min}}}   
\def\thetamax{\theta_{\hbox{\sevenrm max}}}   
\def\thetaBr{\theta_{\hbox{\sevenrm Br}}}   
\def\thetaC{\theta_{\hbox{\sixrm C}}}   
\def\thetakB{\theta_{\hbox{\sevenrm kB}}}   
\def\dover#1#2{\hbox{${{\displaystyle#1 \vphantom{(} }\over{
   \displaystyle #2 \vphantom{(} }}$}}

\def\emph#1{{\it #1}}

{\catcode`\@=11                                                  
\gdef\SchlangeUnter#1#2{\lower2pt\vbox{\baselineskip 0pt\lineskip0pt    
\ialign{$\m@th#1\hfil##\hfil$\crcr#2\crcr\sim\crcr}}}}           
\def\gtrsim{\mathrel{\mathpalette\SchlangeUnter>}}               
\def\lesssim{\mathrel{\mathpalette\SchlangeUnter<}}    
\def\teq#1{$\, #1\,$}                         
\begin{document} 

\newcommand{\Ai}{{\rm Ai}}

\newcommand{\vol}[2]{$\,$\bf #1\rm , #2.}                 

\newcommand{\figureoutpdf}[5]{\centerline{}
   \centerline{\hspace{#3in} \includegraphics[width=#2truein]{#1}}
   \vspace{#4truein} \figcaption{#5} \centerline{} }
\newcommand{\figureoutpdfscale}[5]{\centerline{}
   \centerline{\hspace{#3in} \includegraphics[width=#2\textwidth]{#1}}
   \vspace{#4truein} \figcaption{#5} \centerline{} }
\newcommand{\twofigureoutpdf}[3]{\centerline{}
   \centerline{\includegraphics[width=2.8truein]{#1}
        \hspace{0.1truein} \includegraphics[width=2.8truein]{#2}}
        \vspace{-0.1truein}
    \figcaption{#3} }    
\newcommand{\twofigureoutpdfadj}[3]{\centerline{}
   \centerline{\includegraphics[width=3.2truein]{#1}
        \hspace{-0.3truein} \includegraphics[width=3.2truein]{#2}}
        \vspace{-0.1truein}
    \figcaption{#3} }    
\newcommand{\fourfigureoutpdf}[5]{\centerline{}
   \centerline{\includegraphics[width=2.3truein]{#1}
        \hspace{0.0truein} \includegraphics[width=2.3truein]{#2}}
        \vspace{0.0truein}
   \centerline{\includegraphics[width=2.2truein]{#3}
        \hspace{0.0truein} \includegraphics[width=2.2truein]{#4}}
        \vspace{-0.1truein}
    \figcaption{#5} }    

\title{RESONANT INVERSE COMPTON SCATTERING SPECTRA\\ FROM HIGHLY-MAGNETIZED NEUTRON STARS}  

	 \author{Zorawar Wadiasingh}
   \affiliation{Centre for Space Research,
   	North-West University, Potchefstroom, South Africa 
     {\rm \hspace{3pt} zwadiasingh@gmail.com}}

   \author{ Matthew G. Baring}
   \affiliation{Department of Physics and Astronomy, MS 108,
      Rice University, Houston, TX 77251, U.S.A.
      {\rm baring@rice.edu}}

   \author{Peter L. Gonthier}
   \affiliation{Hope College, Department of Physics,
          27 Graves Place, Holland, MI 49423, U.S.A.
      \rm gonthier@hope.edu}
      
    \author{Alice K. Harding}
    \affiliation{Astrophysics Science Division,
         NASA's Goddard Space Flight Center, Greenbelt, MD 20771, U.S.A.
      \rm alice.k.harding@nasa.gov}

\begin{abstract} 
Hard, non-thermal, persistent pulsed X-ray emission extending between 10
keV and \teq{\sim 150}keV has been observed in nearly ten magnetars. For
inner-magnetospheric models of such emission, resonant inverse Compton
scattering of soft thermal photons by ultra-relativistic charges is the
most efficient production mechanism.  We present angle-dependent
upscattering spectra and pulsed intensity maps for uncooled,
relativistic electrons injected in inner regions of magnetar magnetospheres, calculated using collisional integrals over field loops. Our computations employ a new formulation of the QED Compton scattering cross section in strong magnetic fields that is physically correct for
treating important spin-dependent effects in the cyclotron resonance,
thereby producing correct photon spectra. The spectral cut-off energies
are sensitive to the choices of observer viewing geometry, electron
Lorentz factor, and scattering kinematics. We find that electrons with
energies \teq{\lesssim 15} MeV will emit most of their radiation below
250 keV, consistent with inferred turnovers for magnetar hard X-ray
tails. More energetic electrons still emit mostly below 1 MeV, except
for viewing perspectives sampling field line tangents. Pulse profiles
may be singly- or doubly-peaked dependent upon viewing geometry,
emission locale, and observed energy band. Magnetic pair production and
photon splitting will attenuate spectra to hard X-ray energies,
suppressing signals in the {\it{Fermi}}-LAT band. The resonant Compton spectra
are strongly polarized, suggesting that hard X-ray polarimetry
instruments such as X-Calibur, or a future Compton telescope, can prove
central to constraining model geometry and physics.
\end{abstract}

\keywords{radiation mechanisms: non-thermal --- magnetic fields --- stars:
neutron --- pulsars: general --- X rays: theory}

 \accepted{}

\section{INTRODUCTION}
 \label{sec:intro}
It is now generally accepted that there exists a class of young isolated neutron stars
characterized by their strong inferred dipolar magnetic field, typically up to three
orders of magnitude larger than for canonical radio pulsars and above the quantum critical
or Schwinger field \teq{B_{\rm cr} = m_e^2 c^3/e \hbar \approx 4.41 \times 10^{13}} Gauss,
at which the cyclotron energy of the electron equals its rest mass energy. These
\emph{magnetars}, which include nearly 30 soft gamma-ray repeaters (SGRs) and anomalous
X-ray pulsars (AXPs), evince long pulse periods \teq{P\sim 2-12} s and high
period derivatives \teq{{\dot P}} for their persistent X-ray pulsations, from which high
surface polar fields $B_{\rm p} \sim 10^{13}-10^{15}$ G and short characteristic (i.e.
magnetic dipole spin-down) ages \teq{\tauEM = P/(2\dot{P})} are inferred (e.g. Vasisht \&
Gotthelf 1997).  
{The timing ephemerides permit estimates of
\teq{B_p \sim 6.4\times 10^{19} \sqrt{P \dot{P}}} in the vacuum orthogonal rotator case (Shapiro \& Teukolsky 1983),
though inferred field strengths are also impacted by plasma loading of the 
magnetosphere, where currents supply Poynting flux (e.g., see Harding, Contopoulos \& Kazanas 1999). }
{Locally, fields higher than \teq{10^{15}}G are possible, as is suggested by a proton 
cyclotron line interpretation of the
13 keV absorption feature in the NuSTAR spectrum of a burst from 1E 1048.1-5937 (An et al. 2014).
These high fields may masquerade as substantial non-dipolar
(perhaps toroidal) perturbations.   A comprehensive list of associations,
timing, and spectral properties of magnetars
may be found in the McGill magnetar catalog (Olausen \& Kaspi
2014) and its contemporaneous on-line version.}\footnote{\tt
http://www.physics.mcgill.ca/\~{}pulsar/magnetar/main.html}

The bolometric luminosities of magnetars predominantly come from the soft and hard X-ray
bands, with mostly thermal surface
emission between 0.2 and 5 keV, and non-thermal magnetospheric emission at higher energies
that exhibits approximately power-law spectra.
Most magnetars are radio-quiet or dim, but not all: ephemeral, transient radio activity has
now been observed from four such sources (e.g. see Rea et al. 2012; Pennucci et al. 2015,
and references therein).
For the majority of magnetars, their persistent X-ray emission is extremely bright,
being commensurate with a large equivalent isotropic luminosity (i.e. that integrated over
all solid angles) $L_{\mbox{x}} \sim 10^{35}$ erg s${}^{-1}$ (e.g. Tiengo et al. 2002; 
Vigan\'{o} et al. 2013). In
most cases, this exceeds the rotational energy loss rate (spin-down luminosity) \teq{-
\Edotrot = 4 \pi^2 I \dot{P}/P^3} by one or two orders of magnitude, assuming that the
equations of state and moments of inertia \teq{I} for magnetars are not substantially
different from those invoked for rotation-powered pulsars, i.e. \teq{I\sim 10^{45} \mbox{
g cm}^{2}}. Accordingly, sources of power for magnetar activity as alternatives to rotation
were first proposed by Duncan \& Thompson (1992) for SGRs and later for AXPs by Thompson
\& Duncan (1996); they envisaged structural reconfigurations of magnetic fields in the
crustal and surface regions.  The picture of dynamic structural evolution is supported by
the fact that many AXPs and SGRs exhibit active episodes of transient flares followed by
recovery phases lasting months (e.g. Kaspi et al. 2003; Rea \& Esposito 2011; Lin et al.
2012).  These are presumed to be associated with violent rearrangements of currents and
fields, and subsequent dissipation of magnetic energy from field lines threading the
neutron star crust. 
{We remark that there are a handful of prominent exceptions to this highly super-spin-down
luminosity character, including SGR J1550-5408, SGR 1627-41, SGR J0501+4516 and 
SGR J1935+2154, all in the range of \teq{L_{\mbox{x}} \lesssim 3\vert\Edotrot\vert}.
Understanding why some magnetars are brighter in quiescence than others is clearly an 
important issue.}

Magnetars are complex objects and cannot be completely isolated from the conventional
pulsar population based on the prescription of emission energetics, \teq{P-\dot{P}}
derived dipole fields and \teq{\tauEM} ages: a clean magnetar-pulsar dichotomy is not
sustainable.  A classic exemplar is the ``low-field" magnetar SGR 0418+5729, with a field
estimate of \teq{B_p \sim 1.2 \times 10^{13}} Gauss (Rea et al. 2013) and sporadic
outburst activity that is the hallmark of magnetars. Yet it too, may present a challenge
to stereotypes in that there is suggestive evidence of a variable proton cyclotron
absorption feature (Tiengo et al. 2013), implying that it possesses local magnetic field
components of \teq{10^{14}-10^{15}}G.   Then there are tens of rotation-powered high-field
pulsars such as PSR B1509-58  (Abdo et al. 2010) and PSR J1846-0258 
(Kuiper, Hermsen \& Dekker 2018), whose polar fields
approach or exceed \teq{10^{13}}G, the latter of which has been shown to have
magnetar-like outbursts {(Gavriil et al. 2008)}, as has PSR J1119-6127 (G{\"o}{\u
g}{\"u}{\c s} et al. 2016). 
{Interestingly following their outbursts, the quiescent power-law spectrum 
of J1846-0258 developed a transient thermal-like component  (Kuiper \& Hermsen 2009), while the 
quiescent thermal spectrum of J1119-6127 at temperature \teq{\sim 0.2} keV heated to the higher value 
of \teq{\sim 1}keV, with an additional transient power-law component (Archibald et al. 2016), 
so that each had a transient magnetar-like spectrum. Also, 
the radio pulsations of J1119-6127 were observed to turn off during X-ray bursts 
following the initial outburst activity (Archibald et al. 2017). Some of the high-field pulsars 
are radio-loud and have GeV pulsations, but many have no detected radio pulsations (Kuiper \& Hermsen 2015).  
Their rotation-powered spectra are distinctly different from those of magnetars, possessing 
hard non-thermal X-ray/gamma-ray components (with the exception of PSR J1119-6127) 
with \teq{\nu F_{\nu}} spectral peaks around 1--10 MeV and emission extending to 0.1--1 GeV 
(in the case of B1509-58 and J1846-0258, Kuiper, Hermsen \& Dekker 2018). 
It is not clear that the spectra of magnetars and high-field rotation-powered pulsars 
have the same origin. In fact, a model for high-field pulsar emission as synchrotron radiation 
from the outer magnetosphere can adequately account for their spectral properties and 
light curves (Harding \& Kalapotharakos 2017). It is possible that their spectra only 
resemble those of magnetars following magnetar-like outbursts.
Yet, it should be noted that the hard X-ray pulse profiles of PSRs B1509-58 and J1846-0258 
are quite broad, more reminiscent of those of magnetars than the narrow profiles of young rotation-powered pulsars. Moreover,
low surface temperatures in high-B pulsars could be a significant factor in suppressing
resonant Compton upscattering signals in hard X-rays, and conversely, facilitating them in heated or activated magnetar-like phases.} 
One concludes that the boundary between pulsars and magnetars by various measures is not crisp, but blurred. For a
comprehensive list of young high-field pulsars in proximity to
the magnetar domain, the reader is referred to the ATNF pulsar catalog (Manchester et al.
2005), and in particular its current on-line version.\footnote{{\tt
http://www.atnf.csiro.au/people/pulsar/psrcat/}}

The persistent soft X-ray emission is typically fit with an absorbed blackbody of
temperature kT \teq{\sim 0.5} keV plus a power-law component \teq{dN/dE \propto E^{-
\Gamma}},  at suprathermal energies that is usually fairly steep, with index \teq{\Gamma_s
\sim 1.5-4} (e.g., Perna et al. 2001; {Vigan\'{o} et al. 2013}). Hard X-ray (20--150 keV) tails
have been observed for about nine magnetars by INTEGRAL along
with RXTE, XMM-Newton, ASCA and NuSTAR data in several AXPs (Kuiper et al. 2004, 2006; den
Hartog et al. 2008a,b; Vogel et al. 2014) and SGRs (Mereghetti et al. 2005; Molkov et al.
2005; G\"otz et al. 2006; Enoto et al. 2010; 2017). {For four magnetars, they have also been detected 
by {\it Fermi}-GBM (ter Beek 2012).}  The spectra from these high-energy tails
extend up to 150 keV, and are typically much flatter than the soft X-ray non-thermal
components, possessing power law indices in the range \teq{\Gamma_h \sim 0.7 - 1.5}. 
Moreover, the pulsed portions of the hard X-ray components, with indices
\teq{\Gamma_h^{p} \sim 0.4 - 0.8}, are typically even flatter than the phased-averaged
spectra, and the pulsed fractions approach $100\%$ at higher energies (e.g. den Hartog et
al. 2008a,b). Pulse profiles for all magnetars are ubiquitously broad with a
shape that is single or double-peaked per cycle, contrasting the narrow
peaks typically found in canonical radio and gamma-ray pulsars. The hard tails are also putatively
constrained by upper limits from non-contemporaneous observations by the COMPTEL
instrument on the \emph{Compton Gamma-Ray Observatory} (e.g. {Kuiper et al. 2006;} den Hartog et al. 2008a,b),
indicating sharp spectral turnovers at energies 200--500 keV.  This feature could be as
low as \teq{\sim 130}keV, as has been suggested (Wang et al. 2014) by an analysis of nine
years of INTEGRAL/IBIS data for the bright AXP 4U 0142+61. The need for such a spectral
turnover is reinforced above 100 MeV by upper limits in \emph{Fermi}-LAT data for around
20 magnetars (Abdo et al. 2010, Li et al. 2017).   We mention that Wu et al. (2013)
reported discovery of pulsed gamma-ray emission above 200 MeV from AXP 1E2259+586 with a
targeted search of public 4-year {\it Fermi}-LAT data archive.  This has not been
confirmed by the analysis of Li et al. (2017) that employs six years of {\it Fermi}-LAT
data, and which identifies a contaminating extended gamma-ray source detected around
1E2259+586 that is probably the GeV counterpart of SNR CTB 109.

Magnetic inverse Compton scattering of thermal
atmospheric soft X-ray seed photons by relativistic electrons is
expected to be extremely efficient in highly-magnetized pulsars,
and thus is a prime candidate for generating the hard X-ray tails.
This is because the scattering process is resonant at
the electron cyclotron frequency {\teq{\omegaB = eB/mc}} 
and its harmonics, so that there the cross section in the
electron rest frame exceeds the classical Thomson value of \teq{\sigt = 8\pi r_0^2/3
\approx 6.65 \times 10^{-25}}cm$^2$ by two or more orders of magnitude (e.g. Daugherty \&
Harding 1986; Gonthier et al. 2000).  The non-thermal {\it soft} X-ray components of many
magnetars have also been modeled using resonant Comptonization by mildly-relativistic
electrons to effect the repeated upscattering of photons (Lyutikov \& Gavriil 2006; Rea at
al. 2008; Nobili et al. 2008a,b). The Lyutikov \& Gavriil model uses a non-relativistic magnetic Thomson cross
section (Herold 1979), neglecting electron recoil, and the fits are of comparable accuracy
to empirical blackbody plus power-law prescriptions.  However, for the hard X-ray tails,
such Comptonization models of repeated scattering by mildly relativistic electrons may
have difficulties reproducing the flat spectra due to the ease of photon escape from the
larger interaction volumes. Provided there is a source of ultra-relativistic electrons
with Lorentz factor $\gamma_e \gg 1$, single inverse Compton scattering events can readily
produce the general character of hard X-ray tails (Baring \& Harding 2007; Fern\'andez \& Thompson 2007).

Previous magnetic inverse Compton scattering studies in the context of neutron star models
of gamma-ray bursts (e.g., Dermer 1989, 1990; Baring 1994) computed upscattering spectra
and electron cooling rates and in the non-relativistic magnetic Thomson limit, extending collision integral
formalism for non-magnetic Compton scattering that was developed by Ho \& Epstein (1989).
In the context of magnetars, Beloborodov (2013a) developed a resonant Thomson 
upscattering model for their hard X-ray tails.  Such
analyses do not suffice for modeling magnetars' hard X-ray signals at low altitudes, where
supercritical fields arise, and thereby violate energy conservation and generate too many 
high energy photons.  In contrast,
Baring and Harding (2007) computed inverse Compton spectra fully in the QED domain,
specifically for uniform magnetic fields, producing output photon spectra considerably
flatter
{than are observed in the pertinent magnetars}, 
and violating {\it Fermi}-LAT and COMPTEL bounds
when \teq{\gamma_e\gtrsim 50}. In particular, they discerned that kinematic constraints
correlating the directions and energies of upscattered photons yielded Doppler boosting
and blueshifting along the local magnetic field direction.  Therefore, the strong angular
dependence of spectra computed for the uniform field case extends also to more complex
magnetospheric field configurations. Consequently, emergent inverse Compton spectra in
more complete models of hard X-ray tails will depend critically on an observer's
perspective and the locale of resonant scattering, both of which vary with the rotational
phase of a magnetar.

The principal task of this paper is to extend the analysis of Baring \& Harding (2007) to
encapsulate non-uniform field geometries and model the hard tails of high-field pulsars
and magnetars. Spectra are herein generated for an array of observer perspectives and
magnetic inclination angles \teq{\alpha}, and will serve as a basis for future
calculations that will treat Compton cooling of electrons self-consistently,
{and will also explore reheating of the surface due to bombardment by these electrons.}
We presume electrons with Lorentz factors $\gamma_e \gg 1$ are confined
to move along field lines, an approximation that is generally accurate for high-field
pulsars due to rapid cyclotron/synchrotron cooling of components of electron momenta
perpendicular to $\boldsymbol{B}$ on very short timescales of
\teq{10^{-20}}--\teq{10^{-16}} seconds. We specialize to scatterings that also leave the
electron in the zeroth Landau level, as noted in Gonthier et al. (2000), and adopted by 
Baring \& Harding (2007) and Baring, Wadiasingh \& Gonthier (2011), hereafter BWG11. This
is appropriate when resonant scattering at the cyclotron fundamental cools electrons
efficiently.   The developments use a fully relativistic, spin-dependent QED cross
section that employs Sokolov \& Ternov (1968, hereafter ST) eigenstates of the Dirac
equation in a uniform magnetic field, the appropriate choice for incorporating
spin-dependent cyclotron widths into scattering cross sections. The full details of the ST
cross section formalism is described in Gonthier et al. (2014); they
supplant the spin-averaged Johnson \& Lippmann (1949, hereafter  JL) cross section
formalism found in previous treatments (e.g. Herold 1979; Daugherty \& Harding 1986;
Bussard, Alexander \& M\'esz\'aros 1986).

Resonant Compton upscattering spectra are computed for integrations over curved electron paths
tied to closed magnetic field lines.  The observer perspective relative to the
instantaneous magnetic axis is fixed.  For the present analyses, we consider uncooled electrons,
so as to isolate the principal character of the spectral emissivities, and facilitate
basic understanding. This restriction will be relinquished in future work that will
incorporate the electron cooling rate calculation as a function of altitude and
colatitude, as found in BWG11.  
The photon production rate computations presented here will thus serve as a
foundation for future phase-resolved spectroscopic models of hard tail emission in
magnetars. Section~\ref{sec:formalism} begins with the kinematic formulae central to
resonant Compton upscattering, and then defines the photon production rate formalism in
general magnetic field morphologies. Section~\ref{sec:geometry} specializes the general
formalism of Section~\ref{sec:formalism} to dipole field geometry, specific observer
perspectives and explores the occultation of emission regions.

The results presented in Section~\ref{sec:spectra} survey the parameter space for emergent
spectra for various observer perspectives and upscattering regions of the magnetosphere. 
For scatterings involving electrons transiting single field lines, resonant emission is very hard, and
the maximum resonant energy varies substantially with pulse phase for
different observing perspectives. The spectrum resulting from such passages by
monoenergetic electrons along dipole field loops resembles an \teq{\erg^{1/2}} form due to
contributions of emission at locales proximate to field line tangents that point to an
observer. This spectrum steepens somewhat when adding up over field line azimuthal angles,
most of which preclude such select tangent viewing geometry, and the resonant spectrum
then assumes an approximately \teq{\sim \erg^0} form, which is reminiscent of full
solid-angle-integrated emission results in uniform fields presented in Baring \& Harding
(2007). These spectra from toroidal surfaces comprising dipolar field lines are expected to steepen
further when integrations over maximum surface altitudes are performed and entire emission
volumes are treated. Pulse phase flux maps for different
observer perspectives are displayed for a variety of angles \teq{\alpha} between the
magnetic and rotation axes, highlighting the prospect of using these to constrain
\teq{\alpha} and the typical altitude of hard X-ray tail emission.   A brief illustration
of spectral results in the magnetic Thomson regime is also offered, revealing how they
are not suitable for emission regions very near the star. Section~\ref{sec:spectra} also
touches upon polarization of the signals, and establishes that polarization degrees in
excess of 50\% can be obtained for single field loop cases at the highest resonant
emission energies.  The prospect of using polarization information to more tightly
constrain magnetar geometry parameters, for example the magnetic inclination angle
\teq{\alpha}, motivates the science case for developing hard X-ray polarimeters.
Section~\ref{sec:discuss} draws together various interpretative elements including how
resonant cooling can limit the Lorentz factors of electrons accelerated in magnetospheric
electric fields, and the potential impact that attenuation mechanisms could have on the
emergent spectra.

\section{RESONANT SCATTERING FORMALISM}
 \label{sec:formalism}

The inverse Compton scattering models here assume that the
ultra-relativistic electrons move along field lines, since magnetars are
slow rotators and velocity drifts are small. The detection of emission
out to around \teq{\sim 150} keV does not guarantee the presence of
\teq{\gamma_e \gg 1} electrons.  Yet, we have previously shown in BWG11
that resonant Compton cooling does not operate efficiently for mildly
relativistic electrons except for subcritical fields.  In contrast, such
resonant cooling does become extremely efficient in supercritical
fields, for Lorentz factors as high as \teq{10 - 10^4}, and this serves
as the basic impetus for considering resonant Compton upscattering
scenarios for the generation of hard X-ray tails. Moreover, the
electrons should occupy the lowest Landau level (transverse quantum
state) due to the rapid cyclotron/synchrotron cooling of components of
electron energy perpendicular to \teq{\boldsymbol{B}}.  This restriction
to the zeroth Landau state simplifies the scattering cross section
profoundly. In addition, the assumption that \teq{\gamma_e \gg 1} is
also highly expedient analytically, yielding a relatively simple form of
the relativistic differential cross section for Compton scattering in
strong fields, since the incoming photon angle in the electron rest
frame (henceforth ERF) is approximately zero for nearly all incoming
photon angles in the magnetospheric or observer frame (OF). Then, the
relativistic cross sections in either Sokolov \& Ternov (Gonthier et al.
2014) or Johnson \& Lippmann formalisms (Gonthier et al. 2000, Daugherty
\& Harding 1986) for the eigenstates, have only one resonance at the
cyclotron fundamental.

\subsection{Upscattering Kinematics}
 \label{sec:kinematics}

To set the scene for the exposition on collision integral calculations
of photon spectra, it is instructive to first summarize key kinematic
definitions. Both the Lorentz transformation from the observer's or
laboratory frame to the electron rest frame, and the scattering
kinematics in the ERF, are central to determining the character of
resonant Compton upscattering spectra and the cooling rates. 
The conventions adopted in this paper are now stated; they
follow those used in Baring \& Harding (2007) and BWG11. The electron
velocity vector in the OF is \teq{\boldsymbol{\beta}_e}, which is parallel
or anti-parallel to \teq{\boldsymbol{B}} due to the exclusive 
occupation of the ground  Landau state.  The dimensionless photon energies (scaled
by \teq{m_e c^2}) in the OF are \teq{\erg_{\rm i,f}} where the subscripts \teq{i,f} denote
pre- and post-scattering quantities respectively. The OF angles
\teq{\Theta_{\rm i,f}} for these photons are defined to possess
{\it zero angles anti-parallel to the electron velocity}, \teq{-\boldsymbol{\beta}_e}, 
along the field direction, corresponding to head-on collisions.  With this
choice, for \teq{\boldsymbol{k}_{\rm i,f}} being the photon
momentum vectors, define the photon angle cosines
\begin{equation}
   \mu_{i,f}\;\equiv\;
   \cos\Theta_{i,f}\; =\; - \frac{ \boldsymbol{\beta}_e \boldsymbol{\cdot} \boldsymbol{k}_{\rm i,j}  }{ 
   | \boldsymbol{\beta}_e | | \boldsymbol{k}_{\rm i,j} |  }  \quad ,
 \label{eq:OF_kinematics}
\end{equation}
The relative sense of  \teq{\boldsymbol{\beta}_e} 
and \teq{\boldsymbol{B}} is irrelevant to the scattering, but is 
relevant later on when spectra directed along a given line of sight to an observer 
are considered.  Boosting by \teq{\boldsymbol{\beta}_e} 
into the ERF then yields pre- and post-scattering photon energies of
\teq{\omega_i} and \teq{\omega_f} (also scaled
by \teq{m_e c^2}), respectively, with corresponding
angles with respect to \teq{-\boldsymbol{\beta}_e} of \teq{\theta_i} and
\teq{\theta_f} in the ERF.  The relations governing this Lorentz transformation 
and associated angle aberration are
\begin{equation}
   \omega_{i,f} \; =\; \gamma_e\erg_{i,f} (1+\beta_e\cos\Theta_{i,f})
   \quad \hbox{and}\quad
   \cos\theta_{i,f} \; =\; \dover{\cos\Theta_{i,f} + \beta_e}{
      1 + \beta_e \cos\Theta_{i,f}}\quad ,
 \label{eq:Lorentz_transform}\\[-5.5pt]
\end{equation}
and are illustrated in Figure~1 of BWG11.  The inverse transformation relations 
are obtained from Eq.~(\ref{eq:Lorentz_transform}) by \teq{\beta_e\to -\beta_e} along 
with definitional substitutions \teq{\theta_{i,f}\leftrightarrow\Theta_{i,f}}
and \teq{\omega_{i,f} \leftrightarrow \erg_{i,f}}. 
It is evident from the angle aberration formula that \teq{\theta_i \approx 0} 
when \teq{\gamma_e \gg 1}, except for the small fraction of the
scattering phase space when \teq{\cos\Theta_i\approx -\beta_e}.  In
such circumstances, the magnetic Compton scattering cross section
exhibits just a prominent resonance at the cyclotron fundamental (e.g.
Daugherty \& Harding 1986, Gonthier et al. 2000),
{i.e. when \teq{\omega_i \to \hbar\omegaB/(m_ec^2) = B/B_{\rm cr}}.
The origin of the resonance is that scattering process becomes essentially 
first order in \teq{\fsc = e^2/\hbar c}, being a cyclotron absorption event 
promptly followed by cyclotronic decay of the virtual electron 
from the first excited Landau level.} 
For the rest of this paper, the
approximation that electrons occupy the zeroth Landau state pre- and
post-scattering is made.

The kinematic scattering relations, derived from energy-momentum conservation, 
differ from the classic non-magnetic Compton scattering formula. 
Particle momenta perpendicular to the local field direction are not 
conserved in QED processes due to the lack of invariance of the 
Dirac Hamiltonian under spatial translations transverse to \teq{\boldsymbol{B}}.  
This departure from symmetry modifies the kinematics of electron-photon 
interactions. General formulae for the ERF relationships among \teq{\omega_{\rm i,f}} 
and \teq{\theta_{\rm i,f}} in magnetic Compton scatterings are found in a multitude 
of previous works, for example, Herold (1979) and Daugherty \& Harding (1986).
In the expedient scenario of ground state-to-ground state transitions and 
\teq{\theta_i \approx 0} cases that are adopted in the paper, 
the pre- and post-scattering energies are related by
\begin{equation}
   \omega_f\; =\; \omega' (\omega_i ,\,\theta_f)\;\equiv\;
     \dover{2\omega_i\, \varrho}{1+\sqrt{1-2\omega_i \varrho^2\sin^2\theta_f}}\quad , \quad
     \varrho\; =\; \dover{1}{1+\omega_i (1-\cos\theta_f)}
 \label{eq:reson_kinematics}
\end{equation}
where \teq{\varrho} is the ratio \teq{\omega_f/\omega_i} that one would ascribe to the
non-magnetic Compton scattering formula (which in fact does result when \teq{\omega_i
\varrho^2\sin^2\theta_f\ll 1}). Algebraically rearranging Eq.~(\ref{eq:reson_kinematics}) 
results in a useful alternative form:
\begin{equation}
     (\omega_f)^2\sin^2\theta_f  - 2\omega_i\omega_f (1-\cos\theta_f)
   + 2(\omega_i - \omega_f) \; =\; 0 \quad .
 \label{eq:res_kinematics_alt}
\end{equation}
A direct algebraic inversion of this yields
\begin{equation}
   \omega_i \; =\;  \dover{ \omega_f ( 2 - \omega_f + \omega_f \cos^2 \theta_f)}{2(1-\omega_f+\omega_f\cos \theta_f)}\quad .
 \label{eq:res_kinematics_inversion}
\end{equation}
This particular version assists in identifying the geometric observing 
conditions for the cyclotron resonance to be selected in a scattering event:
since both \teq{\omega_f} and \teq{\theta_f} in the ERF depend on 
the final photon energy \teq{\erg_f} and angle \teq{\Theta_f} in the observer 
frame via Eq.~(\ref{eq:Lorentz_transform}), then so also does \teq{\omega_i} implicitly.
Note that all three of these identities are purely kinematic in nature,
and must be satisfied by any spin-eigenstate formalism for the electron 
wavefunction that is employed to determine the scattering 
cross section.  From the third form, by inspection of the denominator,
the positive energy \teq{\omega_i>0} restriction yields the immediate consequence
\begin{equation}
   0 \; <\; \omega_f (1-\cos \theta_f) \; \leq \; 1 
   \qquad \mbox{for} \qquad | \cos \theta_f | \; <\; 1 \quad .
 \label{eq:res_kinematics_restriction1}
\end{equation}
However, as will shortly be seen, this is always satisfied.  So also is \teq{\omega_i \geq \omega_f},
which is simply deduced from Eq.~(\ref{eq:res_kinematics_inversion}).

To connect the kinematics to the observer's frame, one convolves one of the above ERF identities with the boost 
relations in Eq.~(\ref{eq:Lorentz_transform}).  The angle cosine limits \teq{\mu_{\pm}} of the incoming 
soft photon angles are governed by the magnetospheric geometric locale of a scattering, and yield 
constraints on the accessible ERF values for the incoming photon energy via
\teq{ \gamma_e \erg_s (1+ \beta_e \mu_-) \leq \omega_i \le \gamma_e \erg_s (1+ \beta_e \mu_+)}.  This
inequality helps define the range of soft photon energies that permits access to the cyclotron resonance.
At any scattering locale, due to the collimation of soft photon momenta within a cone with a radial vector as its axis, 
the normalized angular distribution function \teq{f(\mu_i)} about the local field vector \teq{\boldsymbol{B}} is anisotropic; 
forms for it can be found in BWG11, but are redefined for the context of this paper; see Appendix~A.1.
For the outgoing photons, an observer can select a final photon energy \teq{\erg_f} and a local 
scattering angle \teq{\Theta_f} in the OF for the instantaneous orientation of the magnetospheric configuration.
These then fix the value of \teq{\omega_f} in the ERF via the Lorentz transformation 
\teq{\omega_f = \gamma_e \erg_f (1+ \beta_e \cos \Theta_{f})}, and the value of the ERF scattering angle
\teq{\theta_f} using the aberration formula in Eq.~(\ref{eq:Lorentz_transform}).  
Inserting these into Eq.~(\ref{eq:res_kinematics_inversion}) yields 
a relation that defines at what energies (if any) the observer can detect  
upscattered photons that sample the cross section resonance.  One also has 
\teq{\omega_f = \erg_f/ [\gamma_e (1 - \beta_e \cos \theta_f)]}, which can be combined with the inequality
in Eq.~(\ref{eq:res_kinematics_restriction1}) to yield the bound \teq{\cos\theta_f < (\gamma_e - \erg_f)/(\gamma_e\beta_e - \erg_f)},
which is always satisfied, provided \teq{\erg_f < \gamma_e\beta_e}.  Energy conservation then establishes 
the verity of this bound and therefore also that in Eq.~(\ref{eq:res_kinematics_restriction1}).

\subsection{Scattered Spectra and Directed Emission Formalism}
 \label{sec:DEformalism}
To form upscattering spectra from the magnetosphere for thermal soft
photons, a collision integral calculation is appropriate as a prelude to
more sophisticated and complete Monte Carlo simulations that 
will incorporate fully self-consistent cooling and acceleration. Throughout 
this paper we assume uncooled electrons at fixed Lorentz
factor \teq{\gamma_e} and fixed number density \teq{n_e}, i.e. their
distribution function is \teq{n_e \delta (\gamma-\gamma_e)}. 
Note that anti-symmetric pulse
profiles of most AXP/SGRs suggest distributed upscattering
locales in the magnetosphere and non-uniform spatial distributions of
\teq{n_e}; treatment of these will be deferred to future studies. 
A generic formulation in the OF of the
photon production rate in terms of ERF quantities and kinematics may be
found Eqs.~(A7)--(A9) of Ho and Epstein (1989). This development is
readily applied to fully relativistic resonant Compton QED cross
sections and kinematics, as we have previously presented in Baring \&
Harding (2007) and BWG11, and was explored much earlier in 
Dermer (1990).  The differential photon production rate
\teq{dN_{\gamma}/(dt\,d\erg_f ) }, with \teq{\mu_i = \cos\Theta_i} and
\teq{\mu_f=\cos\Theta_f} for compactness, is then
\begin{equation}
   \dover{dN_{\gamma}}{dt\, d\erg_f} \; =\;  n_e n_s\, c
   \int_{\mu_l}^{\mu_u}d\mu_f   \int_{\mu_-}^{\mu_+}d\mu_i    \, f(\mu_i)\,
   \delta \big\lbrack\omega_f -\omega'(\omega_i,\,\theta_f)\, \bigl\rbrack\,
   \dover{1+\beta_e\mu_i}{\gamma_e(1+\beta_e\mu_f)}\,
   \dover{d\sigma}{d(\cos\theta_f) } \,.
 \label{eq:scatt_spec}
\end{equation}
Herein, the angular distribution \teq{f(\mu_i )} of soft photons 
has a normalization reflecting the decline of intensity with distance from the stellar 
surface, details of which will be addressed shortly.
Note that the angle conventions of Ho and Epstein (1989) differ by \teq{\pi} 
(equivalent to \teq{\beta_e\to -\beta_e}) from those presented here and in 
Eq.~(\ref{eq:OF_kinematics}). As in all scattering collisional integrals, 
a relative velocity factor \teq{c(1+\beta_e\mu_i)} in Eq.~(\ref{eq:scatt_spec}) 
between the two species is present.   This result presents the spectrum integrated 
over all scattering angles, a representative indication of the net spectral output.  
When the resonant condition \teq{\omega_i = B} is imposed, this spectral
form exhibits an approximate one-to-one correspondence between upscattered 
energy \teq{\erg_f} and scattering angle \teq{\Theta_f} relative to the field
(Dermer 1990; Baring \& Harding 2007).  A similar formulation using
the head-on scattering restriction is offered in Appendix A of Dermer \& Schlickiser (1993),
for the case of the non-magnetic inverse Compton process in blazars; it too 
exhibits a strong (though different) coupling between final energy \teq{\erg_f} and scattering angle \teq{\Theta_f}.

In general, to connect with observations, this is 
not the optimal construction, since particular scattering angles are selected 
by viewing perspectives and scattering locales.
The above formulation can be readily modified to derive the spectrum of 
emitted radiation directed towards an observer. Electrons are assumed to 
follow some path \teq{S} in the magnetosphere, which for a slow rotator 
will be presumed parallel or anti-parallel to \teq{\boldsymbol{B}}; 
eventually this will be specialized to dipole geometry. Fixing the observer viewing angle 
with respect to the neutron star dipolar axis,
one can represent the angles of the scattered photon in terms of the 
polar angle \teq{\thetaB} and azimuthal angle \teq{\phiB}
about the magnetic field direction at the point of scattering.
Since the scattering cross section employed here is independent of \teq{\phiB} when 
the incident photons are parallel to \teq{\boldsymbol{B}}, the azimuthal angles can be integrated trivially.
The post-scattering solid angle element \teq{d\mu_f\,d\phi_f} is therefore restricted by delta functions in 
two dimensions in order to just encompass the ray to the observer at infinity.  The spectral integrals therefore 
include the factor
\begin{equation}
   \delta (\mu_f - \cos\thetaB)\, \delta (\phi_f - \phiB )\, d\mu_f\, d\phi_f
   \;\to\; \dover{1}{2\pi}\, \delta (\mu_f - \muB)\, d\mu_f\quad ,
 \label{eq:solid_angle_fac}
\end{equation}
where \teq{\muB = \cos\thetaB}.  We have introduced a factor of \teq{1/(2\pi)}, a convention choice, 
for the azimuthal angle delta function, to cancel the factor already integrated in the cross section 
definition \teq{d\sigma/d\cos(\theta_f)}: the azimuthal independence of the differential cross section
yields the operational correspondence
\begin{equation}
   \int d\Omega \,\dover{d\sigma}{d\Omega} 
   \; \equiv\; \int_{\mu_-}^{\mu_+} d\mu_f \int_0^{2\pi} d\phi_f \,\dover{d\sigma}{d\Omega} 
   \;\to\; \int_{\mu_-}^{\mu_+} d\mu_f \,\dover{d\sigma}{d\cos\theta_f} 
 \label{eq:}
\end{equation}
Observe that the Jacobian for the solid angle transformation 
between coordinate angles defined with respect to the local \teq{\boldsymbol{B}} direction and those oriented with respect to 
the line of sight to an observer is unity.

We integrate over angles and energy of a separable soft photon distribution, 
defined as the incoming differential photon number density
\begin{equation}
   n_\gamma (\erg_s, \mu_i) \; =\; n_\gamma (\erg_s) f(\mu_i)
 \label{eq:edit_separate}
\end{equation}
with \teq{\mu_i = \left[ \omega_i/(\gamma_e\erg_s) -1 \right]/\beta_e}.
For monoenergetic electrons, \teq{n_e(\gamma) = n_e \delta (\gamma - \gamma_e)}, the 
upscattered spectrum integrated along the electron path for an arbitrary soft photon distribution, 
directed at some distant observer, is
\begin{eqnarray}
   \dover{dn_{\gamma}}{dt\, d\erg_f} & = & \frac{n_e c}{2 \pi {\cal S}} \int_S ds  
           \int_0^\infty d\erg_s \, n_\gamma(\erg_s) \int_{\mu_l}^{\mu_u}  d\mu_f  \, 
           \delta(\mu_f - \muB ) \int_{\mu_-}^{\mu_+}d\mu_i  \, f(\mu_i) \nonumber\\[-5.5pt]
 \label{eq:scatt_spec2}\\[-5.5pt]
   && \qquad \times \; \delta \big\lbrack\omega_f -\omega'(\omega_i,\,\theta_f)\, 
             \bigl\rbrack\, \dover{1+\beta_e\mu_i}{\gamma_e(1+\beta_e\mu_f)}\, \dover{d\sigma}{d(\cos\theta_f) } \quad .\nonumber
\end{eqnarray}
This form can readily be adapted to treat non-uniform electron distributions that are not mono-energetic, such as are needed 
for more complete studies of radiation-reaction limited resonant Compton scattering in magnetars.  The spectra are normalized by the 
path length \teq{{\cal S}}, which for closed field-line loops in the inner magnetosphere is specified 
in Eq.~(\ref{eq:calS_eval}) below.  This normalization convention will be relinquished 
below when spectra evaluated for surface and volume integrations are depicted.
The path length variable \teq{s} can be chosen to be dimensionless (a convenience)
as long as \teq{{\cal S}} possesses the same dimensions. Hereafter, we will specialize 
this result to \teq{\mu_l = -1} and \teq{\mu_h=1} corresponding to the maximal permitted
range of final OF scattering angles. 
There are two equivalent methods for evaluating the delta functions appearing in Eq.~(\ref{eq:scatt_spec2}). 
One protocol expresses the \teq{\mu_i} and \teq{\mu_f} integrations as integrals over 
\teq{\omega_i} and \teq{\omega_f}, respectively, and is the more illustrative in making a connection with the 
previous work on the uniform field case in Baring \& Harding (2007).  Here, we pursue another development 
that is algebraically simpler, and more useful for directed emission spectra.

The angular \teq{\mu_i} integration is rewritten as an integration over \teq{\omega_i}, with 
\teq{d\mu_i = -\gamma_e \erg_s \beta_e d\omega_i}. The limits on the \teq{\omega_i} 
integration are readily obtained from the Lorentz transformations 
Eq.~(\ref{eq:Lorentz_transform}), i.e. \teq{\gamma_e(1+\beta_e\mu_-)\erg_s \leq \omega_i \leq \gamma_e(1+\beta_e\mu_+)\erg_s}. 
This manipulation leads to the correspondance
\begin{equation}
   \int_{\mu_-}^{\mu_+} (1+\beta_e\mu_i)\, d\mu_i \;\to\; 
   \dover{1}{\gamma_e^2 \beta_e\erg_s^2} \int_{\omega_-}^{\omega_+} \omega_i\, d\omega_i
   \quad ,\quad
   \omega_{\pm} \; =\; \gamma_e(1+\beta_e\mu_\pm)\erg_s\quad .
 \label{eq:mui_omegai_correspond}
\end{equation}
We now interchange the order of integration of \teq{\omega_i} and \teq{\erg_s},
and use the identity \teq{\left[\gamma_e (1+\beta_e \mu_f)\right]^{-1} = \gamma_e (1- \beta_e \cos \theta_f)} 
to derive
\begin{eqnarray}
   \dover{dn_{\gamma}}{dt\, d\erg_f} & = & \dover{n_e c}{2 \pi {\cal S}} \dover{1}{\gamma_e \beta_e} 
   \int_{S} ds   \int_{-1}^{1}  d\mu_f  \, \delta(\mu_f - \muB)    
   \int_{0}^{\infty}d\omega_i  \, \delta \big\lbrack\omega_f -\omega'(\omega_i,\,\theta_f)\, \bigl\rbrack \nonumber\\[-5.5pt]
 \label{eq:scatt_spec4}\\[-5.5pt]
   && \qquad \times \; \omega_i (1- \beta_e \cos \theta_f) \, \dover{d\sigma}{d(\cos\theta_f) }  
      \int_{\erg_-}^{\erg_+} \dover{d \erg_s}{\erg_s^2} \, n_\gamma(\erg_s) \, f(\mu_i) \nonumber
\end{eqnarray}
from Eq.~(\ref{eq:scatt_spec2}), 
where \teq{\mu_i = \left[ \omega_i/(\gamma_e \erg_s) -1 \right] /\beta_e}.  With this step, 
we have exchanged finite limits on the \teq{\omega_i} integration for finite limits
\teq{\erg_{\pm}} for the integral over soft photon energies, with
\begin{equation}
   \erg_{\pm} \; =\; \dover{\omega_i}{\gamma_e (1 + \beta_e \mu_\mp)} \quad .
 \label{eq:erg_pm_def}
\end{equation}
Next, we transform the \teq{\omega_f} delta function to one for \teq{\omega_i}:
\begin{equation}
   \delta \Bigl[ \omega_f - \omega' (\omega_i, \theta_f) \Bigr] 
   \; =\; \biggl\vert \dover{\partial \omega_i}{\partial \omega_f} \biggr\vert\;
   \delta \Bigl[ \omega_i - \hat{\omega}_i (\erg_f, \mu_f) \Bigr] \quad ,
 \label{eq:delta_fn_equiv}
\end{equation}
where \teq{\hat{\omega}_i (\erg_f, \mu_f) } is the relation for \teq{\omega_i} in Eq.~(\ref{eq:res_kinematics_inversion}) 
evaluated at \teq{\omega_f = \gamma_e \erg_f (1+ \beta_e \mu_f) }.   The Jacobian factor can be evaluated 
by taking a derivative of Eq.~(\ref{eq:res_kinematics_inversion}), the result being
\begin{equation}
   \biggl\vert \dover{\partial \omega_i}{\partial \omega_f} \biggr\vert
   \;\to\; \dover{1}{2} \left\{ 1 + \cos\theta_f + \dover{1-\cos\theta_f}{(1 - \omega_f + \omega_f\cos\theta_f)^2} \right\}
   \;\equiv\; \dover{2\hat{\omega}_i - \omega_f - \hat{\omega}_i \omega_f (1-\cos\theta_f )}{ 
        \omega_f \left[ 1 - \omega_f (1-\cos\theta_f ) \right]}
 \label{eq:omega_if_deriv}
\end{equation}
where the angle aberration formula in Eq.~(\ref{eq:Lorentz_transform}) can be 
used to set \teq{\cos\theta_f = (\beta_e+\mu_f)/(1+\beta_e\mu_f)}. 
Observe that for \teq{\mu_f = 1} (i.e. forward scattering \teq{\cos\theta_f=1} in the ERF), 
this Jacobian factor is unity, while for \teq{\mu_f = -1} (backscattering \teq{\cos\theta_f=-1} in the ERF), 
the derivative algebraically approaches \teq{1/(1+2 \hat{\omega}_i)^2};
appreciable departures of the Jacobian from unity arise only for large ERF recoil regimes.  
The second evaluation in Eq.~(\ref{eq:omega_if_deriv}) is included to highlight the 
fact that its numerator conveniently cancels with an identical factor that appears 
in the scattering cross section in 
{Eq.~(\ref{eq:diff_csect_ST}).}

At this point the evaluations of the two delta functions are trivial, and the spectrum 
collapses to a simpler double integral over \teq{ds} and \teq{d \erg_s}.  The factor 
\teq{1-\beta_e\cos\theta_f} reduces to 
\teq{\gamma_e^{-2}\,(1+\beta_e\mu_f)^{-1}}.  The resulting spectrum is
\begin{equation}
   \dover{dn_{\gamma}}{dt\, d\erg_f} \; =\;  \dover{n_e c}{2 \pi {\cal S}} \, \dover{1}{\gamma^3_e \beta_e} 
   \int_{S} \dover{\hat{\omega}_i \, ds}{1+ \beta_e \muB} \, 
   \biggl\vert \dover{\partial \omega_i}{\partial \omega_f} \biggr\vert \, \dover{d\sigma}{d(\cos\theta_f) }  
      \int_{\erg_-}^{\erg_+} \dover{d \erg_s}{\erg_s^2} \, n_\gamma(\erg_s) \, f(\mu_i)
 \label{eq:scatt_spec_fin}
\end{equation}
In this expression, we employ Eq.~(\ref{eq:res_kinematics_inversion}) to represent \teq{\hat{\omega}_i}, insert
\teq{\cos\theta_f = (\beta_e+\muB)/(1+\beta_e\muB)} and \teq{\omega_f = \gamma_e \erg_f (1+ \beta_e \muB) }, and 
have set \teq{\mu_f\to \muB (s)} throughout.  Only \teq{\muB} and the soft photon distribution \teq{f(\mu_i)}
are explicitly functions of the position \teq{s} along the electron path \teq{S}; all other portions of the 
integrand possess only implicit dependence through the variable \teq{\mu_f=\muB}. This double integral 
serves as the basis for our computational results in Section~5.
For resonant regimes that are expected to be the dominant contribution 
for much of the pertinent model parameter space, 
an additional delta function approximation at the peak of the resonance can be made (e.g., see Dermer 1990). This amounts to introducing an equivalent delta function over the path length parameter \teq{s}: 
for a given viewing angle and scattered energy, only certain spatial points satisfy the resonance criteria \teq{\omega_i = B}.
This restriction results in a single integral over \teq{\erg_s} for the spectra. The resonance locales are discussed 
explicitly in Section~4 for a dipole field geometry, an expedient choice for the field morphology adopted in this paper.
The calculations can readily be adapted to arbitrary field configurations, such as those that are encountered 
in dynamic twisted field scenarios (e.g. Beloborodov 2013a) that include toroidal components, 
highlighting the broad utility of Eq.~(\ref{eq:scatt_spec_fin}).
For future explorations of resonant Compton upscattering spectra from self-consistently cooled electron populations, 
one can convolve Eq.~(\ref{eq:scatt_spec_fin}) with an integration over a density distribution 
\teq{n_e(\gamma, \, s)} that is dependent on the path locale parameter \teq{s}.

Specializing this collisional integral for the upscattered photon spectrum to the case of 
thermal soft photons that are uniformly distributed over the neutron star surface, 
we express the soft photon energy distribution as a Planck function in flat spacetime, isotropic 
over a hemisphere at each surface locale.  Such a choice is appropriate 
here as a first approximation to more sophisticated non-Plankian models that treat radiative transfer, 
line formation and vacuum polarization effects in neutron star atmospheres.  Such 
detailed atmosphere models generate both non-blackbody spectral forms, and 
anisotropic zenith angle distributions for the emission (e.g. see Zavlin et al. 1996, 
for normal pulsars and \"Ozel 2002, for magnetar applications).  It is anticipated that such 
anisotropies and departures from Planck spectra will at most introduce only modest 
influences on upscattering spectra: the value of the effective temperature 
and associated soft photon flux will have far greater impact on spectral results presented in this paper.
The Planck spectral form for the differential photon number density is
\begin{equation}
   n_{\gamma}(\erg_s) \; =\; \dover{\Omega_s}{\pi^2 \lambar^3} 
      \, \dover{\erg_s^2}{e^{\erg_s/\Theta} -1} \quad ,
 \label{eq:Planck_spec}
\end{equation}
so that the total distribution in both energy and angles is given by 
\teq{n_{\gamma}(\erg_s)\, f(\mu_i)}.
Here, \teq{\Theta =kT/m_ec^2} is the
dimensionless temperature of the thermal surface photons, and
\teq{\lambar = \hbar/(m_ec)} is the Compton wavelength over \teq{2\pi}. 
Also, \teq{\Omega_s} represents the solid angle of the blackbody photon
population at the stellar surface, divided by \teq{4\pi}.  This fractional solid angle is
introduced to accommodate anisotropic soft photon cases, for example
hemispherical populations (\teq{\Omega_s=1/2}) just above the stellar
atmosphere.  The total number density of soft photons at the surface 
is therefore \teq{2 \Omega_s\zeta (3) \,\Theta^3/(\pi^2 \lambar^3)}, for \teq{\zeta (n)} 
being the Riemann \teq{\zeta} function. The angular portion of the soft photon 
distribution depends on (i) the altitude, which controls the cone of 
collimation of the soft X-rays, and (ii) the vector direction of \teq{\boldsymbol{B}} at the scattering point.
The form for \teq{f(\mu_i)} is essentially adapted from BWG11 for the model  
and loop geometry enunciated in Section 4.2, with a normalization that couples 
to the altitude of the scattering locale:
\begin{equation}
   \int_{-1}^1 f(\mu_i)\, d\mu_i \; =\; 1- \cos\thetaC \;\equiv\; 1 - \sqrt{1- \Bigl( \dover{\rns}{R}\Bigr)^2}
   \quad ,\quad
   R\;\geq\; \rns \quad ,
 \label{eq:fmui_norm}
\end{equation}
which reproduces the inverse square law for \teq{R\gg \rns}.
Here \teq{\thetaC} is the opening angle of the cone of soft photons at altitude \teq{R}, 
and the general shapes of the \teq{f(\mu_i)} function are illustrated in Section~5
of Baring, Wadiasingh \& Gonthier (2011).  The particular 
distributions that satisfy this normalization are posited in Eqs.~(\ref{eq:fmui_equator})
and~(\ref{eq:fmui_pole}).   The integration over the thermal soft photon
energies in Eq.~(\ref{eq:scatt_spec_fin}) involving both \teq{n_{\gamma}(\erg_s)} 
and \teq{f(\mu_i)} is analytically developed in Appendix A.2, where the \teq{\erg_s} 
integration is distilled into series of tractable integrals spanning different parameter regimes.
Note that this analytic development can also be employed for nonmagnetic inverse 
Compton scattering processes for other stellar systems e.g. gamma-ray binaries.

Hereafter, we assume a dipole geometry for the magnetic field in the spectral calculations of this paper.  
Electron paths are defined by field loops parameterized by their foot-point colatitude \teq{\vartheta_{\rm fp}}, 
or equivalently their maximum (equatorial) altitude \teq{\rmax} in units of the neutron star radius \teq{\rns}, the 
two being related by
 \begin{equation}
    \rmax \; =\; \dover{1}{\sin^2 \vartheta_{\rm fp}} \quad .
\end{equation}
Essentially, here all radii are scaled to be dimensionless via \teq{r=R/\rns} so that \teq{\rmax \geq 1}.  Then,
the value of the polar cap angle of the last open field line 
is just \teq{\vartheta_{\rm fp}}, for which \teq{r = \rlc/\rns} at the
light cylinder radius \teq{\rlc}.  Without loss of generality we specialize throughout to the case of electrons 
moving \emph{anti-parallel} to {\teq{\boldsymbol{B}}} along field loops, 
{transiting from one pole to the other}.  The spectral production 
integrals are normalized by the arclength \teq{{\cal S}} of such a loop, 
which is hereafter scaled in units of \teq{\rns}.  This length 
is computed by parameterizing a loop by its colatitude \teq{\theta_{\rm col} \equiv \vartheta},
with \teq{ r(\vartheta) = \rmax \sin^2 \vartheta}, and then forming a path length element  
\teq{ds} (also expressed in units of \teq{\rns})
that satisfies the polar coordinate geometry relation
\begin{equation}
   \biggl( \dover{ds}{d\vartheta} \biggr)^2 \; =\; r^2 + \biggl( \dover{dr}{d\vartheta} \biggr)^2 
   \quad \Rightarrow\quad
   ds \; =\;   \rmax \sin\vartheta\, \sqrt{1 + 3\cos^2 \vartheta } \, d\vartheta 
   \; =\; \dover{ \sqrt{4 \rmax - 3 r } }{2\sqrt{\rmax - r} }\, dr \quad .
 \label{eq:arc_length_element}
\end{equation}
The total dimensionless arclength \teq{{\cal S}} of the loop between its foot-points can then be found analytically
via elementary integration:
\begin{equation}
   {\cal S} \; \equiv\;  2 \int_1^{\rmax}  \dover{ds}{dr} \, dr
   \; =\; \sqrt{(\rmax-1)(4 \rmax-3) } + 
     \dover{\rmax}{\sqrt{3}} \, \mbox{arctanh} \,\sqrt{\dover{3(\rmax -1)}{4\rmax -3}}  \quad .
 \label{eq:calS_eval}
\end{equation}
The factor of two accounts for the identical contributions from ascending and descending portions of a loop.
The field loop parameter \teq{\rmax} will be employed to label spectra in the graphical depictions below;
\teq{\vartheta_{\rm fp}} could serve as an alternate choice.

The magnetic Compton differential cross sections that we employ in the scattering integral of 
Eq.~(\ref{eq:scatt_spec_fin}) are full QED forms for polarized photons developed in 
Gonthier et al. (2014; see also Mushtukov, Nagirner \& Poutanen 2016).  
These incorporate Sokolov \& Ternov (ST) spinor formalism
and spin-dependent cyclotron decay widths, and so 
go beyond magnetic Thomson cross sections used in previous treatments of resonant 
upscattering.  Computed in the electron rest frame, they pertain to 
ground-state to ground-state transitions for incident photons 
parallel to \teq{\boldsymbol{B}}, corresponding to kinematic domains 
below the magnetic pair creation threshold, \teq{\omega_i \sin \theta_i \leq 2}.
Away from the \teq{\omega_i=B} cyclotron resonance, where the decay widths
contribute negligibly to the cross section, we use the 
cross sections given in Eq.~(39) of Gonthier et al. (2014;
see also Herold 1979; Daugherty \& Harding 1986; Bussard, Alexander \& M\'esz\'aros 1986):
\begin{equation}
   \dover{d \sigma^{\perp,\parallel} }{d \cos \theta_f}  \; =\;  \dover{3 \sigt}{16} \dover{ \omega_f^2 \,  
   e^{-\omega_f^2 \sin^2 \theta_f /2 B}}{\omega_i(2 \omega_i - \omega_f - \zeta)}
          \; T^{\perp,\parallel} \; \biggl\{ \dover{1}{(\omega_i - B)^2} +  \dover{1}{(\omega_i + B - \zeta)^2} \biggr\} 
   \;\; ,\quad
    \zeta \; =\; \omega_i \omega_f (1 - \cos \theta_f) \quad .
 \label{eq:diff_csect_ST}
\end{equation}
The subscripts \teq{\perp ,\parallel} denote the polarizations of the 
scattered photon; since the incident photon propagates along the field, 
the cross section is independent of its linear polarization.  
The polarization-dependent factors (spin-averaged) are 
\begin{equation}
   T^\perp \; =\; \omega_i (\omega_i - \zeta)
   \quad ,\quad
   T^\parallel \; =\; (2+ \omega_i) (\omega_i  - \zeta) - 2 \omega_f \quad ,
 \label{eq:Tave_defs}
\end{equation}
and do not depend on the choice of electron wavefunctions (spinors).  
We adopt a standard linear polarization convention:
\teq{\parallel} (O-mode) refers to the state with the
photon's {\it electric} field vector parallel to the plane containing
\teq{\boldsymbol{B}} and the photon's momentum vector, while \teq{\perp}
(X-mode) denotes the photon's electric field vector being normal to this plane.
Our protocol is to use Eq.~(\ref{eq:diff_csect_ST}) away from the resonance, namely when 
\teq{|\omega_i/B -1| \geq 0.05},

In the cyclotron resonance, we adopt an approximate form for the 
spin-dependent differential cross section from Section IIIE of Gonthier et al. (2014).
This uses an expansion in terms of the small parameter 
\teq{\delta \equiv 2(\omega_i - B)}, eliminating terms of order \teq{O(\delta^2)}
and higher.  For a spin-averaged, cyclotron decay width \teq{\Gamma}, 
the approximation is
\begin{equation}
  \left(\dover{d\sigma^{\perp,\parallel}  }{ d\cos \theta _f } \right)_{\rm res}
  \;\approx\;  \dover{3 \sigt}{16} \dover{ \omega_f^2 \, 
       e^{-\omega_f^2 \sin^2 \theta_f /2 B}}{\omega_i(2 \omega_i - \omega_f - \zeta)\, \epsilon^3_\perp}
 \sum_{s=\pm 1} \dover{{\cal N}^{\perp,\parallel}_s }{{\cal D}_s}
  \quad ,\quad
  {\cal D}_s \; =\; 4\left(\omega_i-B\right)^2
         + \left(\epsilon_\perp-s\right)^2\left(1+B\right)^2 \, \dover{\Gamma^2}{\epsilon_\perp^2}  \quad .
\label{eq:dsig_resonance}
\end{equation}
Here \teq{\epsilon_\perp = \sqrt{1 + 2B}}, and \teq{s} is the spin quantum number 
label for the intermediate state.  The numerators are
\begin{eqnarray}
  {\cal N}^{\perp}_s & = & \left(\epsilon_\perp-s\right)^2
     \biggl[ \left(2\epsilon_\perp+s\right) \, \omega_i (\omega_i - \zeta)
     - s \left( \epsilon_{\perp} + s \right)^2 \Bigl\{ \zeta - (\omega_i - \omega_f)/2 \Bigr\} \biggr] 
     - 2 s\, \delta  \Bigl[  (1+2\omega_i ) \, \zeta - 2 \omega_i^2 \Bigr] \quad ,\nonumber\\[2.0pt]
  {\cal N}^{\parallel}_s & = & \left(\epsilon_\perp-s\right)^2
     \biggl[ \left(2\epsilon_\perp+s\right) \, \Bigl\{ (2+ \omega_i) (\omega_i  - \zeta) - 2 \omega_f \Bigr\}
     - s \left( \epsilon_{\perp} + s \right)^2 \Bigl\{ - \zeta + 3 (\omega_i - \omega_f)/2 \Bigr\} \biggr]
\label{eq:calNperp_Npar_final}\\[2.0pt]
  & & \qquad\qquad - 2s\, \delta \, \omega_i \Bigl[  -  \zeta  + (\omega_i - \omega_f) \Bigr]
  -2 s\, \delta \,\cos\theta_f  \Bigl[  (1+\omega_i ) \, \zeta - \omega_i (\omega_i + \omega_f) \Bigr] \quad .\nonumber 
\end{eqnarray}
The spin-averaged, QED \teq{n=1\to 0} cyclotron width \teq{\Gamma}
is taken from  Eq.~(13) of Baring, Gonthier \& Harding (2005; see also Latal 1986).
Asymptotic limits are \teq{\Gamma\approx 2\fsc B^2/3} when \teq{B\ll 1}, and 
\teq{\Gamma\approx  \fsc (1-1/e)} when \teq{B\gg 1}.
A useful empirical approximation to the width \teq{\Gamma} was posited in the Appendix of
van Putten et al. (2016); it reproduces the \teq{B\ll 1} and \teq{B\gg 1} asymptotic limits, 
and possesses a precision of better than around 2\% when compared with the exact form.
For each spin case \teq{s = \pm 1}, the approximation in Eq.~(\ref{eq:dsig_resonance}) 
is numerically accurate to a precision of better than 0.03\% across the
resonance Lorentz profile, i.e., for \teq{|\omega_i/B -1| \leq 0.05}, 
when fields are in the range \teq{0.1 \lesssim B \lesssim 10}, and is still 
extremely good for the range of field strengths \teq{10^{-2} < B < 10^2}.

\section{MODEL GEOMETRY}
 \label{sec:geometry}
 
We now give an idealized yet representative case study of the
formalism for emission directed to an observer for a {magnetar}.
A dipole field geometry for the star is assumed, with dipole moment
$B_{\rm p} \rns^3/2$ (e.g. Shapiro \& Teukolsky 1983), half the value
conventionally used by observational collaborations. Treatment of more
complicated multipole field configurations, outer magnetospheric field
geometry, twisted dipole, and curved spacetime enhancements of the field
are deferred to future work; such added complexity will alter the
beaming characteristics significantly in a model-dependent way. The lack
of a fully self-consistent model for the global field structure for
{magnetar magnetospheres} presents an uncertainty, with force-free MHD models
sustaining complicated non-dipolar morphologies without significantly
altering the spin-down characteristics (e.g. Spitkovsky 2006;
see also Kalapotharakos et al. 2012, 2014, for dissipative MHD
models, and Philippov \& Spitkovsky 2014 {and Chen \& Beloborodov 2017} for particle-in-cell plasma simulations). 
As the focus here is on low altitudes \teq{r\lesssim 20\rns} in closed field regions, 
i.e., well inside the light cylinder radius of \teq{>10^4\rns} for magnetars,
we expect a more thorough MHD treatment of the high-altitude magnetospheric
field geometry not to profoundly modify the general character of the results and conclusions
presented in this paper, motivating the restriction to dipolar morphology.

We define a right-handed Cartesian coordinate system \teq{\{
\hat{\bf{x}} , \hat{\bf{y}} , \hat{\bf{z}} \} } in the corotating frame
of the neutron star, with a star-centered origin, and the
\teq{\hat{\bf{z}}} unit vector collinear with the magnetic field axis. An
observer at infinity's \emph{instantaneous} line of sight in the
corotating frame is defined by angle \teq{\theta_v} relative to
\teq{\hat{\bf{z}}}, i.e. \teq{\hat{\bf{z}}\boldsymbol{\cdot} \hat{\bf{n}}_v
\equiv \cos \theta_v}. Without loss of generality, we define the vector
field of observer line of sights (directed away from the star) to be in
the $x-z$ plane,
\begin{equation}
   \hat{\bf{n}}_v \; =\; \cos \theta_v \,\hat{\bf{z}} +\sin \theta_v \,\hat{\bf{x}} \quad .
 \label{eq:nvec_def}
\end{equation}
Photons are assumed to propagate in straight lines, neglecting general relativity
and vacuum birefringence in the magnetosphere. 
Relativistic aberration is small for magnetars at the low altitudes considered 
here, since they are slow rotators. Curved spacetime will be important for 
photons beamed to the observer from behind the star that then propagate 
through low altitudes, or photons emitted at low altitudes. Such a treatment 
of photon geodesics is deferred to a future Monte Carlo simulation, but is not 
expected to profoundly influence the results presented in this paper since 
much of the spectral generation here arises above two stellar radii.
  
The dipole magnetic field and its unit vector in spherical polar coordinates are 
parameterized by a polar angle \teq{\vartheta}
\begin{equation}
   \boldsymbol{B} \; =\; \frac{B_{\rm p}}{ 2r^{3}} 
      \Bigl( 2 \cos \vartheta \,\hat{\bf{r}} + \sin \vartheta \,\hat{\boldsymbol{\theta}} \Bigr)
   \quad \Rightarrow\quad
   \hat{\boldsymbol{B}} \; =\; \dover{ 2 \cos \vartheta \,\hat{\bf{r}} + \sin \vartheta \,\hat{\boldsymbol{\theta}} }{\sqrt{1+3\cos^2 \vartheta}} \quad .
 \label{eq:Bvec_Bvec_hat}
\end{equation}
where \teq{r \equiv R/\rns} the radius in units of neutron star radii. The magnetic field 
is azimuthally symmetric in definition above, and thus we can parameterize individual 
field loops in terms of an azimuthal angle \teq{\phi_*}, which is defined to be 
zero in the observer $x-z$ meridional plane,
\begin{equation}
    \hat{\boldsymbol{B}} \; =\; 
    \sin \zeta \cos \phi_* \,\hat{\bf{x}} + \sin \zeta \sin \phi_* \,\hat{\bf{y}} + \cos \zeta \,\hat{\bf{z}} \quad ,
\end{equation}
where \teq{\zeta} is the angle between the local direction of \teq{\boldsymbol{B}} 
and  \teq{\hat{\bf{z}}} axis.  It is easily found by consideration of transformations 
between spherical polar and Cartesian coordinates, parameterized in terms of \teq{\vartheta},
\begin{equation}
   \cos \zeta \; =\; \dover{3 \cos^2 \vartheta - 1}{ \sqrt{1+3\cos^2 \vartheta} }
   \quad \hbox{or}\quad
   \sin \zeta \; =\; \dover{3 \cos \vartheta  \sin \vartheta}{ \sqrt{1+3\cos^2 \vartheta} } \quad .
 \label{eq:cos_sin_zeta}
\end{equation}
The angle between the observer line of sight and local direction of \teq{\boldsymbol{B}} 
for a particular loop, parameterized by $\phi_*$, where $\phi_* = 0$ and $\phi_* = \pi$ 
denote ``meridional" and ``anti-meridional" field loops respectively.  The angle is routinely found
\begin{equation}
   \cos \ThetaBn \equiv \hat{\bf{n}}_v \boldsymbol{\cdot} \hat{\boldsymbol{B}} 
   \; =\; \cos \theta_v \cos \zeta + \sin \theta_v \sin \zeta \cos \phi_* \quad .
 \label{eq:cosBn}
\end{equation}
This relation will eventually forge the connection between the final scattering angle 
in the corotating frame and the observer direction.  The direction of electron propagation 
is irrelevant to these geometrical considerations, instead being germane to the scattering kinematics.

\subsection{Resonant Interaction Criteria}
 \label{sec:res_criteria}

Since resonant contributions are generally dominant when defining spectra 
of emission directed to an observer, the energies and location along a field loop 
where the resonant condition \teq{\omega_i =B} in the ERF is accessed are 
crucial to understanding the predominant locale of resonant Compton spectral generation. 
The essential connection is that the final photon scattering angle in the OF is directed
towards the viewer, i.e. \teq{\Theta_f = \ThetaBn} (\teq{\Rightarrow \mu_f = \cos\ThetaBn}) at an interaction point along a 
given field loop for electrons moving anti-parallel to {$\boldsymbol{B}$} along a field loop. 
Given the inversion relationship for the ERF scattering kinematics in Eq.~(\ref{eq:res_kinematics_inversion}),
at each point along a field loop, the parameters that determine whether resonant interactions 
are sampled are \teq{B_{\rm p}, \; \rmax, \; \erg_f} and \teq{\gamma_e}.  
Using \teq{ r(\vartheta) = \rmax \sin^2 \vartheta} in the magnetic field forms 
in Eq.~(\ref{eq:Bvec_Bvec_hat}), inserting the Doppler boost relation
\teq{\omega_f = \gamma_e \erg_f (1+ \beta_e \mu_f) }
and the aberration formula in Eq.~(\ref{eq:Lorentz_transform})
into Eq.~(\ref{eq:res_kinematics_inversion}) for the ERF kinematics then yields
 \begin{eqnarray}
 \dover{\erg_f \lbrack  \erg_f \sin^2 \ThetaBn - 2 \gamma_e (1 + \beta_e \cos \ThetaBn) \rbrack}{
    2 \lbrack \erg_f \gamma_e (1- \beta_e)(1- \cos \ThetaBn) -1 \rbrack } 
 \; \equiv\;  \hat{\omega}_i \; =\;  \vert\boldsymbol{B}\vert
 \; \equiv\; \frac{B_{\rm p}  \sqrt{1 + 3 \cos^2 \vartheta}}{2 r^3_{\rm max} \sin^6 \vartheta} \quad .
 \label{eq:wi_ef_res}
 \end{eqnarray}
Therein, for fixed \teq{\gamma_e}, specifying \teq{\erg_f} and \teq{\ThetaBn} for the 
scattered photon uniquely determines the value \teq{\omega_i} of the photon in the ERF prior to scattering.
Concomitantly, by virtue of Eq.~(\ref{eq:Lorentz_transform}), for photons emanating from a
particular location on the stellar surface, \teq{\Theta_i} is uniquely specified, and the value of the 
soft photon energy is selected.
Given the identity for \teq{\cos \ThetaBn} in Eq.~(\ref{eq:cosBn}), Eq.~(\ref{eq:wi_ef_res}) defines 
an algebraic equation for \teq{\cos\vartheta} that identifies select points along a field loop 
at which resonant scattering directs photons towards the observer; 
in general, this equation has to be solved numerically.  
This result is subject to the additional constraint in Eq.~(\ref{eq:res_kinematics_restriction1}), 
which is satisfied whenever \teq{\erg_f < \gamma_e\beta_e}, i.e. energy is conserved: 
this nuance is discussed at the end of the first part of Section~\ref{sec:formalism}.

\begin{figure}[ht!]
\centering
\includegraphics[scale=0.3]{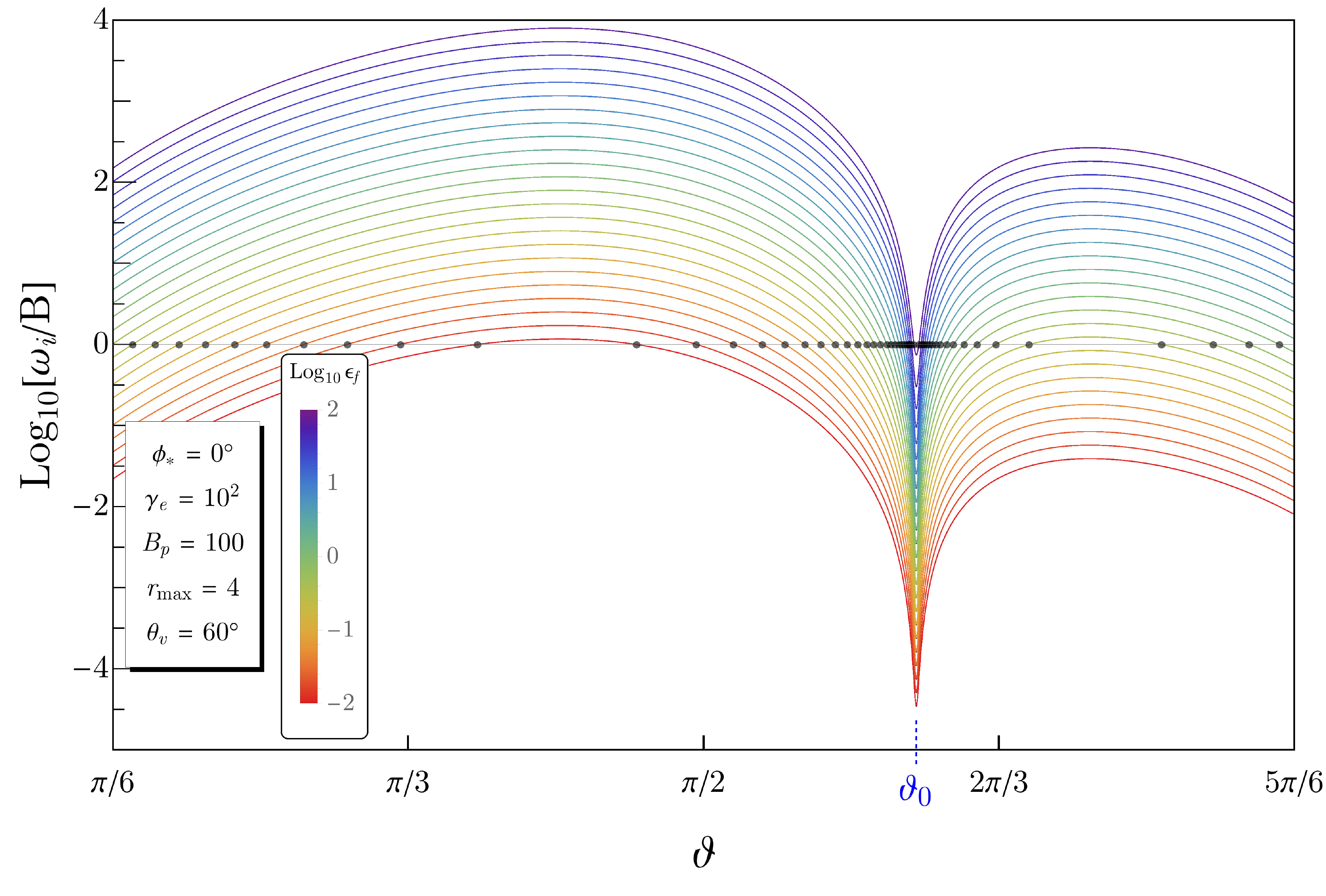}
\caption{Curves representing the ratio of \teq{\omega_i/B} as a function of colatitude 
\teq{\vartheta} along a meridional field line (\teq{\phi_*=0^\circ}) with locus $r_{\rm max} \sin^2\vartheta = r$ with $\rmax =4$.  
The range of colatitudes displayed, \teq{\pi /6 \leq \vartheta\leq 5\pi /6} spans the complete 
domain between the footpoint colatitudes \teq{\vartheta_{\rm fp}} for this \teq{\rmax}.
The curves are color-coded as listed in the inset according to the final photon 
energy \teq{\erg_f}, in units of $m_e c^2$, ranging from keV X-rays (red, bottom) 
to almost GeV energy gamma-rays (purple, top).  
The polar field strength is \teq{B_p=100} and the observer viewing angle is
\teq{\theta_v = 60^{\circ}}, while the electron Lorentz factor is \teq{\gamma_e=10^2}.
Resonance interaction points (black) where the curves intersect the horizontal 
\teq{\omega_i=B} line are marked; these are solutions of Eq.~(\ref{eq:wi_ef_res}).  Only a finite range of final energies 
have access to resonant interactions.  The colatitude \teq{\vartheta_0} of the cusps is given by 
Eq.~(\ref{eq:cos_theta_0}) and is discussed in the text.}
 \label{fig:omegai_B_curves}    
\end{figure} 

Numerical values for the ratio \teq{\hat{\omega}_i/ \vert\boldsymbol{B}\vert} 
along a magnetic field line, as a function of colatitude \teq{\vartheta}, are illustrated 
in Fig.~\ref{fig:omegai_B_curves} for a fixed Lorentz factor of \teq{\gamma_e=10^2} and an 
instantaneous observer viewing angle \teq{\theta_v=60^{\circ}}.  They are obtained 
by taking the ratio of the left and right hand expressions in Eq.~(\ref{eq:wi_ef_res}), and are specifically 
for a meridional field loop, i.e. one coplanar with the plane defined by the rotation axis 
and the line of sight to the observer.  They are color-coded for the final photon energy, 
and highlight the resonant interaction points, \teq{\hat{\omega}_i =\vert\boldsymbol{B}\vert}, 
for each curve specified by \teq{\erg_f} using the black dots.
Observe that for a given viewing angle, if $\erg_f$ is too high or low, 
no resonant interactions are accessed.  When they are, there are generally two, three or four such positions.
The maximum value of \teq{\erg_f} for resonant interactions, which serves as an effective 
cut-off energy to the dominant portion of the emission spectrum, occurs for backscattering 
events in the ERF, i.e. when \teq{\cos \theta_f = -1}, which corresponds to 
\teq{\ThetaBn = \pi } using the aberration formula.  Manipulating the left hand identity in 
Eq.~(\ref{eq:wi_ef_res}), this maximum is then defined by
\begin{equation}
   \erg_f^{\rm max} \;=\; \dover{ \vert\boldsymbol{B}\vert \gamma_e (1 + \beta_e)}{1 + 2 \vert \boldsymbol{B}\vert}\quad ,
 \label{eq:efmax}
\end{equation}
a result that is highlighted in Eq.~(15) of Baring \& Harding (2007).  In highly supercritical fields, 
the resonant scattering is deep in the Klein-Nishina domain and \teq{\erg_f^{\rm max} \propto \gamma_e}, 
as expected.  For subcritical fields \teq{ \vert\boldsymbol{B}\vert \ll 1}, since the resonance is accessed 
when \teq{\vert\boldsymbol{B}\vert \sim \gamma_e\erg_i}, the familiar Thomson dependence 
\teq{\erg_f^{\rm max} \sim \gamma_e^2\erg_i} emerges.  Note that this beaming occurs only along 
meridional (\teq{\phi_* =0}) and anti-meridional field loops (\teq{\phi_* =\pi}) or for azimuthal 
angles \teq{\phi_*} within \teq{1/\gamma_e} of these special cases.  This restriction thus represents 
a spatially small portion of the emitting region along a field loop. Shadowing for certain instantaneous 
observer angles, discussed in Section 4.2, can curtail some beaming contributions significantly, 
particularly for anti-meridional loops and when viewing angles are moderately large.

A noticeable feature of the curves in Fig.~\ref{fig:omegai_B_curves} is that they all possess prominent cusps 
at the same colatitude \teq{\vartheta_0}.  These features are deep local minima 
of the functional expression for \teq{\hat{\omega}_i}.  If one takes the \teq{\gamma_e\gg 1} limit 
of the left hand side of Eq.~(\ref{eq:wi_ef_res}), the Doppler shift formula 
\teq{ \hat{\omega}_i \approx \gamma_e \erg_f (1+ \cos \ThetaBn)} is quickly reproduced, being
equivalent to Thomson kinematics \teq{\omega_i \approx \omega_f} in the ERF.
Thus the local minima can be approximately defined by the root \teq{\cos\ThetaBn = -1}, 
which applies to all values of \teq{\erg_f} and \teq{\gamma_e}.  This select criterion is tantamount 
to requiring that the tangent to the local field line coincides with the line of sight to the observer.  Using the identity 
in Eq.~(\ref{eq:cosBn}), it is quickly discerned that for meridional loops with \teq{\phi_* =0}, 
\teq{\zeta = \theta_v \pm \pi} is established.  This simple result can be inserted into either form in 
Eq.~(\ref{eq:cos_sin_zeta}), and the result squared and then inverted to define an equation for the 
approximate value of \teq{\vartheta_0}: 
\begin{equation}
   \cos^2\vartheta_0 \; \approx \;  \dover{1}{6} \Bigl( 2 + \cos^2\theta_v - \cos\theta_v \sqrt{ 8 + \cos^2\theta_v} \Bigr)
   \;\; \Leftrightarrow \;\; \sqrt{1 + 3 \cos^2\vartheta_0} \; \approx \; \dover{1}{2} \Bigl( \sqrt{ 8 + \cos^2\theta_v} - \cos\theta_v \Bigr) \quad .
 \label{eq:cos_theta_0}
\end{equation}
The choice of sign in the solution of the quadratic is fixed by the magnetic and viewing geometry.
Further when taking the square root of this expression, the negative root is accessed by 
this same geometry: the colatitudes of tangent lines directed to an observer with a 
\teq{0 < \theta_v < \pi/2} viewing angle are generally in the range \teq{\pi /2 < \vartheta_0 < \pi}.   
Eq.~(\ref{eq:cos_theta_0}) applies to the \teq{\phi_*=\pi} case also, where \teq{-\zeta = \theta_v \pm \pi} 
is established using \teq{\cos\ThetaBn = -1}.  In general, as \teq{\phi_*} changes from 
zero or \teq{\pi} and the plane of the field line rotates, the mathematical form for the approximate root 
becomes more complicated.  For the choice of \teq{\theta_v=\pi /3} in Fig.~\ref{fig:omegai_B_curves}, the 
negative square root of Eq.~(\ref{eq:cos_theta_0}) yields \teq{\vartheta_0 \approx 0.62 \pi}, in close 
agreement with the position of the cusps in this Figure.

\begin{figure}[h!]
\centering
\includegraphics[width=0.99\textwidth]{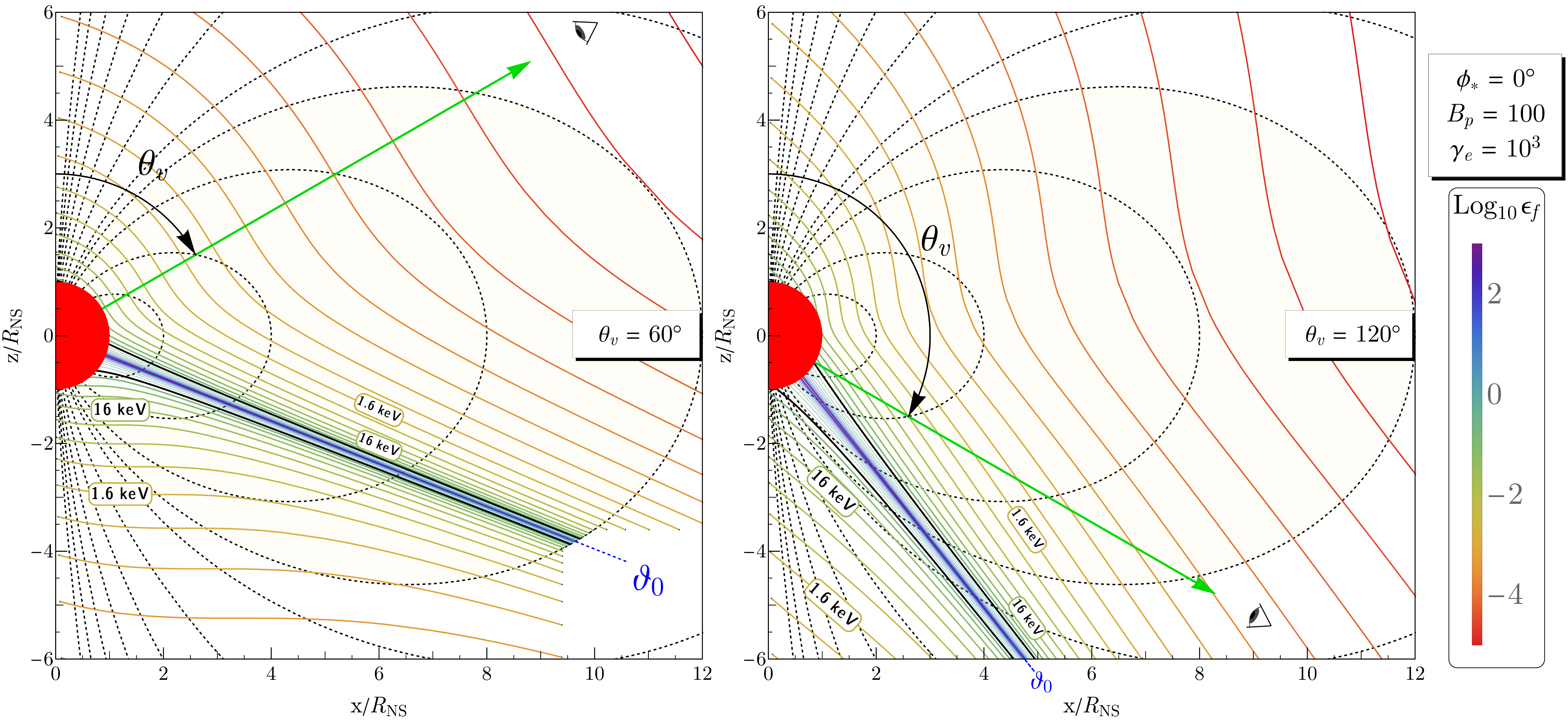} 
\caption{Resonance interaction loci along meridional field loops (\teq{\phi_*=0^{\circ}})
for observer viewing angles \teq{\theta_v = 60^{\circ}} (left panel) 
and \teq{\theta_v = 120^{\circ}} (right panel).  Contours of constant 
\teq{\Psi \equiv B_p/[2\erg_f \gamma_e]} are depicted, 
with a range \teq{10^{-4} < \Psi < 4 \times 10^3}, 
being color-coded according to the 
value of the dimensionless final photon energy \teq{\erg_f}, as indicated in the legend on the right; 
the black loci constitute the value of  \teq{\erg_f=10^{-0.5}\approx 160}keV (i.e., \teq{\Psi \approx 0.16}).
The colatitudes of these loci as functions of the altitude are solutions of Eq.~(\ref{eq:wi_ef_res}).  
The field strength \teq{B_p=100} and electron Lorentz factor \teq{\gamma_e=10^3} are kept fixed.
As the magnetic field drops with increasing altitude, the interaction points converge 
towards a radial dotted blue line with the colatitude \teq{\vartheta_0} that satisfies Eq.~(\ref{eq:cos_theta_0}), 
which is here \teq{\vartheta_0\approx 0.62\pi = 111.6^\circ} for the left panel, and 
\teq{\vartheta_0\approx 0.78\pi = 141.6^\circ} for the right panel.
}  
 \label{fig:resonance_points}    
\end{figure}

\begin{figure}[h!]
\centering
\includegraphics[width=0.99\textwidth]{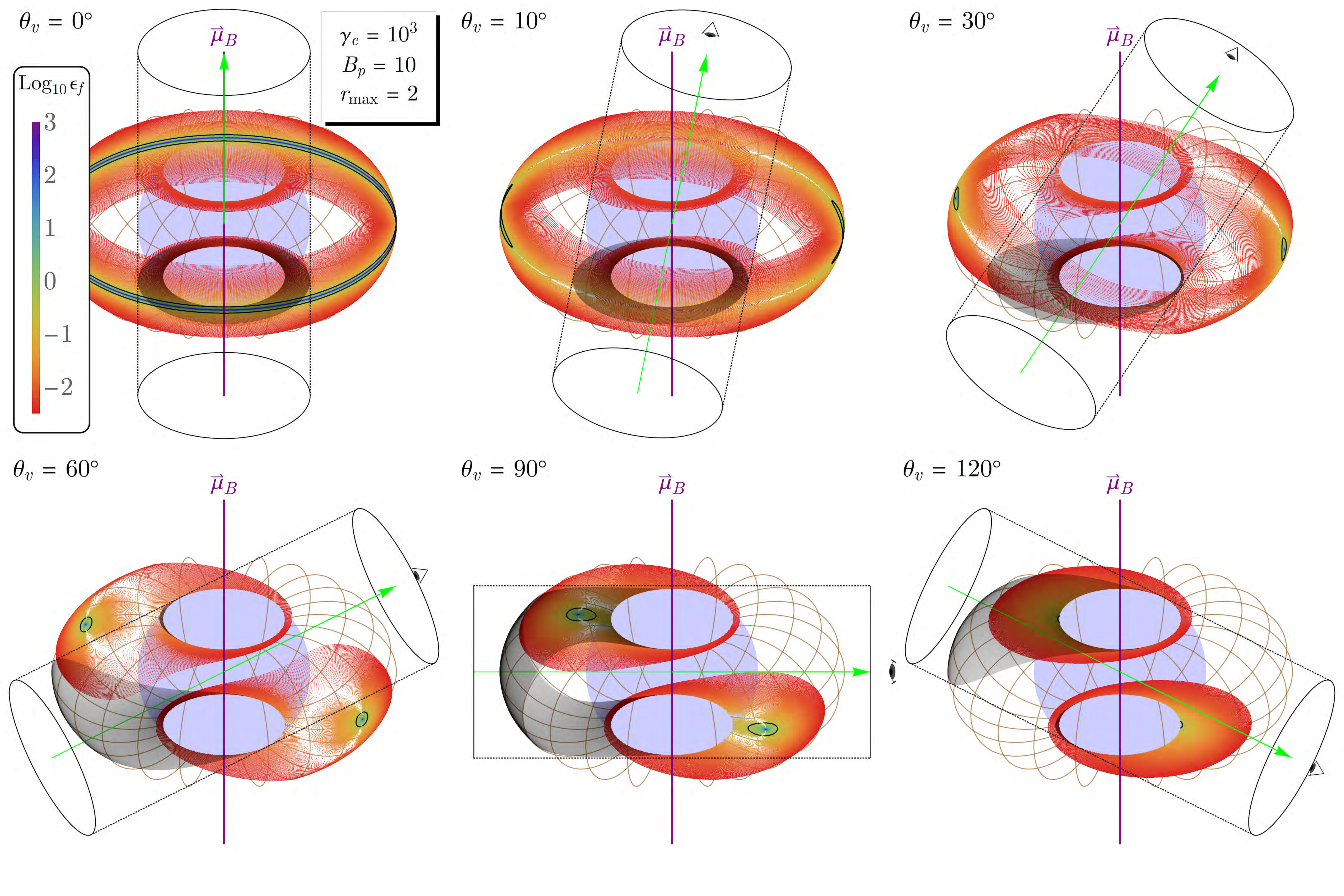} 
\caption{3D orthographic projections (with a linear spatial scale)
of resonant interaction points for a toroidal bubble of field loops of extent \teq{\rmax=2}, color coded 
for final scattering energy \teq{\erg_f} in the OF, plotted here for \teq{B_{\rm p} =10} and uncooled \teq{\gamma_e = 10^3}. 
The six panels are for different viewing angles \teq{\theta_v} relative to the magnetic dipole moment unit vector
\teq{\muBhat}, ranging from \teq{0^\circ} to \teq{120^\circ}, as labelled.
The black curves bound emission that is greater than \teq{\sim 160} keV (green, blue, violet colors), separating 
such from softer emission (yellow, orange, red colors), indicating that 
most of the COMPTEL-violating high energy emission is confined to small surface locales. 
The gray region denotes that of shadowing by the star with respect to the line-of-sight.
}  
 \label{fig:ortho3D_resonance}    
\end{figure}

Another prominent feature of Fig.~\ref{fig:omegai_B_curves} is the concentration of resonant interaction 
points (black dots) near the cusps for large \teq{\erg_f} cases.  This segues the discussion 
to the solution for the \teq{ \hat{\omega}_i \; =\;  \vert\boldsymbol{B}\vert} resonance criterion.
Focusing again on the \teq{\gamma_e\gg 1} domain, equating the left and right hand sides of 
Eq.~(\ref{eq:wi_ef_res}) yields 
\begin{equation}
   K(\vartheta) \; \equiv\; \dover{  \sin^6 \vartheta \, (1 + \cos \ThetaBn )}{ \sqrt{1 + 3 \cos^2 \vartheta}}  
   \; =\; \dover{\Psi}{\rmax^3}
   \quad \hbox{for}\quad 
   \Psi \;\equiv\; \dover{B_{\rm p}}{2 \erg_f \gamma_e}\quad .
 \label{eq:Kfunc_Psi_def}
\end{equation}
If either \teq{\gamma_e} or \teq{\erg_f} are sufficiently large, then \teq{\Psi\ll 1} and this resonance 
condition solves via \teq{\cos\ThetaBn \approx -1}, i.e. \teq{\vartheta\approx \vartheta_0} as before.
This explains the clustering of black dots in Fig.~\ref{fig:omegai_B_curves} in the vicinity of this colatitude.
The same result is realized for high altitude loops with \teq{\rmax\gg 1}.  This circumstance 
is illustrated in Fig.~\ref{fig:resonance_points}, which exhibits solutions to Eq.~(\ref{eq:Kfunc_Psi_def})
for a meridional configuration (i.e., the \teq{x - z} plane corresponding to
\teq{\phi^{\ast}=0^\circ}) of field lines for \teq{\gamma_e=10^3} (which differs from the value
in Fig.~\ref{fig:omegai_B_curves}), and for two contrasting viewing angles, 
\teq{\theta_v = 60^{\circ}} and \teq{\theta_v = 120^{\circ}}.  Even though these two viewing angles are 
symmetrically placed relative to the magnetic equator, the evident asymmetry is incurred because 
electrons are flowing in only one direction along field lines. The solutions define contours of constant 
\teq{\Psi} for the resonance condition, and a broad range of \teq{\Psi} are represented in each panel, color-coded by their 
values of \teq{\erg_f}; this implies that an infinite variety of \teq{B_p} and \teq{\gamma_e} 
choices correspond to each \teq{\erg_f} contour, with the restriction that \teq{\erg_f < \gamma_e}.
For each value of \teq{\erg_f} there generally exist two contours of resonance locales
in distinct portions of the magnetosphere, and these two curves intersect a given field line often 
at four locations, a property that is evinced by the black dots in Fig.~\ref{fig:omegai_B_curves}.  
The exception to this arises for high \teq{\erg_f} contours whose footpoints usually lie 
at colatitudes more remote from the poles than those of select field loops.
The radial direction at colatitude \teq{\vartheta_0} defines a {\it separatrix} for 
each member of a pair of resonance loci, and the contours asymptotically become almost parallel to this 
radial line at high altitudes.  Observe that the contour/separatrix map for the \teq{\phi_*=\pi} case can be 
obtained by a rotation through angle \teq{\pi} in the \teq{x-z} plane.  
Note also that for non-meridional viewing configurations, the contour morphology 
changes significantly, and the separatrix can disappear,
a property that can be inferred from the orthographic projections 
depicted in Fig.~\ref{fig:ortho3D_resonance}.

Returning to the \teq{\phi_{\ast}=0} case, for gamma-ray energies \teq{\erg_f >  1} 
(green, blue and purple), the proximity of the two resonance 
points on each field line loop is obvious, and becomes more marked as \teq{\rmax} increases.
In this asymptotic domain,
the separation of the pairs of resonance points at \teq{\vartheta = \vartheta_{\pm}}
can be specified via a more refined analysis of the solutions of Eq.~(\ref{eq:Kfunc_Psi_def}), 
noting that they lie in proximity to the local 
minimum defined by \teq{\vartheta_0} that gives \teq{K(\vartheta_0)=0}.  Expanding the 
\teq{K(\vartheta )} function about this value to quadratic order in a Taylor series,
it is quickly inferred that \teq{K'(\vartheta_0)=0} at the extremum \teq{\vartheta_0}, 
so that \teq{K(\vartheta ) \approx (\vartheta - \vartheta_0)^2\, K''(\vartheta_0)/2} to leading order.
With this construction, the colatitudes of the two resonance points on each meridional 
field loop are approximately given by
\begin{equation}
   \vartheta_{\pm} \; =\; \vartheta_0 \pm \Delta \vartheta
   \quad \hbox{with} \quad
   \Delta\vartheta \; =\; \sqrt{\dover{2 \Psi}{ K'' (\vartheta_{0}) \rmax^3}} \quad .
 \label{eq:resonance_points}
\end{equation}
The expression for \teq{K''(\vartheta_0)} can be routinely derived in closed analytic form, 
but is rather involved in general.  For the special meridional and anti-meridional cases,
one can determine after a modicum of algebra that 
\begin{equation}
   K''(\vartheta_0 ) \; \approx\; \dover{  9 \sin^6 \vartheta_0 \, (1+\cos^2\vartheta_0 )^2}{ (1 + 3 \cos^2 \vartheta_0 )^{5/2}}
   \quad , \quad \phi_\ast \; =\; 0, \pi\quad .
 \label{eq:Ktwoprimes_final}
\end{equation}
The construction leading to Eq.~(\ref{eq:resonance_points}) is generally robust 
as long as \teq{K''(\vartheta_0)} is not very small.  Thus one infers that it works best when 
\teq{\vartheta_0} is not too near \teq{0} or \teq{\pi}, i.e. that the viewing angle 
is not too close to the magnetic poles.  Returning to general \teq{\phi_*},
clearly \teq{\Delta\vartheta \propto r^{-3/2}_{\rm max}} describes 
the asymptotic behavior of the resonance loci at high altitudes. 
Moreover, since \teq{\Delta\vartheta} is often small for a substantial range of 
energies \teq{\erg_f} above 10 keV, photons scattering in the cyclotron resonance 
are then beamed along a local field line that is aligned virtually towards the observer. 
Note also that \teq{\Delta\vartheta\propto \erg_f^{-1/2}} corresponds approximately to 
\teq{\erg_f (1 - \cos\theta_f) = }const., which is a kinematic coupling between viewing angle 
relative to \teq{\boldsymbol{B}} and photon energy that was highlighted in Baring \& Harding (2007) 
for the case of uniform field geometry.  This proportionality for the 
approximate locales of resonance points has consequences concerning the shape 
of the resonance spectrum that will be detailed in Section~\ref{sec:spec_loops}.

To combine the two pieces of information encapsulated in Figs~\ref{fig:omegai_B_curves} 
and~\ref{fig:resonance_points}, it is instructive to form three-dimensional orthographic 
projections of the resonance interaction locales that generate upscattered photons of a 
specified energy \teq{\erg_f} directed to a particular observer line of sight.  Six examples of these are depicted 
in Fig.~\ref{fig:ortho3D_resonance}, contrasting different viewing angles spanning \teq{\theta_v= 0^\circ} and
\teq{\theta_v = 120^\circ}, mostly in \teq{30^\circ} increments.  These represent different 
rotational phases in magnetars that are oblique rotators.  As is evident in Fig.~\ref{fig:resonance_points}, many resonance 
points form two loci for a fixed azimuthal angle \teq{\phi_*} for the field line.  Summing 
over a complete range of azimuths \teq{0 \leq \phi_* \leq 2\pi}, the resonance locales 
form 2D surfaces.  The positioning of these surfaces depends on the choice of the 
scattered photon energy \teq{\erg_f}.  The higher this energy, the closer the two resonance 
locales reside along a given field line in Fig.~\ref{fig:omegai_B_curves}, a property also indicated 
by the analytic solution in Eq.~(\ref{eq:resonance_points}).  Thus, increasing \teq{\erg_f} 
will bring cyclotron resonance surfaces corresponding to a particular viewing perspective
closer together.  For some energies \teq{\erg_f}, there are 
additional surfaces of resonance.  To capture this complexity of information, 
the 3D orthographic projections in Fig.~\ref{fig:ortho3D_resonance} color code \teq{\erg_f} surfaces to 
render it relatively simple to discern the assemblage of resonance interaction localities.  
For example, for the highest photon energies \teq{\erg_f\sim \erg_f^{\rm max} \sim 10^3} (blue/purple locales), 
for all instantaneous viewing angles \teq{\theta_v},
only small portions of the magnetosphere generate scatterings sampling the cyclotron 
resonance for this \teq{\gamma_e=10^3} example: usually a spot, but in the symmetric case 
of \teq{\theta_v=0^{\circ}}, a ring.  This is because of the extreme 
constraints of Doppler boosting and relativistic aberration in the scattering interaction kinematics.

Much larger magnetospheric volumes can access resonant interactions for photon energies 
\teq{\erg_f < 1} in the X-ray band as the Doppler constraints become less restrictive.  Black loci 
demarcating \teq{\erg_f=160}keV
are included in each of the panels to volumetrically divide these two \teq{\erg_f} energy domains
(green to purple above 160 keV, and yellow and red below),
and at the approximate centers of these ``spherical ellipses'' is a point whose magnetic colatitude is 
\teq{\vartheta_0}; the vector to this point defines the separatrix in the \teq{\phi_{\ast}=0} plane.  
When \teq{\theta_v > 0}, as \teq{\phi_{\ast}} increases from zero, the separation of the resonance contour pairs 
is reduced, eventually shrinking to zero at moderately small values of \teq{\phi_{\ast}} so as to 
close the \teq{\erg_f \gtrsim 160}keV ellipses.  The \teq{\vartheta = \vartheta_0} separatrix thus does not 
exist for most field line azimuthal angles, since the root \teq{\cos\ThetaBn = -1} cannot be realized:  
the tangent to the local field line then cannot point towards the observer.
Note that the value of \teq{B_p} for these projections is lower than that in Fig.~\ref{fig:resonance_points}, 
and does not lead to a change in the value of \teq{\vartheta_0}; the orthographic projection
morphology for a \teq{B_p=100, \gamma_e=10^4} case should closely resemble 
that in Fig.~\ref{fig:ortho3D_resonance}.  Observe also that 
there are large solid angles that are uncolored or ``white''' in the projections,
particularly for viewing angles larger than around \teq{60^{\circ}}; these 
correspond to very low energies, \teq{\lesssim 2}keV, that would be swamped by the surface 
and atmospheric emission signals {(e.g., see Fig.~\ref{fig:f10_spec5} below)}.
The orthographic projections in Fig.~\ref{fig:ortho3D_resonance} also clearly exhibit 
dark {\bf shadow regions} where the line of sight to an observer is occulted 
by the star.  For some observer viewing perspectives, the emission regions 
shadowed can have a profound impact on the spectra observed since emission 
is strongly sensitive to the final scattering angle.   For the dipole field morphology 
employed here for flat spacetime, the boundaries of these zones can be 
computed using simple geometric considerations.

\begin{figure}[h!]
\centering
\includegraphics[width=0.99\textwidth]{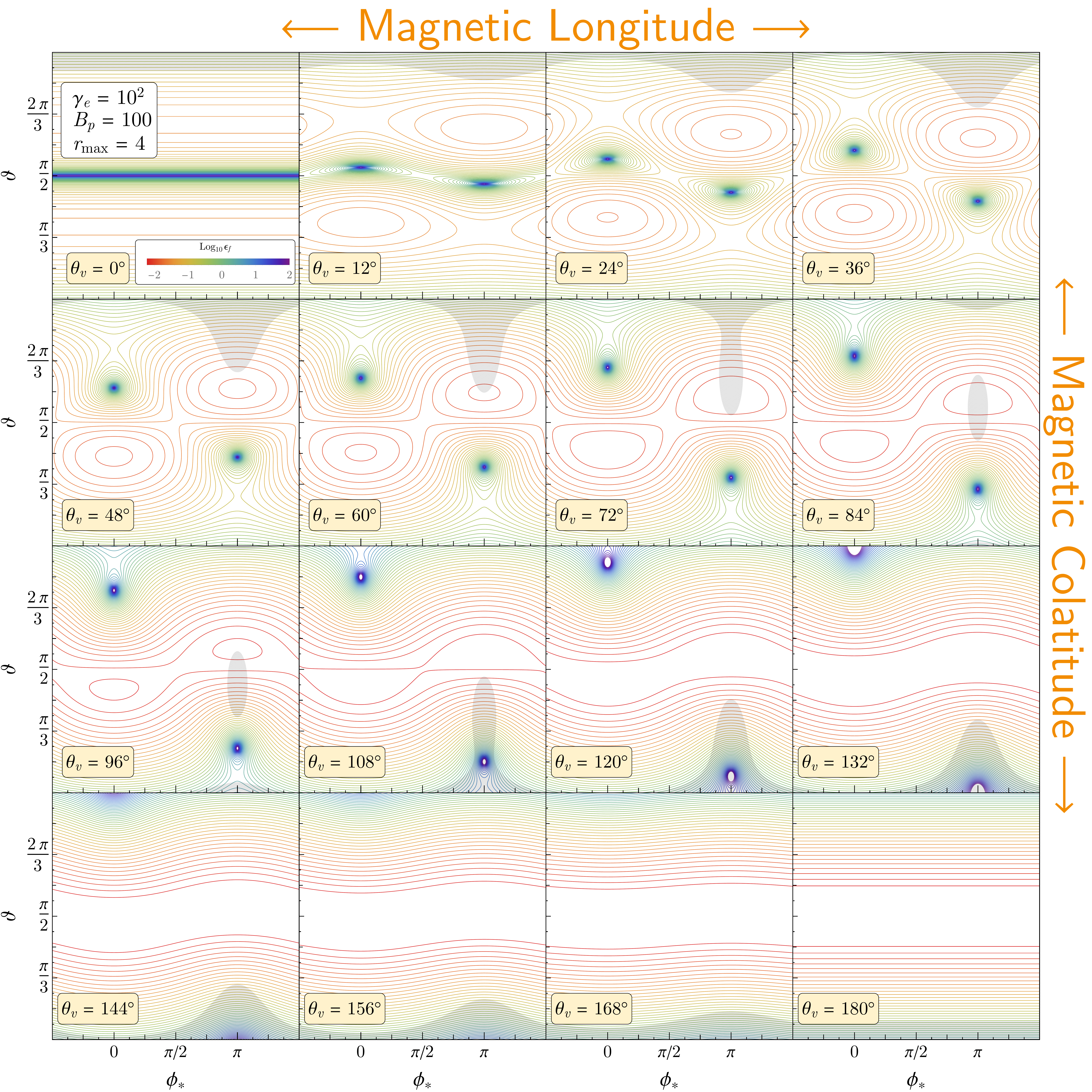} 
\caption{Suite of resonant interaction contour plots
serving as a complement to the 3D orthographic projections in Fig.~\ref{fig:ortho3D_resonance}.
The viewing angle $\theta_v$ relative to \teq{\muBhat} for each panel is fixed at the value indicated therein, 
and the other parameters \teq{B_p=100}, \teq{\gamma_e = 10^2} and \teq{r_{\rm max}=4} 
are identical for all panels, and present a slightly different case from Fig.~\ref{fig:ortho3D_resonance}.
Again, the contours are for fixed final scattering energy $\erg_f$ in the OF and are 
color-coded as indicated by the legend in the upper left.  
plotted here for $B_{\rm p} =100$ and uncooled $\gamma_e = 10^2$. 
The gray region again denotes that of shadowing by the star with respect to the line of sight to the viewer.
The coupling between instantaneous viewing angle \teq{\theta_v} and rotational phase  
\teq{\Omega t/2\pi} is discussed in Section~\ref{sec:modulation}.
}
 \label{fig:resonance_theta_phi_space}    
 \end{figure}

To complement this three-dimensional illustration of the resonant energy \teq{\erg_f} geometry, an 
alternative representation of such information can be provided by projecting the spherical surface 
onto the 2D polar angle/azimuth plane.  This is done using \teq{\vartheta - \phi_{\ast}} coordinates 
in Fig.~\ref{fig:resonance_theta_phi_space}, again with the electrons flowing outward from 
the upper hemisphere; bi-directional flows, for example of pairs accelerated in an
electric field, will generate different resonant interaction phase space plots. 
The case exhibited therein is for somewhat different 
values of the parameters, namely \teq{B_p=100} and \teq{\gamma_e = 10^2},
and for a higher altitude surface, \teq{r_{\rm max}=4}.   These choices lower the value of
\teq{\erg_f^{\rm max}} in Eq.~(\ref{eq:efmax}) by a factor of just over 10 relative to those in
Fig.~\ref{fig:ortho3D_resonance}.  Also, the sixteen panels progress through 
a denser sequence of viewing angles than in Fig.~\ref{fig:ortho3D_resonance}.  These span 
perspectives over the pole, where the hardest resonant emission comes from equatorial 
field lines and the system is azimuthally symmetric (yielding unpulsed emission), 
to instantaneous lines of sight in the magnetic equator where no low-altitude field line tangents point 
to the observer, so that then all resonant emission is softer than around 1 MeV.
As with the orthographic projections, displaying this 2D angular phase space 
clearly illustrates that hard emission above 1 MeV in resonant Compton upscattering 
is confined to only a small solid angle in the magnetosphere, the hallmark of strong 
Doppler boosting.  It is anticipated that due to the field line curvature, such emission 
will likely be attenuated by magnetic pair creation or photon splitting, a prospect 
addressed in the Discussion section, but not detailed numerically in this paper.   Photons 
directed into the remaining solid angle phase space will suffer at most modest or minimal 
such attenuation, and are of energies approximately consistent with those of the observed hard 
X-ray tails.  

To interpret this phase diagram further, since \teq{\phi_{\ast}=\Omega t}
constitutes the rotational phase in a spinning magnetar, the time evolution of the 
sampling of these resonant energy maps is a sinusoidal trace for \teq{\theta_v(t)} 
that is dependent on the inclination angle \teq{\alpha} between the magnetic and 
rotation axes, given specifically in Eq.~(\ref{eq:thetav_t_alp}).  Since \teq{\theta_v}
is fixed for each panel, this evolution effectively amounts to a repetitive rastering 
in a sequence between a subset of the panels.  For most \teq{\alpha ,\zeta} parameters, 
\teq{\theta_v} will not exceed around \teq{135^{\circ}} and meridional and 
anti-meridional configurations at select phases will access the blue/green ``hot spots.'' 
Thus, one expects that the 
maximum energy of resonant emission should be quite sensitive to the pulse phase, 
as long as \teq{\alpha} exceeds around \teq{25^{\circ}}.  This variation, explored 
more in Fig.~\ref{fig:f5_efmax} below, serves to define 
a potentially useful observational diagnostic on the magnetic inclination 
angle \teq{\alpha} for select magnetars; see also the pulse phase properties 
presented in Section~\ref{sec:toroid}.  Finally, observe that the solid angle of 
shadowing (again depicted as gray regions) is substantially diminished in this higher 
altitude case relative to its prevalence in Fig.~\ref{fig:ortho3D_resonance} for 
resonance zones nearer the star.

These two illustrations thus serve to outline two key elements of the overall character of 
resonant Compton upscattering 
spectra in inner magnetospheres: (i) that the flux and spectral shape will be strongly dependent on 
pulse phase for a magnetar, and (ii) adding up over substantial volumes should generate 
a profusion of X-rays, as opposed to hard gamma-rays, in the emergent signal.
The first of these will be explored briefly in Section~\ref{sec:modulation},
the second in Section~\ref{sec:spectra}.

\subsection{Modulation in Oblique Rotators}
 \label{sec:modulation}

When the magnetic moment of the star is inclined with respect to the rotation axis, 
the effective zenith angle of the line of sight to an observer at any point in the magnetosphere 
changes as a function of the rotational phase \teq{\Omega t}.  
A magnetic field loop's longitude or azimuthal angle \teq{\phi_*} then defines 
a periodic sweep of an instantaneous viewing angle \teq{\theta_v} as the star spins about 
its rotation axis.  This is precisely analogous to the path of the sun in the sky to an observer
at a specific longitude and latitude on Earth.  Thus, only twice a period does the line of 
sight to a viewer lie coplanar with the field loop, and at low altitudes often one of these is occulted by the star.
In general, at these instants this plane is not coincident with the contemporaneous
plane defined by the magnetic and rotation axes.  The exception to this arises for meridional 
field loops, those specified by phase \teq{\phi_{\ast}=0} and defined by being co-planar 
with the magnetic dipole and rotation axes exactly twice a rotational period.  It is these 
meridional loops that are emphasized more in the illustrations of this subsection, motivated by 
simplicity of description and interpretation. Azimuthal symmetry about the magnetic field axis 
is presumed for particle number density and Lorentz factors, and consequently the emission 
energies/spectra of the photons. In principle, such quantities can vary with azimuth in the magnetosphere, and an 
example is provided by localized field twists and current bundles that are a defining feature of some models 
(e.g. Beloborodov \& Thompson 2007; Nobili, Turolla \& Zane 2011).  Here we adopt the simpler 
azimuthally symmetric case to more clearly illustrate the general character of resonant Compton upscattering signals.

We define a new coordinate system such that the axis of rotation of the star is 
\teq{\hat{\boldsymbol{\Omega}}= \hat{\bf{z}}_v} for an angular rotation vector 
\teq{{\bf{\Omega}} = \Omega \, \hat{\bf{\Omega}}}. The magnetic obliquity angle 
\teq{\alpha} then allows us to specify the instantaneous magnetic dipole moment unit vector 
\teq{\muBhat}  such that \teq{\hat{\bf{\Omega}} \cdot \muBhat  = \cos \alpha},
\begin{equation}
   \muBhat \; =\; \sin \alpha \cos (\Omega t) \, \hat{\bf{x}}_v 
        + \sin \alpha \sin (\Omega t) \, \hat{\bf{y}}_v + \cos \alpha \, \hat{\bf{z}}_v  \quad ,
 \label{eq:muBhat_def}
\end{equation}
where \teq{\Omega t} is an arbitrary time coordinate for rotation phase. In this coordinate system
(observe that subscripts \teq{v} are employed to distinguish these observer frame triad vectors 
from those in the magnetar rest frame), 
we define the viewer direction is at angle \teq{\theta_{v0}} with respect to the magnetic axis 
\teq{\muBhat} at time \teq{t=0}. The observer vector itself is fixed and independent of time, 
\begin{equation}
   \hat{\bf{n}} \; =\; \sin (\alpha + \theta_{v0}) \, \hat{\bf{x}}_v + \cos (\alpha + \theta_{v0}) \, \hat{\bf{z}}_v \quad .
 \label{eq:hatBn_ident}
\end{equation}
Note that trigonometric double angle identities confirm that \teq{\muBhat (t = 0) \cdot   \hat{\bf{n}}  
= \cos \theta_{v0}}. For general rotational phases, we define the effective viewing angle
\teq{\theta_v \in \{0, \pi \}} as the instantaneous angle between the viewer and magnetic moment,
\begin{equation}
   \cos \theta_{v}\; \equiv\; \muBhat  \cdot  \hat{\bf{n}} 
   \; =\; \sin \alpha \cos (\Omega t) \sin \zeta + \cos \alpha \cos \zeta
   \quad ,\quad
   \zeta \; =\; \alpha + \theta_{v0} \quad ,
 \label{eq:thetav_t_alp}
\end{equation}
with \teq{\sin \theta_v = \sqrt{1- \cos^2 \theta_v}}, since \teq{0\leq \theta_v \leq \pi}.
At phase \teq{ \Omega t = 0} one has \teq{\theta_v = \theta_{v0}}, and we note that 
\teq{\zeta} is the fixed angle between the observer direction and the 
rotation axis, adopting notation familiar to pulsar astronomers.  The other extremum
at phase \teq{\Omega t = \pi} establishes \teq{\cos \theta_v = \cos (2 \alpha + \theta_{v0})},
i.e., \teq{\theta_v = 2\alpha + \theta_{v0}} or \teq{\theta_v = 2(\pi - \alpha ) - \theta_{v0}},
whichever is the smaller.  This algebra defines cases applicable to meridional field loops, 
and those with non-zero azimuthal angles \teq{\phi_*} (magnetic longitudes) relative to the meridian
will possess a mathematical form for \teq{\theta_v(t,\, \alpha)} different from Eq.~(\ref{eq:thetav_t_alp}).  
Varying \teq{\theta_v} with rotational phase then yields fluctuating spectra.  
Analytic pulse profiles as a function of energy can also be constructed 
in the assumed azimuthally-symmetric magnetosphere.  This is most expediently 
achieved if one neglects time-of-flight and light aberration effects induced by the 
rotation for different photon interaction locales, good approximations in 
the inner magnetospheres of magnetars.

\begin{figure}[h!]
\centering
\includegraphics[width=0.999\textwidth]{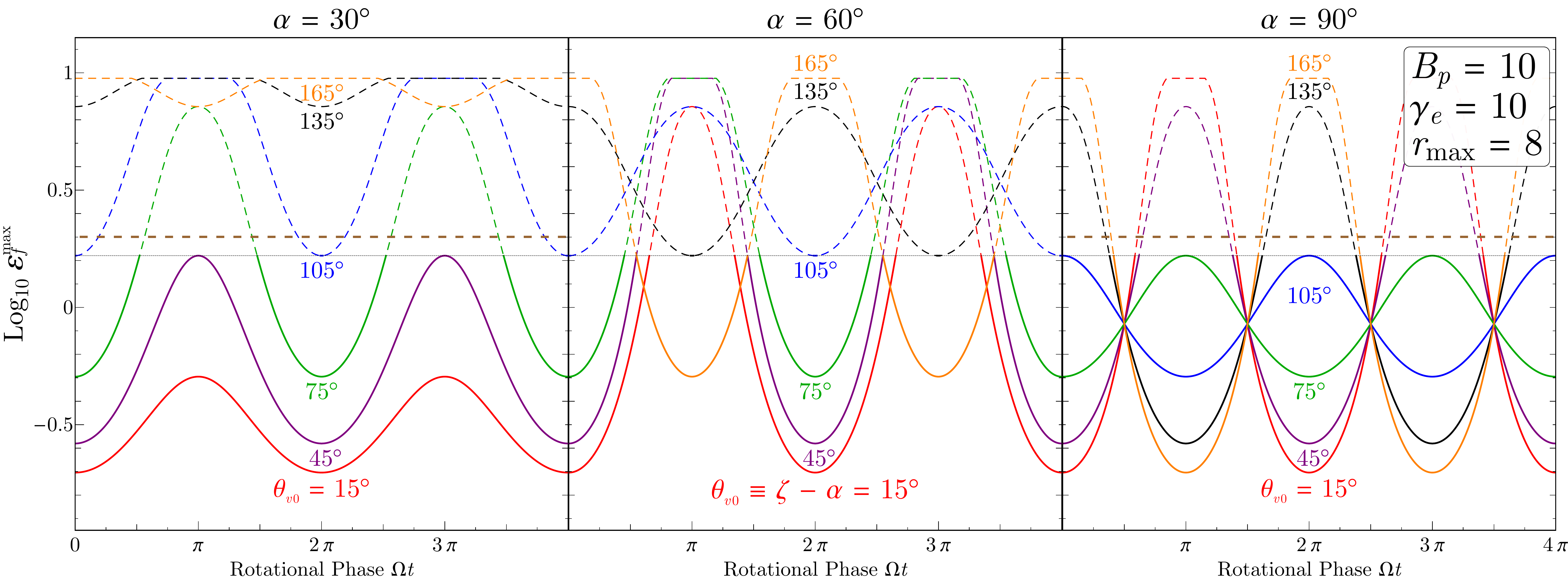} 
\caption{The maximum resonant cut-off energy $\erg_f^{\rm max}$ from Eq.~(\ref{eq:efmax}) 
as a function of rotation phase, for oblique rotators with different magnetic axis 
inclination angles \teq{\alpha = 30^{\circ}, 60^{\circ}, 90^{\circ}} (panels as labelled)
to the rotation axis.  For each panel, profiles are depicted for several choices of 
the viewing angle at the zero phase instant (\teq{\Omega t=0}), 
with \teq{\theta_{v0} \equiv \zeta - \alpha} values labelled.
The cut-off energies are modulated by the stellar rotation, 
and correspond to monoenergetic electrons with an azimuthally-symmetric 
density distribution about the magnetic axis.  The electrons are presumed to 
populate meridional magnetic loops that extend to altitudes \teq{\rmax=8},
propagating from the south magnetic footpoint to its northern counterpart.
The heavyweight solid portions of the curves are for the maximum soft photon energy being 
constrained such that $B_{\rm local} / \gamma_e < \erg_s^{\rm max} = 10^{-2} $, i.e. sampling the bulk of the surface X-rays. 
In contrast, the dashed portions of the curves are unconstrained; see text for a discussion.  
Also shown in the left and right panels are horizontal {dash-dot} lines that represent 
the absolute magnetic pair creation threshold of \teq{2m_ec^2}.
 \label{fig:f5_efmax}  }  
\end{figure}

The phase dependence of the spectra is induced by the rotational 
modulation of the \teq{\ThetaBn} value  as \teq{\theta_v} changes.
In particular, the maximum observed scattered energy beamed 
toward an observer varies with rotational phase when \teq{\alpha \neq 0}. 
This modulation signature is approximately represented by the
variations in the resonant interaction criterion in Eq.~(\ref{eq:efmax}),
mediated by the modulation of instantaneous viewing angles 
in Eq.~(\ref{eq:thetav_t_alp}) that select different resonant interaction locales as portrayed 
in Fig.~\ref{fig:resonance_points}. 
This variation of the characteristic cut-off energy \teq{\erg_f^{\rm max}} in the observer frame 
is illustrated in the triptych of Fig.~\ref{fig:f5_efmax}, for three different magnetic axis inclination angles \teq{\alpha}
and for a neutron star with \teq{B_p=10}.
In each \teq{\alpha} configuration, these maximum energy oscillation profiles are 
computed for six different viewing angles \teq{\theta_{v0}}, and are applicable to the 
meridional field loop.  For non-meridional 
field loops, the modulation profiles generally possess different amplitudes and normalizations, different 
distortions from purely sinusoidal character, and have their extrema shifted in phase
relative to those in the Fig.~\ref{fig:f5_efmax}.  The curves presented are for monoenergetic electrons
of Lorentz factor \teq{\gamma_e=10}, streaming along field lines from near the south 
magnetic pole to near the north one (i.e. closer to the direction of \teq{\muBhat}).
For values where \teq{\sin (\alpha + \theta_{v0}) = 0}, there is no variation of 
\teq{\theta_v} with rotation phase [see Eq.~(\ref{eq:thetav_t_alp})], 
since the viewing direction is parallel to the rotation axis (\teq{\zeta =0}).  Such a 
circumstance is not specifically depicted in Fig.~\ref{fig:f5_efmax} for the chosen 
\teq{\theta_{v0}} values, though small amplitude oscillations for 
\teq{\theta_{v0}} choices sampling the 
neighborhood of \teq{\alpha + \theta_{v0} = \pi} are apparent 
in each of the panels.

All the \teq{\erg_f^{\rm max}} traces exhibit the \teq{2\pi} periodicity in 
\teq{\Omega t} that is manifested in Eq.~(\ref{eq:thetav_t_alp}).  Thus, 
it is evident that local extrema in this energy are separated by \teq{\pi} 
in rotational phase.  The dynamic range in \teq{\erg_f^{\rm max}} between 
these extrema is controlled by the factor \teq{B/(1+ 2B)} in Eq.~(\ref{eq:efmax}), 
where \teq{B} is the local value such that field line is tangent to the observer's
line of sight, i.e. \teq{\cos \ThetaBn = -1}.  Specifically, the range of the modulation 
is fixed by the ratio \teq{B_{\rm p}/\rmax^3} for a particular field loop.  
When \teq{B_{\rm p}} is large
or \teq{\rmax} is small, the factor \teq{B/(1+ 2B)} approaches \teq{1/2} 
at low altitudes.  In contrast, the minimum cutoff energy is realized when 
resonant scattering occurs at a high altitude, thereby sampling low local \teq{B}.
This range is curtailed in 
several instances by a ``capping" at large \teq{\erg_f^{\rm max}}, where
the profiles possess flat portions.  This feature is due to a shadowing effect, 
where \teq{\cos \ThetaBn = -1} tangency cannot be realized outside the star, 
and the maximum energy along a magnetospheric field loop
is established at the southern magnetic footpoint. 

Observe that the \teq{\erg_f^{\rm max}} normalization or \teq{y}-axis 
domain is also controlled by the \teq{\gamma_e(1+\beta_e)} factor 
in Eq.~(\ref{eq:efmax}).  Therefore, raising the value of \teq{\gamma_e} to 100 
would move the curves up in energy by a factor of around ten, but no more.
This then provides insight into the modulation behavior that should arise when 
electrons rapidly cool due to resonant scatterings, thereby precipitating 
spatially-dependent values for \teq{\gamma_e}.  Such variations of Lorentz factor 
with magnetic colatitude along field loops should yield distortions to the traces 
like those exhibited in Fig.~\ref{fig:f5_efmax}, essentially being a convolution of forms 
representing different \teq{\erg_f^{\rm max}} domains and amplitudes at different rotational 
phases.  The result will be distinctly non-sinusoidal phase profiles for \teq{\erg_f^{\rm max}},
perhaps more pronounced than those in the Figure.  Note also that if there are electrons 
and/or positrons moving in two directions along field loops, more complicated 
modulation profiles will result, sometimes exhibiting additional extrema and 
pronounced Fourier power at \teq{2\Omega} frequencies.

Each \teq{\erg_f^{\rm max}} profile consists of two portions, encapsulated in
the solid (and heavyweight) and the dotted phase ranges in the Figure.
The entire profile samples all possible ranges of soft photon energies \teq{\erg_s=\erg_i}
that can contribute to resonant interactions.  However, for the spectra 
computed in Section~\ref{sec:spectra} below, the \teq{\erg_s} distribution 
is the narrow Planck spectrum that approximates the signal emanating from 
a neutron star atmosphere. Many \teq{\erg_i} values that generate 
the \teq{\omega_i=B} resonance condition via the Lorentz boost in
Eq.~(\ref{eq:Lorentz_transform}) do not lie anywhere near the 
peak of the Planck spectrum.  Accordingly, the sampling of the 
Planckian X-ray spectrum, or otherwise, can profoundly influence the 
normalization of resonant Compton upscattering spectra. 
Thus it is informative to introduce an additional restriction that \teq{\erg_i} 
not exceed a maximum value, which we choose to be \teq{\erg_s^{\rm max} = 10^{-2}}.
This chosen energy \teq{\erg_s^{\rm max}} is somewhat above that seen 
in thermal components in magnetars, but comparable to the energies 
seen in the steep, non-thermal, soft X-ray tails at a few keV. 
With this division, the heavyweight portions of the modulation profiles approximately
correspond to \teq{B_{\rm local} \lesssim \gamma_e \erg_s^{\rm max}},
so that the peak of the Planck spectrum can be sampled, and the 
upscattered spectrum is very luminous.  The remaining lightweight dotted 
portions are where the scattering is resonant, but that the 
soft photons participating are deep in the exponential tail of the 
Planck form, and so the normalization of the hard X-ray signal is 
much lower, generally by several orders of magnitude.
For the illustrated \teq{B_p=10}, \teq{\gamma_e=10} case, these two 
domains are partitioned by the lightweight horizontal line in 
Fig.~\ref{fig:f5_efmax} at \teq{\erg_f^{\rm max} \sim 2\gamma_e^2\erg_s^{\rm max}}.
This value is controlled purely by the resonant Compton kinematics,
and is therefore independent of the observational and stellar configuration 
parameters \teq{\alpha} and \teq{\theta_{v0}}, and the pulse phase.
In addition, the magnetic pair creation threshold is marked in the two
side panels, above which pair opacity can act to significantly attenuate 
upscattered photons: the potential for this is addressed in Section~\ref{sec:attenuation}.

It is evident that for these modest uncooled Lorentz factors \teq{\gamma_e\gtrsim 10}, 
this meridional case yields variations that mostly violate COMPTEL upper bounds 
at a few hundred keV for magnetars.  The exception is the \teq{\alpha =30^{\circ}} 
example when \teq{\theta_{v0} \lesssim 30^{\circ}}, for which tangents to 
field lines point in the direction of an observer only in equatorial zones 
at altitudes near \teq{\rmax}.  The local field is then subcritical and 
low enough to reduce \teq{\erg_f^{\rm max}} below 300 keV.  
Yet, this meridional specialization presents the hardest emission that can be beamed along field 
lines towards an observer.  For off-meridional loops, even azimuths \teq{\phi_{\ast}} 
of a few degrees are sufficient to reduce \teq{\erg_f^{\rm max}} values to 
be more or less consistent with the COMPTEL constraints, as softer spectra result.  For all field line 
azimuths, increasing \teq{\gamma_e} will increase the domain of \teq{\erg_f^{\rm max}}
values, and so the conflict with COMPTEL data is exacerbated.  
It is clear that future 3D Monte Carlo photon transport simulations replete with 
Compton cooling and photon absorption due to magnetic 
pair creation and photon splitting are necessary in order 
to generate a more complete picture of the cut-off energy and spectral shape 
as a function of rotation phase.  With the advent of such developments,
it is anticipated that phase-resolved spectroscopy, or equivalently energy-dependent 
pulse profiles, imbued with a wealth of information embedded in amplitudes and 
non-sinusoidal shapes for time 
traces, should provide constraints on the magnetic inclination \teq{\alpha}
of the rotating magnetar and the altitude (i.e., \teq{\rmax}) of the scattering region
(see Section~\ref{sec:toroid}), and also the action of attenuation processes 
at hard X-ray and soft gamma-ray energies.

\section{COMPTON UPSCATTERING SPECTRA}
 \label{sec:spectra}

In this Section computations of representative spectra from individual field 
loops are illustrated, highlighting important features while exploring the phase
space of key parameters; they serve as templates for future
integrations over substantial magnetospheric volumes with prescribed lepton injections. 
In this resonant Compton upscattering exposition, which does not treat electron cooling
self-consistently, there are six parameters: the electron Lorentz
factor \teq{\gamma_e}, the instantaneous viewing angle \teq{\theta_v},
the stellar surface temperature \teq{T=\Theta m_ec^2/k} (presumed
uniform across all colatitudes), the polar dipole field strength
\teq{B_{\rm p}}, the field loop extent \teq{\rmax}, and the
azimuthal location with respect to the line-of-sight, given by
\teq{\phi_{\ast}}. The spectra are computed using
Eq.~(\ref{eq:scatt_spec_fin}) with the Planck form in
Eq.~(\ref{eq:Planck_spec}) inserted, modeling a uniformly bright spherical
surface threaded by a dipole magnetic field.  The spin-dependent Sokolov \& Ternov
cross section and widths are employed as described in
Section~\ref{sec:formalism}, patched with the spin-averaged cross section
away from the resonance, using the hybrid procedure adopted in
BWG11. Analytic developments and approximations for the
integral over the soft photon angular distribution \teq{f(\mu_i)} at the site 
of scattering are presented in the Appendix; these facilitate efficient numerical
computation of the spectra from the field loops. Portions of field line arcs
that are shadowed are excluded from the integral
Eq.~(\ref{eq:scatt_spec_fin});  for magnetic dipoles in flat spacetime, the boundary of such occulted
regions can be determined in terms of \teq{\vartheta} via a 
root solving protocol for a sixth order polynomial in \teq{\sin^2\vartheta}.

In this paper, to streamline the information conveyance and
reduce the number of variables, we fix the stellar temperature to be
\teq{T = 5\times 10^6}K in all the spectra presented in this Section,
corresponding to \teq{0.43} keV.  This is a typical 
temperature for magnetar thermal X-ray emission. 
The actual local surface temperature is probably 
around a factor of 1.4 higher, because of redshifting during propagation in the 
general relativistic metric.  Yet, at the typical altitudes the scattering events
are sampled in the figures of this section, gravitational redshifting has 
mostly occurred, and so the local effective temperature of the X-rays from 
the stellar surface is close to that observed at infinity.
Resonant Compton cooling calculations (see Figure 4
of BWG11) have exhibited a strong temperature dependence for the
electron cooling rates.  Higher temperatures are vastly more efficient
for cooling, due to the increased total number of photons, 
\teq{n_{\gamma}\propto T^3}, in the Planck
population.   Moreover, the Lorentz factors \teq{\gamma_e} where
resonant cooling is at a maximum decline with surface temperature as
\teq{1/T}, a property dictated by resonant Compton kinematics, namely
\teq{\gamma_e \Theta \sim B}. Hence, it is expected that computations of
spectra with higher temperatures than employed here will yield much
higher fluxes, and will correspond to slightly higher altitudes on
average for resonant interactions, where the fields are lower.  This can
then yield substantial changes to the detailed shape of spectra when all
other parameters are kept fixed.  Notwithstanding, by varying a number
of different parameters in the ensuing figures, the breadth of spectral
character encompassed by surface temperature variations is effectively
captured in our presentation.  We note that future work where electron
cooling and upscattering emission is to be treated coincidently, it will
be important to include a non-uniform temperature profile. A commonly
invoked prescription for the colatitudinal variation, underpinned by
consideration of anisotropic thermal conductivities in strong field environs, is that of
Greenstein \& Hartke (1983), which has a pole hotter than the equator
in neutron stars of moderate magnetization.  In the context 
of magnetars, coronal outflow models (Hasco\"et et al. 2014) 
for the quiescent hard X-ray emission have particles that radiate along 
closed field loops and return to the neutron 
star surface, thereby producing a hot spot at the footpoints of the active loops; such 
would also be expected in our scenario here.
Such return currents also generate hotter surfaces away from the equator.
We note that NASA's new NICER mission, recently deployed on the International 
Space Station, will afford measurements of such surface temperature 
non-uniformities, for millisecond pulsars in particular.

\subsection{Emission from Uncooled Electrons on Individual Field Lines}
 \label{sec:spec_loops}

\begin{figure}[h!]
\centering
\includegraphics[width=0.99\textwidth]{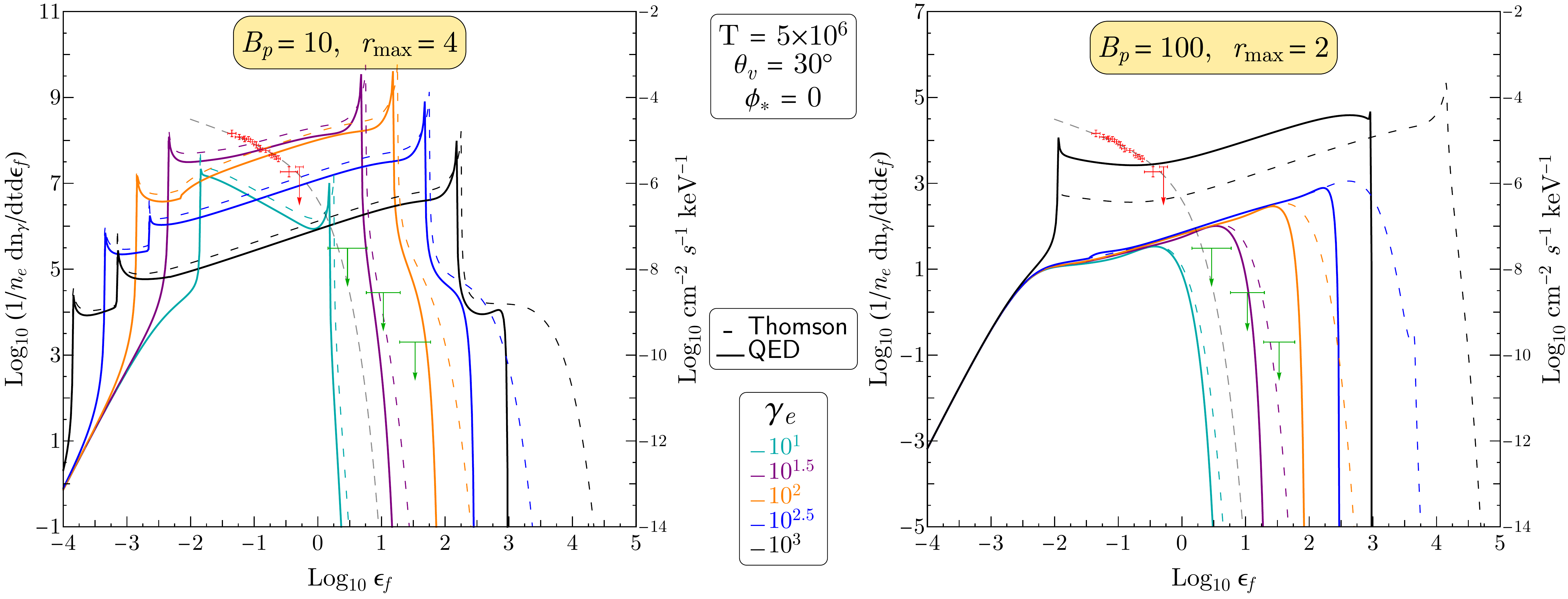} 
\caption{Spectra I: meridional field loops. Both left and right panels illustrate spectra computed 
for meridional $\phi_* =0$  field loops for a viewing angle of $\theta_v = 30^\circ$ with 
fixed Lorentz factors ranging from $\gamma_e = 10^1 - 10^3$.  Solid curves 
represent spectra computed with the full ST cross section in QED,
i.e., Eq.~(\ref{eq:dsig_resonance}); lighter weight dash-dot curves define spectra 
determined using elastic kinematics and the magnetic Thomson cross section 
instead (see text).  The left panel illustrates 
higher-altitude and lower-field directed spectra computed for $B_{\rm p} = 10$ and $\rmax = 4$ 
where the resonant interactions are readily sampled for Lorentz factors $>10^2$. 
The right panel's local $B$ is much higher, by a factor of $\sim 80$, with $\rmax =2$ and 
$B_{\rm p} = 100$ so resonant interactions near equatorial regions for the given stellar 
temperature are not realized unless Lorentz factors are much higher. Overlaid on the 
computed spectra, with arbitrary normalization, are observational data points for AXP 4U 0142+61 
(den Hartog et al. 2008b) along with a schematic \teq{\erg_f^{-1/2}} power-law with 
a \teq{250} keV exponential cutoff (gray dashed curve).
 \label{fig:f6_spec1} } 
\end{figure}

\begin{figure}[h!]
\centering
\includegraphics[width=0.99\textwidth]{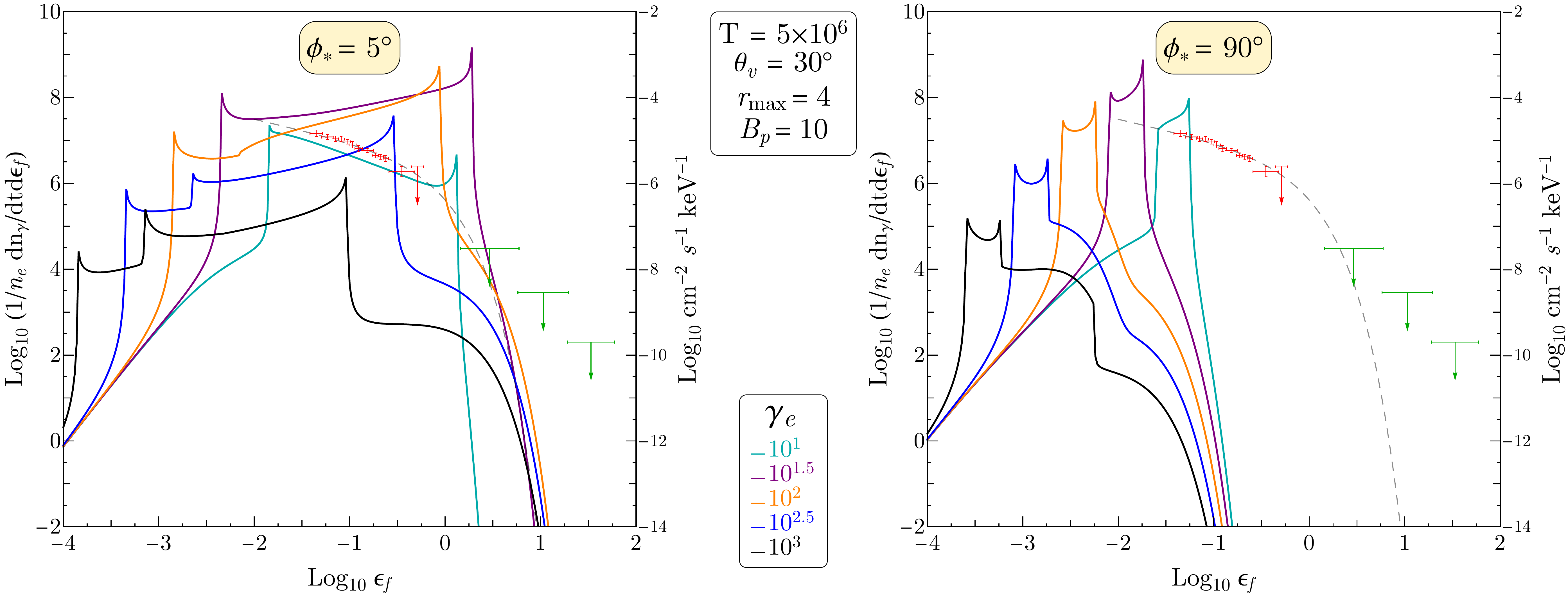} 
\caption{Spectra II: Off-Meridional field loops. Adopting most of the parameters of the 
left panel of Figure~\ref{fig:f6_spec1}, with \teq{B_p=10} and \teq{\rmax =4}, 
both left and right panels here illustrate spectra computed for off-meridional field loops 
\teq{\phi_* = 5^\circ} and  \teq{\phi_* = 90^\circ} respectively, for a viewing angle of 
\teq{\theta_v = 30^\circ} with fixed Lorentz factors ranging from \teq{\gamma_e = 10^1 - 10^3}. 
For loops somewhat away from the meridian (or anti-meridian), the hardest emission 
is never beamed towards the observer.  In these off-meridional examples,  
resonant interactions are realized only for low final scattering energies \teq{\erg_f}
and in polar regions of the loop.
 \label{fig:f7_spec2}  } 
\end{figure}

\begin{figure}[h!]
\centering
\includegraphics[width=0.99\textwidth]{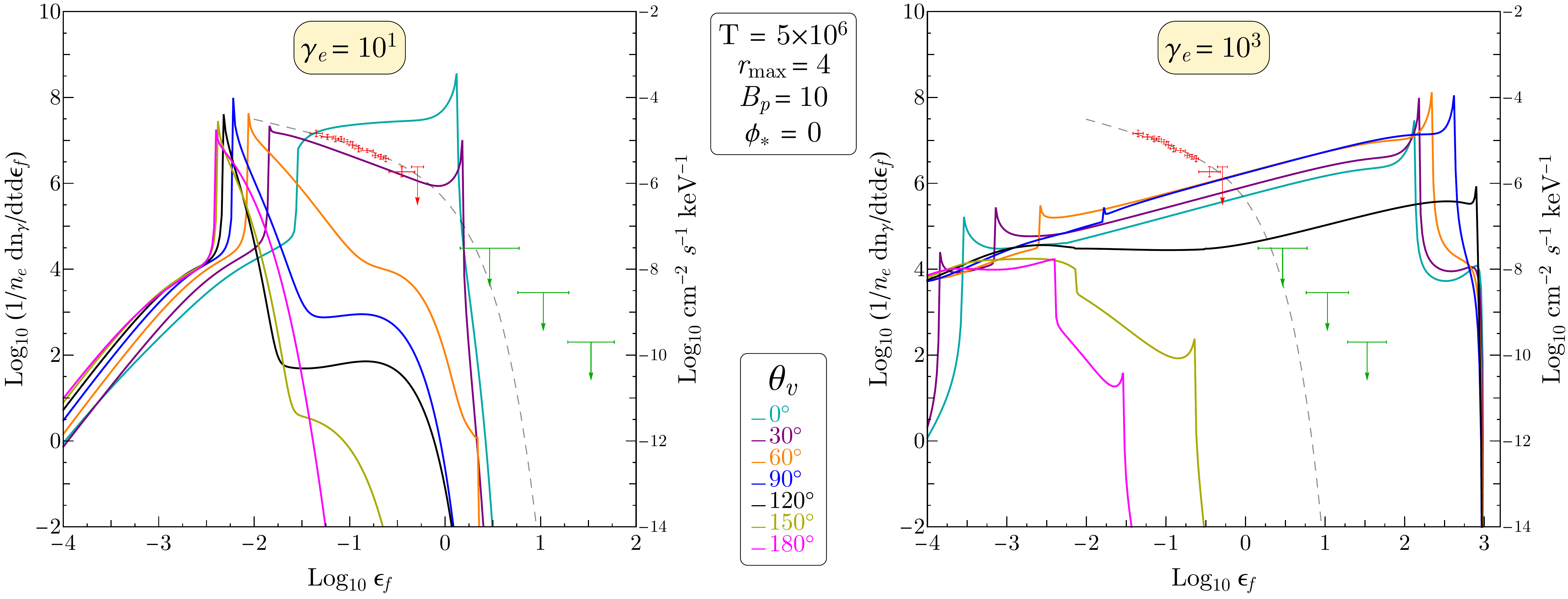} 
\caption{Spectra III: Spectra as a function of viewing angle for meridional field loops.
Both left and right panels present seven curves of varied viewing angle 
$\theta_v = 0^\circ , 30^{\circ}, 60^{\circ} \dots 180^\circ$ with respect to the magnetic axis, for meridional 
$\phi_* =0$ with $\rmax =4$ and $B_{\rm p} = 10$. The left panel is fixed at 
$\gamma_e = 10$ while the right panel is for $\gamma_e =10^3$. The relatively 
low Lorentz factor for the left panel is insufficient to fully sample resonant interactions 
in the peak of the Planckian incoming soft photon distribution, while the Lorentz factor 
of $10^3$ for the right panel violates COMPTEL bounds for most viewing angles. 
 \label{fig:f8_spec3} } 
\end{figure}

\begin{figure}[t]
\centering
\includegraphics[width=0.99\textwidth]{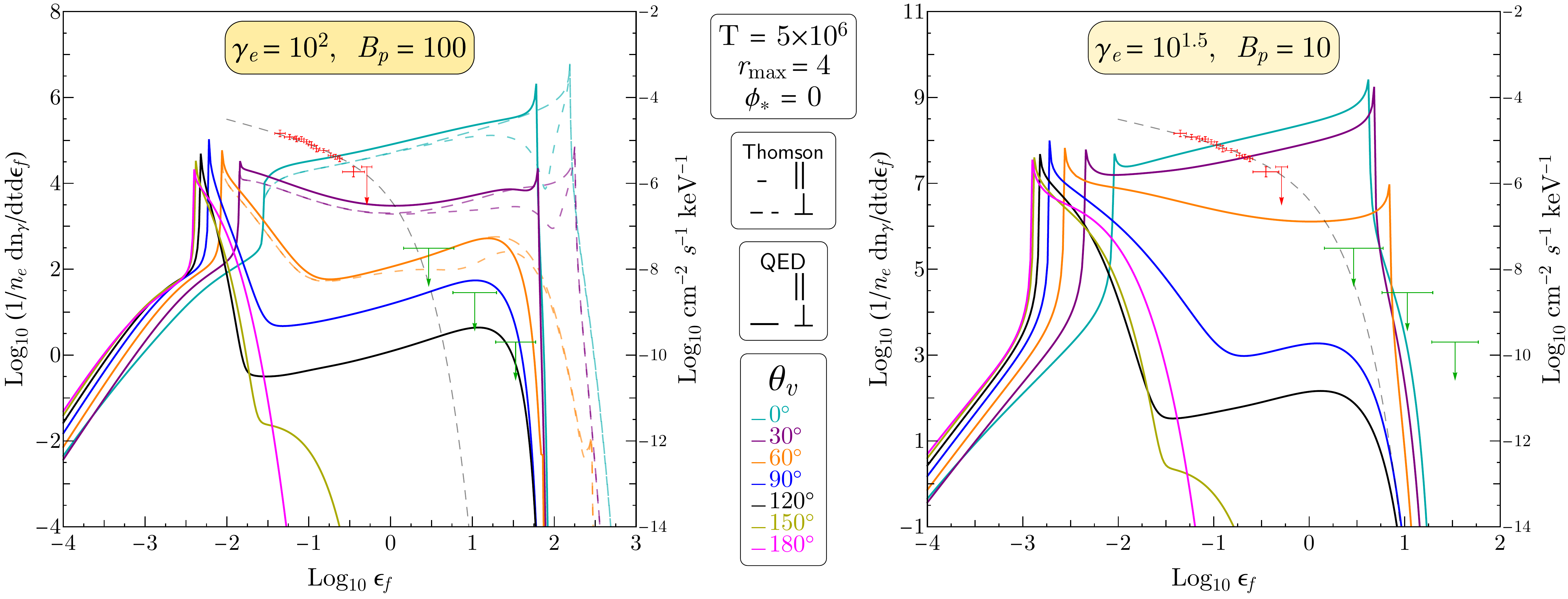} 
\caption{Spectra IV: Meridional field loops at (left) \teq{\gamma_e = 10^{2}} and (right) 
\teq{\gamma_e = 10^{1.5}} as a function of viewing angle, both with \teq{\rmax =4} 
but differing in local \teq{\boldsymbol{B}}.  Solid curves 
represent spectra computed with the full Sokolov \& Ternov (ST) cross section in QED,
i.e., Eq.~(\ref{eq:dsig_resonance}); dashed (\teq{\perp} mode) and dot-dashed (\teq{\parallel} mode)
curves in the left panel define spectra determined using the magnetic Thomson cross section 
instead (see text).  The left panel has relatively high 
local field at the resonant interaction point, since \teq{B_{\rm p} = 100},
while the right panel illustrates the same parameters as those in Figure~\ref{fig:f8_spec3} 
but with \teq{\gamma_e=10^{1.5}}.  It is apparent that polarization 
\teq{\perp} (solid curves) exceeds \teq{\parallel} (dotted curves) in the 
\teq{0.05-1} MeV hard X-ray band for most viewing angles where head-on resonant 
interactions are sampled. 
\label{fig:f9_spec4}  } 
\end{figure}

A suite of results for the upscattering spectra that encompass a substantial range 
in character is presented in Figs.~\ref{fig:f6_spec1}--\ref{fig:f9_spec4}.  These
spectra for uncooled electrons can qualitatively be understood by a careful inspection 
of Figs.~\ref{fig:omegai_B_curves}--\ref{fig:ortho3D_resonance}, which depict
information concerning the resonant interaction points, and the relevant geometric and 
kinematic relations in Section~\ref{sec:res_criteria}.
For meridional field loops with viewing angles that 
readily sample the Doppler boosting and beaming,
the combination of \teq{\gamma_e} and local \teq{B \sim B_{\rm p}/\rmax^3} 
essentially controls the onset of resonant interactions in a manner similar to 
that for the cooling rates
in BWG11, in that $\gamma_e \sim B/\Theta$ is required to access 
the Wien peak of the Planck spectrum.  Moreover, for resonant interactions, 
the \teq{\erg_f\ll \gamma_e} specialization of Eq.~(\ref{eq:wi_ef_res}) indicates that
the scaling \teq{\hat{\omega}_i \sim B_{\rm p}/\rmax^3 \sim \gamma_e \erg_f (1 + \cos \ThetaBn)} operates.
Accordingly, increasing values of \teq{\gamma_e} must compensate for large \teq{B_{\rm p}/\rmax^3}
in defining similar spectral character.
This kinematic scaling for \teq{\hat{\omega}_i} essentially controls the spectral index of 
the resonant spectra for uncooled electrons, a core property that will be discussed shortly.
Not all spectra evince frequency ranges where resonant interactions are accessible:
for values of local \teq{B} that are large, resonant interactions in the Wien peak are 
often not fully sampled, as is evident from computed spectra presented in 
the right panel of Fig.~\ref{fig:f6_spec1} and the left panels of Figs.~\ref{fig:f8_spec3} and~\ref{fig:f9_spec4} below.
In particular, for the \teq{B_p=100} example on the right of Fig.~\ref{fig:f6_spec1}, since 
\teq{\rmax=2}, resonant conditions \teq{\gamma_e\Theta \sim B} are never sampled 
along the chosen field loop for all \teq{\gamma_e \geq 10}, but would be if \teq{\rmax} were increased by a 
factor of a few.  By the same token, in the left panel of Fig.~\ref{fig:f6_spec1}, once \teq{\gamma_e} 
drops below around \teq{30}, resonant interactions at \teq{\gamma_e\Theta \sim B} are not 
accessed in this \teq{\rmax=4} example.  In this domain, the upscattering kinematics 
are similar to those for the non-magnetic inverse Compton process, and so the maximum 
energy scales roughly as \teq{\gamma_e^2\Theta}, a trend that is evident in the 
right panel of Fig.~\ref{fig:f6_spec1}.  Note that results are not depicted for \teq{\gamma_e} 
values lower than 10 because then the photons are no longer roughly beamed along \teq{\boldsymbol{B}} 
in the ERF, and a more general, complicated form for the scattering cross section 
that involves multiple resonant harmonics of the cyclotron fundamental must 
be employed instead of that outlined in Section~\ref{sec:formalism}.

\begin{figure}[h!]
\centering
\includegraphics[width=0.99\textwidth]{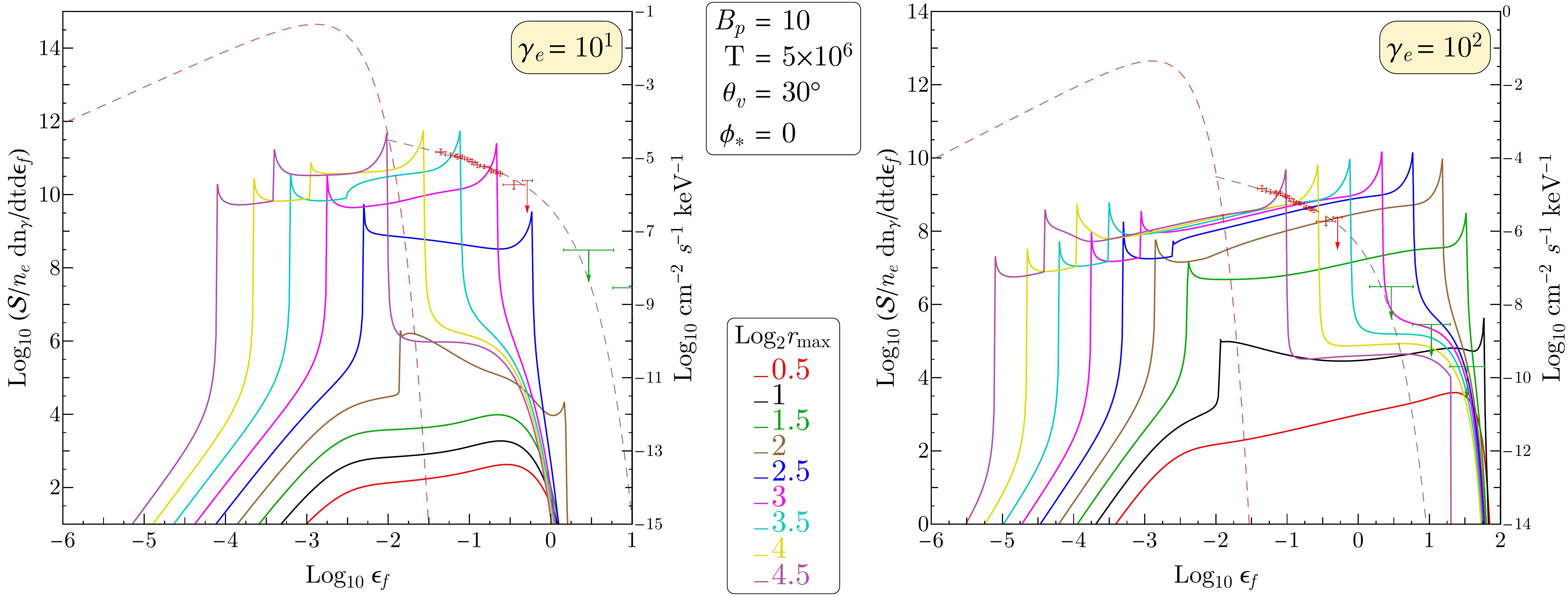} 
\caption{Spectra V: Upscattering spectra for meridional field loops (\teq{\phi^{\ast}=0})
depicting nine choices of the maximum loop altitude parameter 
\teq{r_{\rm max} = \{2^{0.5},...,2^{4.5}\}}.  The two panels are for two different 
Lorentz factors, and spectra are realized for a viewing angle of 
\teq{\theta_v = 30^\circ}. Clear variation in the cut-off energies and normalization are evident, 
as well as a transition from weakly-resonant to fully-resonant from low to 
moderate and higher altitudes. Other parameters are similar to those in the 
previous four spectral figures: \teq{B_p=10} and \teq{T = 5\times 10^6} K.
The superposition of these curves gives an indication of spectra that might result from 
toroidal volumes: see text.  
{As with Figs. 5--9, observational data points for AXP 4U 0142+61 are overlaid, 
along with a schematic \teq{\erg_f^{-1/2}} power-law with 
a \teq{250} keV exponential cutoff (gray dashed curve).  In addition, the Planck spectrum 
matching the soft X-ray data for this magnetar is indicated by the brown dashed curve in each panel.}
\label{fig:f10_spec5} } 
\end{figure}

A distinctive feature in these four figures is that
the boundaries for resonant interactions in spectra are characterized by the ``horns" or cusps. 
These appear both at low values \teq{\erg_f} in the soft X-rays/EUV, where they 
would be dominated by the surface emission signal 
{(depicted in Fig.~\ref{fig:f10_spec5})}, and also in the hard X-ray 
and gamma-ray domains: see the left panel of Fig.~\ref{fig:f6_spec1} and the right panels 
of Figs.~\ref{fig:f7_spec2}, \ref{fig:f8_spec3} and \ref{fig:f9_spec4}.  The cusps mark the 
kinematic extremities of the range of \teq{\erg_f} energies for which resonant scatterings 
arise: they correspond to the \teq{\cos\ThetaBn \approx -1} criterion.  At energies \teq{\erg_f} 
between the cusps, the spectra are quite smooth due to the intrinsic spread in 
soft X-ray photon energies and angle cosines \teq{\mu_i}.  In this window,
the scattering mostly corresponds approximately to Thomson kinematics,
and there is a one-to-one correspondence between the final scattering angle 
along a field line beamed towards the observer, defined by \teq{\ThetaBn},
and the final scattering energy, \teq{\erg_f}.  From Eq.~(\ref{eq:wi_ef_res}), this coupling is
\teq{B\sim \gamma_e \erg_f (1 + \cos \ThetaBn)}, which is in concordance
with the uniform field case explored in Baring and Harding (2007). 
The low \teq{\erg_f} cusps are realized for large \teq{\pi -\ThetaBn}, 
where the angle of scattering in the ERF is modest or small.
From Fig.~\ref{fig:omegai_B_curves} one can discern 
that they correspond to interaction colatitudes well removed from the \teq{\vartheta_0} value,
sampling two disparate \teq{\vartheta} values.  Moreover, these two values 
correspond to two different field strengths, so that the resonant kinematics 
embodied in Eqs.~(\ref{eq:Lorentz_transform}) and~(\ref{eq:res_kinematics_inversion}) indicate that the two 
low energy cusps should appear at different \teq{\erg_f}, as is apparent in the spectral figures.
The hard X-ray/gamma-ray cusps are realized near \teq{\vartheta_0} when backscattering of the incoming 
photon in the ERF occurs, and for meridional field loops this arises 
at a scattered energy of \teq{\erg_f^{\rm max}} given by Eq.~(\ref{eq:efmax}).
It follows that \teq{\erg_f^{\rm max}\propto \gamma_e}, behavior that is evident 
in the left panels of Figs.~\ref{fig:f6_spec1} and~\ref{fig:f7_spec2}.
The prominence of the horns at \teq{\erg_f^{\rm max}} is partly a consequence of 
Klein-Nishina kinematics at the highest energies introducing a peak to the Jacobian 
in Eq.~(\ref{eq:omega_if_deriv}), which appears in Eq.~(\ref{eq:scatt_spec_fin}).  
But mostly, their conspicuousness is significantly enhanced 
by the integration over a comparatively long field line arc when the viewer line of sight 
is almost tangent to the field; in contrast, such a circumstance is not evinced by the large solid angle
integrations presented in Baring \& Harding (2007).
The narrow peaks of the horns are weighted images of the resonant differential cross section 
in Eq.~(\ref{eq:dsig_resonance}).  The low flux wings to these cusps outside the main 
resonant interaction range, specifically the left wing in EUV/soft X-rays, and the high frequency 
wing in hard X-rays/gamma-rays, are imprints of a convolution of the resonant cross section 
and the Planck distribution of the soft X-ray surface photons.  
We also note a nuance --- for the cusp energy \teq{\erg_f^{\rm max}},
the pairs of resonance locales illustrated in Fig.~\ref{fig:resonance_points}
are approximately defined by symmetric solutions \teq{\vartheta = \vartheta_0 \pm \Delta \vartheta}
to Eq.~(\ref{eq:Kfunc_Psi_def}) that represent two sides of a Lorentz cone; thus their cusp contributions 
coincide in energy and the peaks are not two-pronged, contrasting the situation
for the low-energy cusps.

The quasi-power-law dependence between widely-separated horns/cusps is another essential 
characteristic of the spectra presented in this Section.  For most of the range in \teq{\erg_f}, 
the spectra that sample resonant interactions possess a characteristic scaling 
\teq{dn/(dt d\erg_f) \sim \erg_f^{1/2}}, i.e. are extremely flat. This approximate 
power-law dependence is a consequence of kinematics and magnetospheric 
geometry, and can be simply derived by considering the analytics
associated with the description of Figs.~\ref{fig:omegai_B_curves} 
and~\ref{fig:resonance_points}.  Specifically, \teq{\hat{\omega}_i = \gamma_e \erg_f (1 + \cos \ThetaBn)}
defines the tight coupling between the energy \teq{\erg_f}
of the scattered photon and the scattering angle \teq{\ThetaBn}.
For near-meridional cases where \teq{\phi_{\ast}} is very small,
i.e., \teq{\phi_{\ast} \lesssim 1/\gamma_e},
the high-energy cusp corresponds to the Doppler-shifted \teq{\cos\ThetaBn \approx -1}
solution where the viewer's line of sight aligns with the field line at the point of 
scattering (tangent point).  As the interaction points
move along a field loop away from this point (an extremum in \teq{\cos\ThetaBn}), the correlation between arclength 
\teq{s} and angle \teq{\ThetaBn} or \teq{\vartheta} is inherently quadratic, 
i.e. \teq{\Delta s \propto (\pi - \ThetaBn)^2}.  The pairs of resonant interaction points 
defined by \teq{\vartheta = \vartheta_{\pm}} then diverge in a one-to-one correspondence 
with declining energy \teq{\erg_f}, a property that is evident in Fig.~\ref{fig:resonance_points}.
Specifically, inspection of Eq.~(\ref{eq:wi_ef_res}), and in particular Eq.~(\ref{eq:Kfunc_Psi_def}), 
reveals the approximate correlation \teq{(\vartheta -\vartheta_0)^2 \propto (\pi - \ThetaBn)^2 \propto \erg_f^{-1}} as 
the scattering locales move away from the tangent point along a field arc.  It then
follows from Eq.~(\ref{eq:wi_ef_res}) that \teq{\hat{\omega}_i \propto \erg_f(\vartheta - \vartheta_0)^2} so that
\teq{\partial \hat{\omega}_i/\partial\vartheta \propto \erg_f(\vartheta - \vartheta_0) \propto \erg_f^{1/2}}
for \teq{\vartheta \to\vartheta_{\pm}}.
Now the \teq{\erg_f} dependence in the spectrum in Eq.~(\ref{eq:scatt_spec_fin}) 
appears almost totally due to two factors, namely \teq{\hat{\omega}_i} and the differential 
cross section:
\begin{equation}
   \dover{dn_{\gamma}}{dt\, d\erg_f} \; \propto\; 
   \int_{S} ds \, \hat{\omega}_i  \dover{d\sigma}{d(\cos\theta_f) }  
   \;\propto\;  \int_{S} d\vartheta \, \hat{\omega}_i \, \delta(\hat{\omega}_i - B)
   \;\propto\; \sum_{\pm} \int_{S} d\vartheta \,\hat{\omega}_i\, \dover{ \delta (\vartheta - \vartheta_{\pm})}{\vert \partial \hat{\omega}_i/\partial\vartheta \vert}
   \;\propto\; \erg_f^{1/2}\quad .
 \label{eq:scatt_spec_propto}
\end{equation}
In this manipulation sequence, the extremely narrow resonant differential cross
section (polarization-averaged and spin-summed) in Eq.~(\ref{eq:dsig_resonance})
is replaced by the delta function \teq{\delta(\hat{\omega}_i - B)}, a common
protocol (e.g. see Baring \& Harding 2007); 
this is then recast as a delta function in
terms of the resonant interaction colatitudes \teq{\vartheta_{\pm}}.  Also,
\teq{ds\propto d\vartheta} is a proportionality that possesses only weak
dependence on \teq{\erg_f}. The \teq{\erg_f^{1/2}} spectral dependence then
quickly follows.  The remaining factors in Eq.~(\ref{eq:scatt_spec_fin}) are
largely immaterial to this determination.  First, the Jacobian \teq{\vert
\partial \omega_i /\partial \omega_f \vert} is approximately unity in this
\teq{\erg_f} range because it corresponds mostly to forward scatterings in the
ERF, which inherently access Thomson scattering kinematics.  Second, the
integration over the soft photon spectrum is only weakly dependent on
\teq{\erg_f}, since generally the bulk of the Planck distribution is accessed
for upscattered photon energies between the cusps; this is no longer the case
outside the resonant interaction domain, and also in the neighborhood of the
cusps.  Furthermore, near the high-energy cusp, when \teq{B\gtrsim 1}, the
scattering kinematics depart from the true Thomson domain, as was observed by
Baring \& Harding (2007). This is also true for the \teq{\gamma_e\lesssim 30}
examples in these figures, where Klein-Nishina reductions are more prevalent and
lead to a general steepening of spectra in several curves.

To provide a benchmark for how important relativistic quantum influences are, it
is insightful to provide spectral computations using just magnetic Thomson
scattering formalism.  These inherently non-relativistic cross sections are
often invoked for expediency in studies of inverse Compton scattering in neutron
star contexts: see for example Dermer (1990) for old gamma-ray burst models, and
Beloborodov (2013a) for application to magnetar X-ray tail emission.  The
magnetic Thomson physics can be derived using QED techniques in the
\teq{\omega_i \approx \omega_f \ll 1} regime (e.g. see Herold 1979), but remain
an essentially classical electrodynamics result derivable using dipole radiation
formalism (Canuto, Lodenquai \& Ruderman 1971), possessing a single fundamental
resonance at the cyclotron energy. Hard X-ray tail spectra are also presented in
Fig.~\ref{fig:f6_spec1} for runs where the magnetic Thomson differential cross
section was employed; these are the dash-dot curves. These magnetic Thomson
spectra were determined using Eq.~(23) of BWG11 for the polarization-averaged
cross section that neglects spin dependence in the cyclotron resonance. In both
panels, the non-relativistic, spin-averaged cyclotron decay width \teq{\Gamma =
2\fsc B^2/3} was used
to cap the resonance by forming a Lorentz profile.  In addition, elasticity
\teq{\omega_f=\omega_i} in the scatterings was presumed in the ERF, so that
electron recoil energies were neglected, following normal Thomson scattering
protocols.

Comparison with the QED-originating curves clearly reveals that the magnetic
Thomson spectra do not cut off at \teq{\erg_f < \gamma_e} in cases where
\teq{\gamma_e kT/m_ec^2 >1}, and therefore violate energy conservation, as
expected. The cutoffs are generally around \teq{\erg_f^{\rm max} \sim \gamma_e
B}, for a local field strength \teq{B}, and so this problem becomes worse at low
altitudes and for higher magnetar \teq{B_p} values.  These violations are
particularly obvious when \teq{\gamma_e \gtrsim10^2}. Observe that the results
for lower Lorentz factors in the right panel are quite close to those in the
left panel of Fig.~\ref{fig:f6_spec1}; this is because at the resonant
scattering locale along the \teq{\rmax =4} loop, the field strength is
substantially sub-critical, and scattering is approximately in the magnetic
Thomson domain. The Thomson illustrations often over-estimate the resonant
spectrum relative to the QED-computed emission, for example in the left panel.
The relative normalizations of the Thomson and full QED spectra are subject to
two main influences that compete against each other in opposite senses.  The
first is that the magnetic Thomson cross section does not incorporate magnetic
Klein-Nishina reductions as does the exact QED form (e.g. see Fig.~2 of Gonthier
et al. 2000), and thereby tends to increase the emissivity.  The second is that
the non-relativistic cyclotron decay width \teq{\Gamma = 2\fsc B^2/3} is
substantially larger than the full QED value when \teq{B\gtrsim 0.3 B_{\rm cr}}, and since the resonant spectra scale as the
equivalent width \teq{1/\Gamma}, the Thomson approximation can yield
under-estimates of the spectra  -- see the \teq{\gamma_e = 10^3} example in the
right panel. Thus, in summation, the normalization and cut-off energies may be
significantly inaccurate in the magnetic Thomson case; these quantities control
the yield of any pair creation that might be precipitated (see
Section~\ref{sec:attenuation}) by super-MeV emission. Yet, the resonant spectral
index for the magnetic Thomson invocation is fairly similar to that realized in
correct QED computations.

Fig.~\ref{fig:f7_spec2} illustrates the spectral character of emission from
off-meridional field loops.  Increasing \teq{\phi_{\ast}}
reduces the maximum resonant \teq{\erg_f} and the minimum upscattered energy produced by the 
cyclotron resonance, a trend that is evident in the 3D orthographic 
projections of Fig.~3. In essence, for large \teq{\phi_*}, i.e. far off the meridian, the only resonant interactions 
sampled are those corresponding to the black dots on the red curves 
in Fig.~\ref{fig:omegai_B_curves} in quasi-polar regions of a field loop.
For such circumstances, the value of \teq{\ThetaBn} sampled by the resonant interactions is generally 
well-removed from \teq{\pi}, so that the quasi-Thomson scattering kinematics generates resonant 
\teq{\erg_f} ranges much restricted relative to those for meridional cases:
as \teq{\phi_{\ast}} increases, the cusps move closer to each other in energy.  To detail 
this variation a little, observe that as the azimuthal angle of a loop increases from 
the meridional case of zero, in the vicinity of the \teq{\ThetaBn = \pi} resonance locale,
the line of sight direction establishes a correlation 
\teq{1 + \cos \ThetaBn \sim 1-\cos\phi_{\ast}} with the field loop longitude.  This applies 
once the viewer direction \teq{\hat{\bf{n}}_v} lies outside the Lorentz cone for the scattering, 
i.e. \teq{\phi_{\ast}\gtrsim 1/\gamma_e}.  Then the approximate kinematic coupling 
\teq{B\sim \gamma_e \erg_f (1 + \cos \ThetaBn)} at resonance, deduced from Eq.~(\ref{eq:wi_ef_res}),
translates to an approximate correlation 
\begin{equation}
   \gamma_e \erg_f ( 1-\cos\phi_{\ast} ) \; \sim\; B
   \quad \Rightarrow\quad
   \erg_f^{\rm max} \;\propto\; \dover{1}{1-\cos\phi_{\ast}}
   \quad \hbox{for}\quad
   \dover{1}{\gamma_e} \;\lesssim\; \phi_{\ast}\; \ll\; 1\quad . 
 \label{eq:ergf_azimuth_form}
\end{equation}
Thus, \teq{ \erg_f^{\rm max}\sim \phi_{\ast}^{-2}} declines as \teq{\phi_{\ast}} increases and the 
local field line tangent tilts with respect to the line of sight.  The illustration 
in the right panel of Fig.~\ref{fig:f7_spec2} clearly portrays that off-meridional loops, which are the 
most common arcs encountered in the magnetosphere, generate upscattering spectra at lower frequencies
when \teq{\phi_{\ast} \gtrsim 5^{\circ}}
that do not violate the COMPTEL upper limits for magnetar hard X-ray tails.

The effect of the variation of viewing angles \teq{\theta_v} is illustrated in Figs.~\ref{fig:f8_spec3} and~\ref{fig:f9_spec4}, 
and is important when considering phase-resolved spectroscopy in oblique rotators. 
The general pattern is an almost monotonic decline in both the overall flux and the 
maximum resonant energy \teq{\erg_f^{\rm max}} as \teq{\theta_v} increases
from zero to around \teq{150^{\circ}}.
This is because the electrons were launched from near the lower ``south'' pole in 
Figs.~\ref{fig:resonance_points} and~\ref{fig:ortho3D_resonance}, so that 
Doppler boosting and beaming is preferentially sampled for smaller viewing angles 
\teq{\theta_v \lesssim 60^{\circ}}.  Specifically, observer viewing angles that never sample 
field lines that are directed approximately towards the observer never attain the highest \teq{\erg_f}.
For the meridional specialization exhibited in Figs.~\ref{fig:f8_spec3} and~\ref{fig:f9_spec4}, this distinction depends on
the colatitude of the field line footpoint, or equivalently \teq{\rmax}.   The value of the polar 
magnetic field controls whether visual alignment with field lines can be coincident 
with resonant collisions.   
For those systems that do attain visual alignment geometry at 
particular pulse phases, the modest differences in the curves 
(see right panels of Figs.~\ref{fig:f8_spec3} and~\ref{fig:f9_spec4}) essentially come about due to the 
different locations for which the alignment is realized, thereby sampling different values of the local magnetic field. 
An additional influence is that different angles \teq{\mu_i} for the incoming soft photons 
are then sampled at these emission points.  Both these elements also impact the 
cases where resonant scatterings are never sampled, depicted in the left panels 
of Figs.~\ref{fig:f8_spec3} and~\ref{fig:f9_spec4}.
The variation of spectra with \teq{\theta_v} that is highlighted in these 
two figures informs the claim in Section~\ref{sec:modulation}
that phase-resolved observations of magnetars will exhibit not just flux 
variations, but also hardness or \teq{\erg_f^{\rm max}} modulations.
Such signatures will be realized not only for magnetars, but perhaps also for the tens of rotation-powered high-field pulsars {following magnetar-like outbursts}.
Note that for these inner magnetospheric emission regions, the viewing angle dependence 
of the fluxes that is illustrated here is consistent with the broad pulse profiles observed in magnetars.
These figures also clearly confirm that with the modulational variation of viewing angle, modest Lorentz factors of 
\teq{\lesssim 30} must be realized in a (subsequent) self-consistent cooling analysis,
so as to not violate the COMPTEL upper bounds on emission {\it at any pulse phase}.

Another dimension to the results that is illustrated in Fig.~\ref{fig:f9_spec4} is
provided by the polarization dependence of the inverse Compton spectra.  
For all viewing angles, the resonant upscattering signal is highly-polarized above around \teq{0.03\erg_f^{\rm max}},
with the result that the \teq{\perp} mode exceeds the \teq{\parallel} one.  The polarization degree is 
only significant at higher energies because then the scatterings are of large angles in the ERF.
When forward scatterings in the ERF are sampled at lower \teq{\erg_f} energies, a 
quasi-Thomson domain, the polarization drops to zero.  This character is in general 
concurrence with the previous uniform field results of Baring and Harding (2007),
{and can be inferred primarily from the \teq{\omega_i\approx B} contributions of the \teq{T^\perp} and 
\teq{T^\parallel} factors appearing in Eq.~(\ref{eq:Tave_defs}).}
Such energy-dependent polarization signatures that are also sensitive to electron 
Lorentz factor, field loop altitude and azimuth, afford the prospect of powerful 
pulsar geometry diagnostics in the age of X-ray polarimetry, particularly if 
phase-resolved measurements are attainable.
Future hard X-ray polarimeters such as X-Calibur (Guo et al. 2013) and soft gamma-ray 
Compton telescopes with polarimetric capability such as e-ASTROGAM (see De Angelis et al. 2017)
and AMEGO\footnote{see {\tt https://asd.gsfc.nasa.gov/amego/index.html}.}
will therefore be critical to constraining the rotator geometry, 
activation locales, and radiative dissipation physics in magnetars.
Determining phase-resolved polarization degrees and position angles 
will be an important inclusion in future resonant upscattering studies of hard X-ray tail emission.
Finally observe the examples of spectra computed in the magnetic Thomson approximation 
(see the discussion for Fig.~\ref{fig:f6_spec1}) are also depicted in the left panel.  These illustrate 
not only energy non-conservation, but also the overestimates obtained for polarization 
degrees that are obtained when using magnetic Thomson cross formalism -- this follows 
from the somewhat weaker polarization dependence in full QED magnetic scattering cross sections.

The final dimension of the suite of spectral figures addresses the variation in altitude 
for field loops.  In Fig.~\ref{fig:f10_spec5}, spectra are displayed for an array of meridional field loops with 
different \teq{\rmax} values.  The illustration is for a viewing angle 
of \teq{\theta_v=30^{\circ}}, and for two different electron Lorentz factors, 
\teq{\gamma_e = 10, 100}.  Note that these spectra are now not normalized by the 
field line arclength \teq{{\cal S}}, as before, so that the relative contributions of 
different \teq{\rmax} values can easily be assessed.
The various curves clearly evince a trend of the upper cusp 
photon (``cut-off'') energy declining with increasing \teq{\rmax}, i.e. dropping when the loop field 
is lower, on average.  This is amply described by Eq.~(\ref{eq:efmax}), 
i.e., \teq{\ergfmax \sim 2\gamma_e B/(1+ 2 B)}, noting that this cusp energy is 
generally realized for quasi-equatorial locales and for quasi-polar viewing perspectives. 
Accordingly, it is readily ascertained {from Fig.~\ref{fig:f10_spec5}} 
that {\it contributions from resonant Compton upscattering to hard X-ray tail emission 
above 10 keV can only come from regions where}
{\rm {\teq{4 \lesssim r_{\rm max}\lesssim 15} for \teq{\gamma_e = 10} (left panel)}},
or 
{\rm {\teq{2.5 \lesssim r_{\rm max}\lesssim 30} for \teq{\gamma_e = 10^2} (right panel)}}.
{At altitudes above these values, the weaker fields move the resonant
emission to energies below 10 keV, where it is swamped by the surface
thermal signal.  At very low altitudes \teq{r_{\rm max} \lesssim 2-4}, the field is high enough that
resonant interactions only sample soft photons deep in the exponential portion of the Planck distribution, 
and the upscattering signal is far less luminous. The restriction of altitude domains where bright
\teq{>10} keV upscattering emission is probable would suggest that a
reason the majority of magnetars do not exhibit hard X-ray tail emission
might be that these constrained resonant upscattering zones are not
activated in many sources.}

The envelope of {the spectra in Fig.~\ref{fig:f10_spec5}} provides a 
visual indication of what would be expected when one integrates over a magnetic 
toroidal section. This superposition would be approximately flat for \teq{\gamma_e\lesssim 100}: 
the dependence of  the cut-off energy on altitude then yields an approximate 
steepening by index \teq{1/3}.  For higher Lorentz factors (not shown), this flattening is muted somewhat. 
Given the results of azimuthal integrations addressed in Section~\ref{sec:toroid}, one may 
anticipate a further steepening when summing spectra over quasi-toroidal volumes.
This will naturally be modified when electron cooling is treated, and when some dependence 
of field line activation on footpoint colatitude and azimuth is introduced.  Yet, it is clear 
that unless \teq{\gamma_e\lesssim 10-30}, the upscattering spectra will extend beyond 
500 keV and violate the COMPTEL upper bounds {for magnetars} unless the expected soft gamma-ray attenuation 
is operational.  {In the context of high-field pulsars such as PSRs B1509-58 and J1846-0258, 
detailed consideration of spectral modeling via this inner magnetosphere scenario, 
perhaps focusing on their states after magnetar-like outbursts,
is deferred to future studies.  Yet it is clear that Lorentz factors \teq{\gamma_e\gtrsim 10^2} will probably be 
needed in order to generate the spectral turnovers that match those observed in the 1--10 MeV range
(e.g., see Kuiper, Hermsen \& Dekker 2018), 
modulo attenuation mechanisms as discussed in Section~\ref{sec:attenuation}
and in Harding, Baring \& Gonthier (1997).}

\subsection{Upscattering Emission from Toroidal Surfaces of Field Lines}
 \label{sec:toroid}

We now address the overarching character of the spectral results presented here.
On face value, compared with observations of magnetars 
{as listed in the McGill magnetar catalog (Olausen \& Kaspi 2014)}, 
the \teq{\erg_f^{1/2}} spectrum from individual field loops is too flat. 
Shown in each of Figs.~\ref{fig:f6_spec1}--\ref{fig:f9_spec4} is a representative hard X-ray spectrum 
of AXP 4U 0142+61 (from den Hartog et al. 2008b) with COMPTEL upper bounds;
this spectrum is normalized to roughly match the rate of the model spectra, 
which include a scaling by the electron number density.
In addition, a power-law of \teq{\erg_f^{-1/2}} with an exponential cut-off 
at \teq{250} keV is also shown as a guide to these data points.
This mismatch in spectral slopes is not a major concern because the analysis 
thus far is restricted to uncooled electrons moving along individual field lines.
In reality, simultaneous electron cooling by the resonant scattering process
will modify the shapes of these spectral forms, generally leading to a steepening
from the \teq{\erg_f^{1/2}} nature.  Moreover, total magnetar spectra are convolutions 
of the forms presented here, summed over different field loops represented by 
ranges of \teq{\rmax} and \teq{\phi_{\ast}}.  Such emission 
volume integrations, akin to the magnetic Thomson study of Beloborodov (2013a),
will in part smear out the cusp and edge structures, and will yield a net steepening 
of the spectra.  Note that the activation of the magnetosphere is unlikely to be uniform, 
with examples provided by current ``j" bundle of field loops invoked in twisted magnetosphere models of 
magnetars (e.g. Beloborodov \& Thompson 2007; Nobili, Turolla \& Zane 2011; Beloborodov 2013a).
Therefore, the weighting of azimuths \teq{\phi_{\ast}} in the volume 
integrations is quite model-dependent.

\begin{figure}[h!]
\centering
\includegraphics[width=4.0truein]{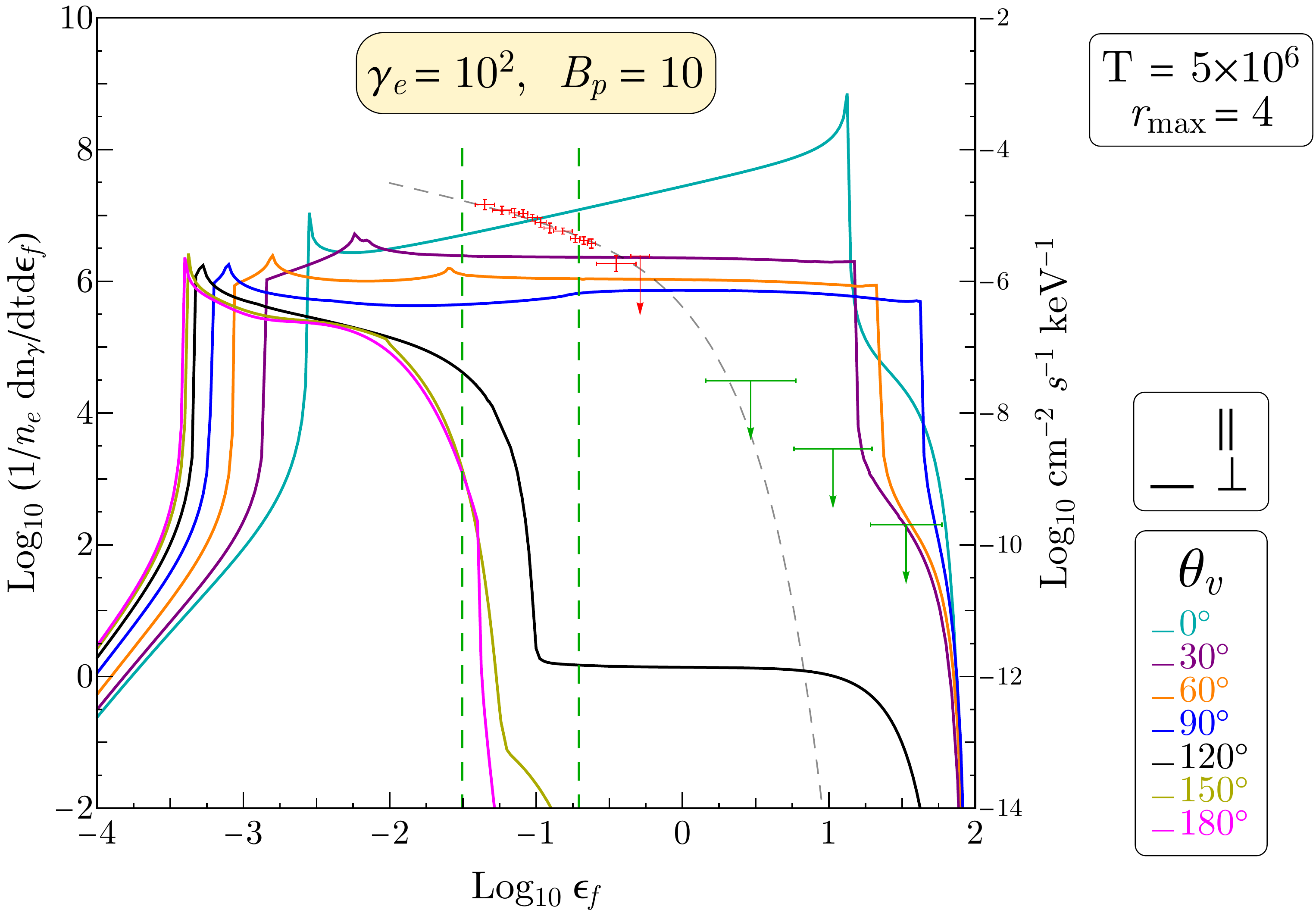} 
\caption{Shown here are special \teq{\phi_*}-integrated spectra, with shadowing, for instantaneous 
\teq{\theta_v = 30^{\circ}}, \teq{B_{\rm p}=10}, \teq{\rmax=4} and fixed and uncooled \teq{\gamma_e=10^2}.
For all \teq{\theta_v > 0} examples, the flat spectrum is reminiscent of uniform field spectra presented in Baring and Harding (2007) 
that accessed all final scattering angles. Compared to meridional spectra shown in 
Figures \ref{fig:f6_spec1}, \ref{fig:f8_spec3} and \ref{fig:f9_spec4}, here the spectrum 
is steeper by a factor $\sim \erg_f^{-1/2}$ due to the strong beaming of the hardest emission 
observed only near meridional field loops.  The \teq{\theta_v=0^{\circ}} example is a symmetric case 
where the spectrum is independent of the azimuth, and so it is flatter, resembling those 
in earlier figures.  The two vertical dashed lines mark the energies 16 keV and 100 keV
that correspond to the pulse phase maps in Fig.~\ref{fig:toroid_phase_alpha_var}.
 \label{fig:toroid} }  
\end{figure}

The character of such summations will be ascertained 
in detail in a future stage of this program, specifically with the inclusion of 
self-consistent electron cooling.  In the interim, 
a rough guide can be ascertained for uniformly-activated surfaces in the magnetosphere.
Since most of the hardest emission is realized for a relatively small spatial 
portion of the magnetosphere, the highest energies of a spectrum integrated over 
field line longitudes is dominated by \teq{\phi_{\ast}\ll 1} regions. 
The sharp fall-off in \teq{\erg_f^{\rm max}} in Eq.~(\ref{eq:ergf_azimuth_form}) 
away from the meridional loops results in a steepening of the spectrum;
the dependence \teq{\phi_{\ast} \propto \erg_f^{-1/2}} thus serves as 
an energy scaling in an integration over magnetic longitudes.  Therefore, 
for a field line {\it toroidal surface} corresponding to a fixed \teq{\rmax}, 
the integrated spectrum an observer would discern between the horns
would scale as
\begin{equation}
   \int_0^{2\pi} \dover{dn}{dt d\erg_f} \, d\phi_{\ast}
   \;\sim\; \kappa\, \erg_f^{1/2}\, \erg_f^{-1/2}
   \;\propto\; \erg_f^0\quad ,
 \label{eq:spec_azimuth_integ}
\end{equation}
for some \teq{\kappa (\erg_f,\, \gamma_e ,\, \Theta)} that is only weakly-dependent on \teq{\erg_f}.
Here it is understood that the spectrum in Eq.~(\ref{eq:scatt_spec_fin}) is differential 
in \teq{\phi_{\ast}}.  Thus the differential spectrum from the toroid is flat, a result that 
is evident in the computation presented in Fig.~\ref{fig:toroid}, 
which is an integration over an entire toroidal surface with \teq{\rmax = 4}. 
Due to the small spatially confined region of azimuthal
angular extent $1/\gamma_e$ around the meridional field loop, such spectra 
should also be realized for bundles of field loops around the meridional line, i.e. 
for incomplete toroidal geometries.  Note that these spectra are again normalized by the 
field line arclength \teq{{\cal S}}, as for the spectral figures of Figs.~\ref{fig:f6_spec1}--\ref{fig:f9_spec4}.

This \teq{\erg_f^0} form is reminiscent of the flat
solid-angle-integrated spectra for uniform \teq{\boldsymbol{B}} that were presented in Baring and Harding (2007).
Such resemblance is not merely coincidental, since integrations over longitudes for 
fixed viewing perspectives
are geometrically similar to integrations over observational solid angles
in uniform field scenarios.  The \teq{\erg_f^{1/2}} dependence emerges only
in the azimuthally-symmetric example with \teq{\theta_v=0}, 
where the value of \teq{\phi_{\ast}} is immaterial.  
Observe also that the high-energy cusp is 
still present in the \teq{\theta_v=0} spectra exhibited in Fig.~\ref{fig:toroid}, 
though its prominence in \teq{\theta_v > 0} cases is 
curtailed somewhat for the \teq{\parallel} polarization, and 
substantially for the \teq{\perp} mode, due to smearing caused by the \teq{\phi_{\ast}} integration.  
One anticipates that while this spectral form is representative of what an 
entire toroidal surface contributes, the superposition of various 
such surfaces with different \teq{\rmax} values, and furthermore
the introduction of electron cooling, should further steepen 
the cumulative spectrum from an active magnetosphere.

{The spectral properties that have been identified so far suggest that there is 
only a weak dependence of the spectral index \teq{\Gamma_h} on the strength of the magnetic field, 
though the maximum energy \teq{\erg_f^{\rm max}} of resonant emission is indeed sensitive to 
the value of \teq{\vert \boldsymbol{B}\vert }.  Interesting correlations between magnetar 
spectral and spin parameters have been highlighted in the papers by Kaspi \& Boydstun (2010)
and Enoto et al. (2010), the most salient for the work here being a steepening of the 
hard tail spectrum (increase of \teq{\Gamma_h}) when the magnetar polar field \teq{B_p} 
is higher.  At present it is not possible to assess whether our resonant Compton model 
generates such a correlation, principally because the spectral slope above 10 keV 
will depend on (i) details of the volumetric integration over \teq{r_{\rm max}} ranges, 
(ii) how self-consistent resonant cooling of electrons limits their Lorentz factors in 
different magnetospheric regions, and (iii) the significance of attenuation processes 
such as photon splitting and pair creation that seed cascading.  Exploring such correlations 
defines an objective for future stages of our program.  A particular nuance of interest will be 
to assess how spectral variations associated with burst storms and following giant flares 
might provide insights into the ranges of \teq{r_{\rm max}} sampled, and therefore 
probe variations in activation volumes during dynamic epochs of a magnetar's history.}

\begin{figure}[ht!]
\epsscale{1.15}
\plotone{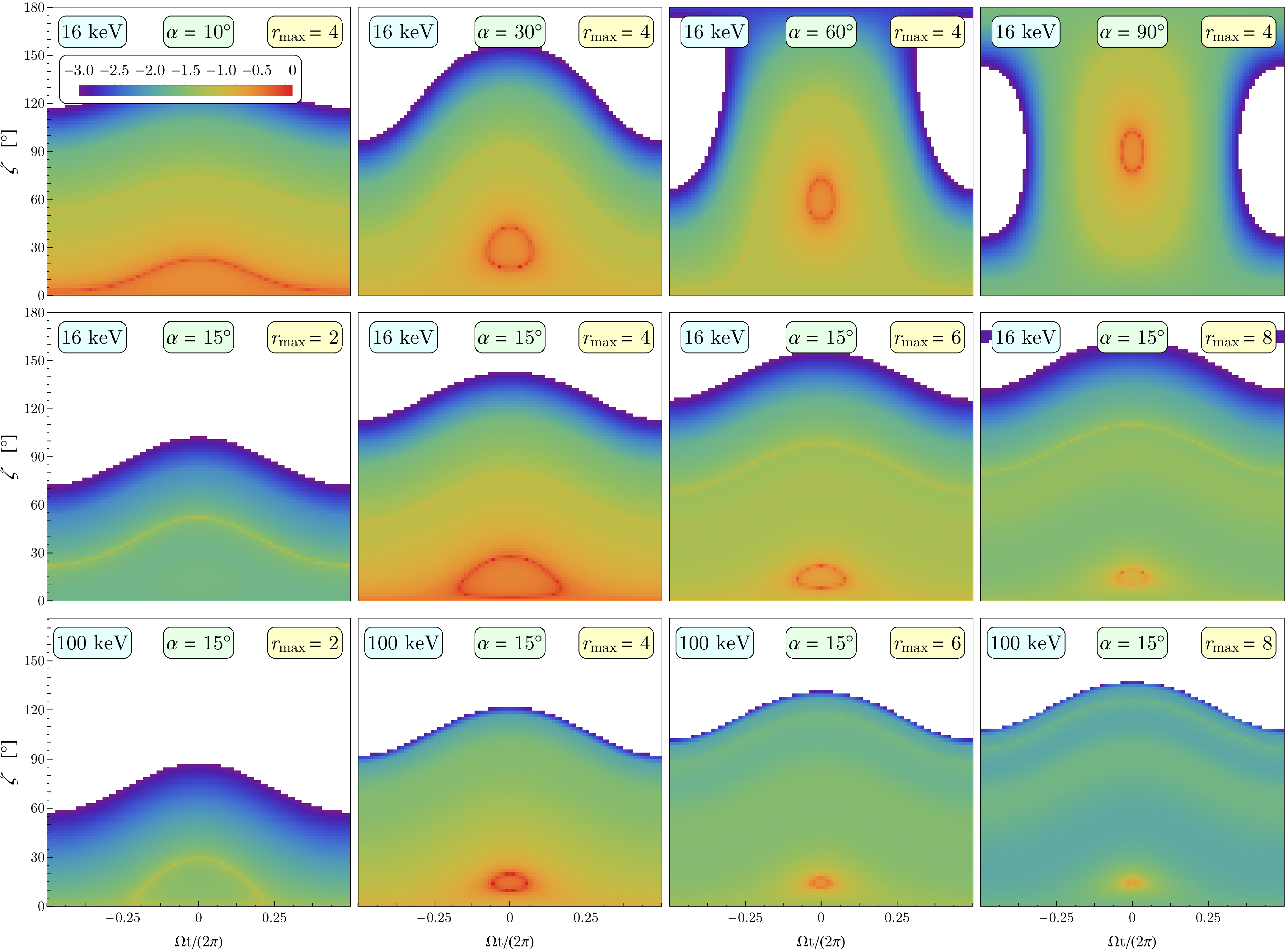}
\vspace{-5pt}
\caption{Normalized photon flux \teq{\zeta - \Omega t/2\pi} phase space maps for 
resonant Compton upscattering.  These represent the 
logarithmically scaled (base 10) intensity at energies \teq{16} keV (top two rows)
and \teq{100} keV (bottom row), color coded as in the legend, as a function of 
spin phase \teq{\Omega t/2\pi} for each value of \teq{\zeta} on the ordinate.  
The intensity maps are for uncooled electrons with \teq{\gamma_e = 10^2}, 
and a uniform surface temperature \teq{T=5 \times 10^6}K. 
The maps are obtained for azimuthally-integrated bundles 
of field lines, i.e. a toroidal surface, with \teq{B_p=10} and \teq{r_{\rm max} = 4}.  The 
panels in the top row sequentially sample magnetic inclinations 
\teq{\alpha = 10^{\circ}, 30^{\circ}, 60^{\circ}, 90^{\circ}}, 
clearly presenting the trend with rotator obliquity.  
The bottom two rows are for a single magnetic inclination angle of \teq{\alpha = 15^{\circ}}, and
depict maps for maximum loop altitudes \teq{r_{\rm max}=2,4,6,8}, as indicated,
for the two different emergent photon energies.
Pulse profiles for a particular observer \teq{\zeta} are represented by horizontal cuts 
of the maps.  Accordingly, symmetric double-peak structure of pulse profiles in 
domains \teq{\zeta \approx \alpha} is readily apparent, being manifested 
as sections of the red rings: these are realized when quasi-polar viewing is 
possible at select phases.  The phase separation 
of the double-peak structure of pulse profiles in 
domains \teq{\alpha \approx \zeta} shrinks at higher \teq{r_{\rm max}} and larger \teq{\erg_f}. 
The normalization across all panels in a row is relative to the brightest flux realized in that row.
 \label{fig:toroid_phase_alpha_var} }  
\end{figure}

The focus of the azimuthal integrations now progresses from spectral elements 
to pulsation properties.  Fig.~\ref{fig:f5_efmax} provides an assessment 
of how the cutoff energy of the resonant spectrum varies with pulse phase.
Since this is not precisely what is measured by a telescope in a narrow 
energy window, to augment such it is incisive to present phase-resolved variations
for toroidal surface integrations.  The coupling between the 
instantaneous viewing angle \teq{\theta_v} and pulse phase \teq{\Omega t/2\pi}
is addressed in Section~\ref{sec:modulation}.  Thus with a fixed magnetic inclination 
\teq{\alpha}, since the spectral dependence ties intimately to the 
value of \teq{\theta_v}, inversion of Eq.~(\ref{eq:thetav_t_alp}) yields 
a periodic (but asinusoidal) modulation of the flux at any given upscattered energy \teq{\erg_f} 
with time \teq{t}, the character of which is controlled by the value of \teq{\zeta}.
To represent these pulse profile variations, in Fig.~\ref{fig:toroid_phase_alpha_var} twelve
\teq{\zeta - \phi} phase space diagrams (where \teq{\phi = \Omega t/2\pi}) are illustrated for an array of 
\teq{\alpha} and \teq{ r_{\rm max}} values, and for \teq{\erg_f=16}keV (top two rows)
and \teq{\erg_f=100} keV (bottom row).
This is a common representation in gamma-ray pulsar 
studies (e.g. see Harding et al. 2008 for the Crab pulsar; Johnson et al. 2014, for 
{\it Fermi}-LAT millisecond pulsars).  The various panels depict the intensity 
on a base-10 logarithmic scale, and one value of \teq{\zeta} can be chosen so as to select
the pulsing flux profile for a particular observer.   The intensity values are 
azimuthally integrated over toroidal surfaces, as in Fig.~\ref{fig:toroid}.  
The uncooled electrons of
Lorentz factor \teq{\gamma_e = 10^2} propagate from the south to north 
magnetic foot point, as always.  The spectra are computed {\it omitting} the 
arclength normalization factor \teq{1/{\cal S}(r_{\rm max})}, so that surfaces 
with different \teq{r_{\rm max}} values can be compared on an equal footing.
We also note that because the
density of information in each panel corresponds to \teq{91 \times 91 = 8231}
pixels, to reduce the computational time, the resonant cross section in the 
Lorentz profile was approximated by a \teq{\delta} function form via a standard prescription. 
This serves to evaluate one of the integrals in Eq.~(\ref{eq:scatt_spec_fin}), and 
thereby reduce the \teq{\phi^{\ast}} integration to a manageable two integrals; 
this introduces differences in the color coding in Fig.~\ref{fig:toroid_phase_alpha_var}
for the flux scale that are imperceptible.  The red portions 
mark the peaks of the pulsation, while the white space corresponds to flux levels 
below the blue/purple level (i.e., \teq{ < 10^{-3}}) that are essentially unobservable.  
The scaled intensities for all each row of panels possess a 
single normalization factor, chosen so that the maximum flux 
for a row and all \teq{\Omega t/2\pi} phases and \teq{\zeta} choices is set to unity.

The maps in Fig.~\ref{fig:toroid_phase_alpha_var}
are phase-symmetric, a property that is
dictated by our uniform activation assumption for the toroidal surface, and 
should not be considered sacrosanct.  Asymmetries may naturally be imposed by a 
variety of influences, including twisted fields, and non-uniform electron phase space densities 
perhaps resulting from spatially-dependent cooling.
For most values of \teq{\{ \alpha, \zeta \}}, there is generally a single broad pulse
that is typically of phase width \teq{\sim 0.5}.  Such moderate pulse 
fraction examples are approximately commensurate with the 
majority of hard X-ray tail data in the magnetar population.  In contrast to this there are also
cases of sharply-peaked fluxes that manifest themselves only for fairly narrow 
ranges of \teq{\zeta} proximate to \teq{\alpha}.  These arise when 
the observer line-of-sight cuts across the magnetic pole at some phases, 
thereby sampling instantaneous viewing angles \teq{\theta_v \approx 0}.  Inspection of 
the orthographic projections in Fig.~\ref{fig:ortho3D_resonance} reveals that in 
such cases, a large portion of the magnetosphere can then emit towards the 
observer in resonant interactions that generate \teq{\sim 16} keV photons (of red color),
thereby bolstering the flux.  For other  \teq{\{ \alpha, \zeta \}} combinations, 
\teq{\theta_v} is much larger on average, and much of the magnetosphere emits 
at lower energies, obscured by the dominant surface emission contribution.  
When \teq{\zeta \sim \alpha}, the pulse profiles obtained 
from horizontal \teq{\zeta} slices are double-peaked, corresponding to two 
samplings of the minimum \teq{\theta_v} during a rotation period.  

Such double-peaked pulse profiles are occasionally seen in the hard X-ray tails 
of magnetars for certain energy bands -- see for example Kuiper et al., 2004, for RXTE/HEXTE 
data above 50 keV for 1E 1841-045; den Hartog et al. 2008a for 4U 0142+61 
in the window 20-50 keV; and An et al., 2013, 2015, for NuSTAR data between 
24 keV and 35 keV on 1E 1841-045.  Generally, such bifurcated peaks are either 
poorly resolved or completely absent in magnetar pulse profiles above 10 keV.  
For cases where two peaks are well distinguished, an important observational diagnostic on 
the resonant Compton upscattering model is suggested: matching the peak 
separation will constrain the \teq{(\alpha ,\, \zeta )} space.  This 
is directly analogous to protocols adopted for caustic descriptions 
of gamma-ray pulsars (e.g. Watters et al. 2009; Pierbattista et al. 2015).
For example, given the phase separation of 0.4 between 
the peaks in 1E 1841-045 of An et al. (2013)  at an energy of 20-35 keV, taking
Fig.~\ref{fig:toroid_phase_alpha_var} at face value suggests that \teq{\alpha \lesssim 20^{\circ}}.
This is quite similar to the value of \teq{\alpha \sim 15^{\circ}} inferred in the analysis 
of An et al. (2015) that was based on the modeling of Hasco{\"e}t, et al. (2014)
{that invoked twisted fields}. {One naturally expects that the different inferences of \teq{(\alpha ,\, \zeta)} values 
might arise for the dipole geometries considered here and twisted field models.
Yet, the scattering locales that dominate the spectral signal in the 20-100 keV band
in almost aligned rotators are generally quasi-equatorial since 
the viewing angle \teq{\theta_v \lesssim 30^{\circ}} for all pulse phases:
see the orange/yellow regions in Fig.~\ref{fig:ortho3D_resonance}.  
In the detailed MHD simulations of closed field line untwisting
in Chen \& Beloborodov (2017), while twist angles \teq{\Delta \phi \sim 1} 
can be realized near the quasi-polar footpoints, values substantially smaller than 
unity are found at colatitudes \teq{\vartheta > 60^\circ} nearer \teq{\rmax} 
as the dissipative untwisting progresses 
and the field tends towards its relaxed dipolar configuration.  Accordingly, the 
hard tail pulse profiles for twisted field scenarios should resemble those presented 
here for dipolar geometries.}

The array of panels in the bottom two rows evinces two clear trends: 
the separation of peaks in double-peak \teq{\zeta \sim \alpha} domains declines 
as \teq{r_{\rm max}} is increased, or as \teq{\erg_f} becomes larger.  Both these 
circumstances correspond to the resonant locales converging towards the 
separatrix that is illustrated in Fig.~\ref{fig:resonance_points}.  The coupling with photon energy 
is again a hallmark of the kinematics of the resonant upscattering process, 
and signals a useful observational diagnostic.  From this Figure, one again infers 
that for \teq{\alpha = 15^{\circ}}, the \teq{r_{\rm max}=4} example could 
accommodate the NuSTAR pulse profile data for 1E 1841-045 in An et al. (2013)
when \teq{5^{\circ}\lesssim \zeta \lesssim 20^{\circ}}.  Yet other values 
of \teq{r_{\rm max}} could not, suggesting a fairly restricted range of footpoint 
colatitudes for the active field lines.  At other energies for this magnetar,
and indeed for other magnetars at all hard X-ray tail energies, prevalent single-peaked structure 
would indicate convolutions of \teq{r_{\rm max}} surfaces being required,
perhaps with \teq{r_{\rm max}\gtrsim 5}, in order 
to match the pulse profiles.  Other pulse broadening mechanisms can be surmised, 
for example distribution of electron Lorentz factors, such as would be realized in 
cooling treatments, thereby blurring the upscattering kinematics.   Also, 
twists to the field structure distribute the resonant locale geometry and the Compton 
kinematics, and so would act to broaden intrinsic pulse structure.  
Furthermore, mechanisms such as photon splitting that can be operational 
above 50 keV can lead to phase-dependent attenuation of the pulse profile, 
thereby modifying its shape.  As a competing influence, reducing the surface 
emission locales from a complete sphere to a hot spot will alter the resonant 
scattering kinematics and perhaps narrow the pulse width slightly.
It is evident that even with the action of broadening and attenuation processes, 
for a given uncooled Lorentz factor, {\it the maximum pulse width or double-peak 
phase separation at a given energy 
provides a lower bound to the lower altitude of emission}.  Thus, we see that 
energy-dependent pulse profile modeling augurs the potential to significantly 
constrain the values of \teq{\alpha}, \teq{\zeta} \underline{and} \teq{r_{\rm max}}, 
a task that will be undertaken in the future using self-consistent cooling/emission models.

{Another apparent property is that when the viewing angle is well removed 
from the magnetic axis, specifically, \teq{\zeta} is not close to \teq{\alpha}, not 
only is the pulse prominence quite limited, but the relative brightness of the 
upscattered emission is low.  This circumstance applies because of our uniformly hot surface assumption, 
and one might expect that if the soft photon supply is constrained to a hot spot centered near the footpoints 
of an equatorial field bundle, the \teq{\zeta}-\teq{\alpha} coupling will be profoundly different.
For our uniform surface illumination case, one could surmise that a geometrical
reason for a magnetar not possessing a detectable hard X-ray tail is that 
\teq{\vert \zeta - \alpha\vert} is sufficiently large, perhaps around \teq{45^{\circ}} 
or more.  Comparing with inferences of these geometry parameters from 
surface emission is a potentially productive path.  G\"uver, G{\"o}{\u g}{\"u}{\c s}
\& \"{O}zel (2015) offer an analysis of the strong pulsation of surface X-rays 
below 10 keV observed by XMM-Newton in 1E 1048.1-5937, concluding that hot spots from 
an orthogonal rotator (\teq{\alpha \sim 90^{\circ}}) can account for the 
observed pulse fraction of \teq{\sim 75}\% with a viewing angle of 
\teq{\zeta \sim 45^{\circ}}.  This magnetar does not exhibit a prominent hard X-ray tail, 
and so the upper right panel of Fig.~\ref{fig:toroid_phase_alpha_var} would 
indicate that \teq{\zeta \lesssim 45^{\circ}} would be favored when \teq{\alpha \sim 90^{\circ}}
in the light of the resonant upscattering model, consistent with the inference of 
G\"uver, G{\"o}{\u g}{\"u}{\c s}
\& \"{O}zel (2015).  Yet, obviously, complete lack of activation 
of the magnetosphere could be the explanation for the paucity of 
steady hard X-ray emission in 1E 1048.1-5937.}

To extend the surface spectral results to complete volumes that span
a range of \teq{r_{\rm max}} values requires detailed numerical 
computations that should nominally include treatment of electron 
cooling.  The total cooling rate scales as the integral of 
\teq{\erg_f\, dn/dt d\erg_f \to \erg_f\, n_{\gamma}(\erg_f)}, 
which for flat spectra like those in Fig.~\ref{fig:toroid}, approximately 
expresses \teq{(\erg_f^{\rm max})^2} times the normalization 
\teq{n_{\gamma} (\erg_f^{\rm max})} at the maximum energy of emission.
For the spectral integration over azimuths, which at each colatitude represents the 
total emission from a particular point along a field line, one can then 
equate \teq{(\erg_f^{\rm max})^2 \, n_{\gamma} (\erg_f^{\rm max})} 
to computed cooling rates.  These can be found in Fig.~10 of BWG11, 
which display \teq{\dot{\gamma}_e} for different altitudes and an array of 
magnetic colatitudes.  From such depictions the following general 
inferences can be made.  If \teq{\gamma_e} exceeds the Lorentz factor
\teq{B/\Theta} at the peak cooling rate, then \teq{\dot{\gamma}_e}
approximately scales as \teq{r_{\rm max}^{-\lambda}} with \teq{\lambda \sim 6-7},
i.e. roughly as \teq{B^2}.  Cooling is in the \teq{B\ll 1} domain at 
modest to high altitudes.  This is then approximately proportional to 
\teq{(\erg_f^{\rm max})^2}, deduced using Eq.~(\ref{eq:efmax}) 
in the magnetic Thomson domain where \teq{\erg_f^{\rm max} \approx 2 \gamma_e B}.
This then sets \teq{n_{\gamma} (\erg_f^{\rm max})} to be approximately 
independent of \teq{\erg_f^{\rm max}}, so that flat spectra should result from 
volume integrations at relatively high altitudes and for \teq{\gamma_e > 10^3}.
For lower Lorentz factors, the opposite trend in cooling is observed when \teq{\gamma_e < B/\Theta}, 
so that \teq{\dot{\gamma}_e} scales as \teq{r_{\rm max}^{-\lambda}} with \teq{\lambda \sim -3}, or perhaps 
with more negative \teq{\lambda} values.  Then 
\teq{n_{\gamma}(\erg_f^{\rm max}) \sim (\erg_f^{\rm max})^{-3}}, a very steep power-law.
The reality will probably lie somewhere in between these two extreme cases, and be influenced by the choice of 
\teq{\gamma_e} and contributions from low altitudes where QED 
modifications to the cooling rate and emission are strong.
To assess this more incisively, more detailed numerical determinations are needed, 
the subject of future work.

The dipolar field configuration is clearly a convenient idealization for magnetars.
Field-line twists offer more complicated morphologies, and na{\"i}vely one might 
anticipate that tangled field line geometries might evince spectral character 
somewhat reminiscent of that for the toroidal surfaces generated by integrations over 
dipolar field line longitudes.  This can only be truly assessed via detailed 
geometric modeling of twisted field scenarios.  Yet it can be surmised that the spectral signatures 
of non-dipolar systems will be modulated by stellar rotation in ways quite different from 
the pure dipolar case.  To excogitate this, consider a simple toroidal field loop 
such as would be precipitated by current bundles tracking along poloidal field lines.
For energetic charges moving along the toroidal loop, the plane of Doppler beaming 
of upscattered radiation is orthogonal to the planes of dipole field line loops of any longitude \teq{\phi_{\ast}}.
Therefore the array of local viewing perspective angles \teq{\ThetaBn} sampled at any rotational phase 
will be quite different for the poloidal and toroidal field line paths.  Thus the phase modulation 
of spectra and \teq{\erg_f^{\rm max}} for the case of toroidal trajectories will differ 
from that for poloidal electron paths.  This will constitute not just a general phase offset of 
the peaks of \teq{\erg_f^{\rm max}} as displayed in Fig.~\ref{fig:f5_efmax}, but distortion 
of symmetric sinusoidal modulations and of the onset and egress of shadowing.  
Accordingly, detailed measurement of energy-dependent pulse morphologies should afford 
diagnostics on the magnetic field geometry. In addition, 
the general polarization character of phase-resolved emission from toroidal field paths should 
differ from the poloidal loop case, both in degree of polarization and also 
the position angle sweep.  Thus, hard X-ray polarimetry should serve to help discriminate 
between these two emission loop geometry cases, as it should do also for more tangled 
twisted field morphologies.

\section{DISCUSSION}
 \label{sec:discuss}

A number of the resonant Compton spectra presented in this paper violate
COMPTEL upper bounds imposed on soft gamma-ray emission from several
magnetars, particularly for Lorentz factors \teq{\gamma_e \gtrsim 30}.
For higher Lorentz factors, they can even violate constraining 
upper limits from the {\it Fermi}-LAT experiment for magnetars (Abdo et al. 2010, Li et al. 2017).
The significance of such violations as a
potential issue for the resonant Compton upscattering model for hard
X-ray tails depends on two elements that are discussed here: (i) how high are the Lorentz
factors generated for electrons in the activation zones, and (ii) how
much are hard X-rays and gamma-rays attenuated in magnetospheres.

\subsection{Radiation Reaction-Limited Acceleration}

In a complete resonant Compton upscattering model, the spectral
templates exhibited in Section~\ref{sec:spectra} need to be convolved with the
spatially-dependent cooling of electrons due to the scattering process. 
While a full exploration of such is beyond the scope of this paper, here
we offer a flavor of what might be expected when cooling is
incorporated.   From the detailed resonant Compton cooling analysis in BWG11, 
it is apparent that cooling rates \teq{\dot{\gamma_e}} due to resonant
scattering in locally uniform fields typically peak at \teq{\gamma_e
\sim B/ \Theta}, where \teq{\Theta = kT/m_ec^2} defines the temperature
of the thermal soft photons. This is dictated by the kinematic criterion
for resonant scatterings, yielding optimal sampling of the peak of the
Planck spectrum for the soft X-rays. At this most-efficient cooling
point, the corresponding cooling rate length scale, \teq{\lambda_c \equiv c 
\gamma_e/\dot{\gamma}_e}, realizes minimal values commensurate with 
\begin{equation}
   \lambda_{\rm c,min} \;\equiv\; \lambda_c\Bigl\vert_{\gamma_e\sim B/\Theta}
   \;\sim\; \dover{1}{2\Theta^3} 
        \, \dover{\lambar}{\fsc}\, \max \Bigl\{ \dover{1}{2} \, ,\; 2 B \Bigr\} \quad .
 \label{eq:cool_mfp_scale}
\end{equation}
This is an approximate evaluation that encapsulates the sub-critical and
super-critical field domains, and can be deduced from Eq. (55) of BWG11.
For \teq{B=10} and \teq{T \sim 5\times 10^6 - 10^7\,}K, the
corresponding cooling length scale  at the stellar surface is of the 
order of  \teq{1-30} cm $\ll \rns$. As this estimate is only modestly
dependent on the field strength when \teq{B\gg 1}, it is quickly
discerned that resonant Compton cooling of relativistic electrons should
generally be prolific throughout the inner regions of magnetar
magnetospheres.

If sufficiently rapid, resonant cooling could ultimately quench any
acceleration process operating above the surface.  If not, or if 
the cooling subsequently renders the charge unable to access the resonance, then it can
still act after the charges have emerged from the electric potential zone
or gap.  In both cases, cooling will substantially impact spectral
formation. The mechanisms of particle acceleration operating in magnetars are not fully understood, and probably differ 
between the two source classes. The acceleration may be
precipitated by electrostatic potential gaps, or by dynamic,
non-potential twisted fields close to the surface.  These are each considered 
in the ensuing discourse. The component {\teq{E_{\parallel}=\boldsymbol{E} \cdot \boldsymbol{B} /\vert \boldsymbol{B}\vert}  
of the electric field {\bf E} that is parallel to \teq{\boldsymbol{B}}} within a potential gap in a pulsar-like
mode putatively scales with the co-rotating electric field,
\teq{E_\parallel \sim E_{\rm rot} = r \Omega \vert \mathbf{B}\vert /c},
the Goldreich-Julian value {(GJ: Goldreich and Julian 1969)}, 
and thus is independent of \teq{\gamma_e}. 
Note that for the purposes of this discussion, 
we omit the non-trivial dependence of this spin-down estimate on the 
inclination angle \teq{\alpha} of the rotator.
If the efficiency of {electron} acceleration relative to this familiar {GJ} 
benchmark {$E_{\rm rot}$} is represented by a dimensionless scaling parameter \teq{\eta}, then the
rate of increase of the Lorentz factor of an ultrarelativistic electron
can be written \teq{{\dot \gamma}_{\rm acc} \sim 2\pi \eta
\omegaB\rns/(Pc) = 2\pi \eta B \rns/(\lambar P)}, where \teq{B} is the
field strength now in units of \teq{B_{\rm cr}}, and \teq{\omegaB =
e\vert \mathbf{B}\vert /m_ec} is the electron cyclotron frequency. Thus,
the length scale for such electrostatic acceleration is 
\begin{equation}
   \lambda_{\rm acc} \;\equiv\; \dover{\gamma_e c}{{\dot \gamma}_{\rm acc}}
   \; \sim\; \dover{\gamma_e \lambar P c}{2\pi \eta B \rns}
   \;\equiv\; \dover{\gamma_e}{\eta B} \, \dover{c}{\Omega \rns} \, \lambar \quad ,
 \label{eq:acc_mfp_scale}
\end{equation}
again for \teq{B = \vert \mathbf{B}\vert /B_{\rm cr}}, as noted in
Eq.~(57) of BWG11. When {\teq{\gamma_e \sim 10^3} and the acceleration is fast, i.e., \teq{\eta \sim 1}},
this length scale is of the order of \teq{10^{-5} - 10^{-3}\;}cm for
\teq{B \sim 1-10^2} and typical magnetar periods; it is therefore
significantly inferior to the resonant Compton cooling lengths. 
There is no mandate
that \teq{\eta} be as large as unity, and in fact it could be quite
small, since \teq{\eta\rns} represents the product of the physical
extent of the electrostatic gap and \teq{E_{\parallel}/E_{\rm rot}}.
In the context of magnetars, since their emission cannot 
be powered by stellar rotation alone, this GJ evaluation is a general guide,
but should not be over-interpreted.  

The importance of resonant Compton cooling for limiting the acceleration
depends critically upon the locale of the electric potential and the
value of \teq{\eta}.  For gamma-ray pulsars, the acceleration length
scale becomes larger in the outer magnetosphere, where the magnetic
field is much lower, more than offsetting the increase in \teq{\eta\rns
\lesssim \rlc = c/\Omega}. This establishes \teq{\lambda_{\rm acc}
\lesssim \rlc} in both outer gap and slot gap models. 
While resonant Compton cooling can have
a partial impact in slowing down acceleration in pulsars (e.g. see
Daugherty \& Harding 1996; Sturner 1995, for the magnetic Thomson
regime), it usually does not shut it down completely.  Mostly, the
dominant mode of cooling of primary electrons in models of many
energetic young pulsars (e.g., excepting the Crab)
and old millisecond pulsars is through the emission of curvature radiation.
In the case of millisecond pulsars, radiation reaction mediated by curvature emission 
actually forces the cessation of acceleration (Timokhin and Harding 2015)
generally at Lorentz factors of \teq{\gamma_e \sim 10^6 - 10^7}
before a pair-formation front can be established.

For the magnetar case, provided that
\teq{\eta \lesssim 10^{-5} - 10^{-3}}, the estimates from
Eqs.~(\ref{eq:cool_mfp_scale}) and~(\ref{eq:acc_mfp_scale}) yield
\teq{\lambda_{\rm c,min} \lesssim \lambda_{\rm acc}}, and the
acceleration will be radiation reaction-limited (RRLA) once the
resonance is encountered.
The opposite circumstance, namely when \teq{\eta \gtrsim 10^{-5} -
10^{-3}}, will generate prompt and unhindered acceleration out to
Lorentz factors that sample the full electric potential.
Outside the ``gap'' the electrons will then quickly cool until
\teq{\lambda_c} exceeds around \teq{10-30 \rns} and \teq{\gamma_e} is
below the cooling peak at \teq{B/\Theta}. The main material difference
between this unimpeded acceleration case and the RRLA domain is that the
maximum Lorentz factor injected into the system will be higher.  In both
cases, the progressive cooling of electrons outside the potential
``gap/zone'' will generate a convolution of spectra like those exhibited
in Section~\ref{sec:spec_loops}. For meridional planes, inspection of
the left panel of Fig.~\ref{fig:f6_spec1} indicates that lower
\teq{\gamma_e} yield spectra with lower \teq{\erg_f^{\rm max}} but
higher normalization, so that this superposition would yield a
substantial overall steepening of the resonant spectrum.  Off-meridional
viewing perspectives may present different convolutions, possibly
leading to pulse phase-dependent spectral index variations. Whether
these are approximately commensurate with the observed spectral indices
will require a complete, self-consistent computation of emission and
electron cooling.

Moving on from static potentials,
the paradigm of dynamic, twisted magnetospheres that generate electric fields and currents,
is presently popular as a model of magnetar activation and dissipation.  
First put forward as a possibility to power magnetar hard X-ray emission by Thompson, 
Lyutikov, and Kulkarni (2002), it has been developed in a number of papers, 
including more recent expositions in Nobili, Turolla, \& Zane (2011),
Parfrey, Beloborodov \& Hui (2013), and Chen \& Beloborodov (2017).  
In perturbed force-free magnetic dipole configurations, 
the current density {\bf j} above the atmosphere 
is approximately parallel to the magnetic field, \teq{\mathbf{j} \times \mathbf{B} \approx \mathbf{0}}. 
Then, for small twists with an angle \teq{\Delta \varphi} deviating from 
the pure dipole geometry at colatitude \teq{\vartheta} and altitude \teq{r}, 
the current can be found in Eq.~(1) of 
Beloborodov and Thompson (2007), from which the transient electric field 
scaling from electrodynamics can be inferred according to the prescription in Eqs.~(13)--(17) of 
Beloborodov and Thompson (2007):
\begin{equation}
   E_{\parallel} \;\sim\;  \dover{4 \pi | \mathbf{j}|}{\omega_p} \;\approx\; \sqrt{ \dover{4\pi m_ec \,\vert \mathbf{j}\vert }{e} }
   \quad \hbox{with}\quad
   \mathbf{j} \;\approx\;  \dover{c \boldsymbol{B}}{4 \pi r} \sin^2 \vartheta \, \Delta \varphi
 \label{eq:twistj_E}
\end{equation}
Here \teq{E_{\parallel}} refers to the component of the electric field parallel 
to the local dipole \teq{\boldsymbol{B}} direction, and in this equation, \teq{\boldsymbol{B}} 
is expressed in Gaussian units.  Thus small departures from ideal MHD conditions 
exist, and these are ephemeral, leading to dynamic untwisting of a magnetosphere
with hot spot formation at the footpoints of the \teq{\nabla \times \mathbf{B}} ``j-bundle''
(Beloborodov 2009).  Since electrostatics determines the 
dynamic charge separation potentials, the electron plasma frequency \teq{\omega_p} appears,
and this is evaluated using \teq{n_e = \vert \mathbf{j}\vert /ec} 
for ultra-relativistic electrons.  Observe that this estimate of \teq{E_{\parallel}} is 
explicitly independent of the magnetic inclination \teq{\alpha}, contrasting the 
situation for gap fields in rotation-powered pulsars.  The acceleration rate \teq{\dot{\gamma}_{\rm tw} \to e E_{\parallel}/m_ec} for 
magnetospheric twists can then be computed using the rate of work done in this {\bf E} field, 
and the coupling of the acceleration lengthscale \teq{\lambda_{\rm tw}} from 
dynamic magnetic field perturbations to the twist angle \teq{\Delta \varphi} in the 
linear domain then quickly follows (for \teq{r\gtrsim \rns}):
\begin{equation}
    \dot{\gamma}_{\rm tw} \; \lesssim\; c\,\sin \vartheta \sqrt{ \dover{B\, \Delta\varphi}{\lambar \rns} }
    \quad \Rightarrow\quad
    \lambda_{\rm tw} \;\equiv\; \dover{\gamma_e c}{{\dot \gamma}_{\rm tw}}
    \; \gtrsim\; \dover{\gamma_e}{\sin\vartheta} \sqrt{ \dover{\lambar \rns}{B\, \Delta\varphi} }\quad .
 \label{eq:twist_acc} 
\end{equation}
Here \teq{B} is now in units of \teq{B_{\rm cr}}, and as throughout, \teq{\lambar = \hbar /m_ec}.
When \teq{\gamma_e \sim 10^3}, this length scale is of the order of \teq{3- 30\;}cm for
\teq{B \sim 1-10^2} and \teq{\Delta \varphi =0.1}.  It is several orders of magnitude 
larger than the GJ comparison in Eq.~(\ref{eq:acc_mfp_scale}), by a factor 
of the order of \teq{(\rns/\rlc ) \sqrt{B\rns /\lambar} }.
Yet, interestingly, this acceleration length is quite comparable to the minimum cooling length 
\teq{\lambda_{\rm c,min}} in Eq.~(\ref{eq:cool_mfp_scale}) for resonant Compton scatterings,
so that upscattering spectral details might be sensitive to the choice of the twist angle.   
Accordingly, detailed considerations of resonant cooling in combination with magnetospheric 
twist acceleration models offer the prospect that useful probes of the values of \teq{\Delta\varphi} 
and other twist parameters may result; this task is deferred to future studies.

\subsection{Hard X-ray and Gamma-ray Attenuation Mechanisms}
 \label{sec:attenuation}

In all likelihood, the most energetic photons produced by resonant
Compton interactions by ultrarelativistic electrons do not actually
escape the magnetosphere, but rather are attenuated.  Two main physical
processes can effect such attenuation.  The first of these is
single-photon magnetic pair creation, \teq{\gamma \to e^+e^-}, which
becomes permissible in strong fields because momentum conservation
perpendicular to \teq{\boldsymbol{B}} is not required; it is absorbed by the global
field.  Then energy conservation yields a formal threshold of
\teq{\erg_\gamma = 2/\sin \thetakB}, in units of $m_e c^2$, where
\teq{\thetakB} is the angle between the photon direction and the
local magnetic field.  The threshold is actually polarization dependent: 
the threshold energy for \teq{\perp} photon  at \teq{(1+\sqrt{1+2B})/\sin\thetakB} 
is greater the \teq{2/\sin \thetakB} value for \teq{\parallel}.
Above these thresholds, pair creation can proceed prolifically.
For energies below the pair production threshold, magnetic photon splitting, 
a third-order QED process, is also expected to significantly attenuate the emergent spectra,
particularly if all three polarization splitting modes allowed by CP symmetry in QED operate:
\teq{\perp \rightarrow \parallel \parallel}, \teq{\parallel \rightarrow \perp \parallel}, and \teq{\perp \rightarrow \perp \perp}. 
Although photon splitting is of higher-order than \teq{\gamma \to e^+e^-}, it can kinematically operate below pair creation 
threshold, and thereby reprocess hard X-rays and soft gamma-rays into lower energy photons. 
Vacuum polarization/dispersion due to the intense magnetic field (e.g. Adler 1971) introduces significant complexity 
into the calculation of attenuation coefficients for photon splitting -- in the weakly dispersive regime, 
only the \teq{\perp \rightarrow \parallel \parallel} is kinematically allowed.  For details 
concerning the physics properties, consult Harding, Gonthier \& Baring (1997), and Baring \& Harding (2001).

One can best assess the pair attenuation situation for magnetars using the escape
energies plotted in Fig.~1 of Baring \& Harding (2001). These are the
critical energies above which the magnetosphere is opaque for light. 
In that Figure, photons were emitted from the surface, parallel to field lines, 
or approximately so, the situation appropriate to light
generated above around 100 keV via the inverse Compton scattering process.  Yet,
the geometry can apply to any dipolar field loops at any altitude,
merely with a corresponding adjustment to the field strength. From this
depiction in Baring \& Harding (2001), one can infer that for magnetic
loops with \teq{r_{\rm max}\gtrsim 4} in our upscattering picture,
photons with energies above around 15 MeV will generally not escape
equatorial regions, though this pair opacity boundary can move up to around
100 MeV for polar zones.  A similar picture is presented in Fig.~13
of Story \& Baring (2014). For lower altitudes \teq{r_{\rm max}< 4}, the
field is higher, and the opacity rapidly rises, so that one anticipates
that even photons with energies around 3-5 MeV will be attenuated in zones away
from the poles.  This character is determined primarily because field line curvature rapidly
establishes significant photon angles \teq{\thetakB} to the field during
propagation, even when these angles are very small at the point of
scattering.   None of these escape energies fits what is needed to
explain the constraining upper limits imposed by the COMPTEL
observations of magnetars, but they do provide an explanation for 
why no photons are detected above 100 MeV in magnetars by 
the {\it Fermi}-LAT telescope.  Obviously, any ensuing pair creation 
and cascading will provide feedback for the formation of spectra and also 
the acceleration process via the partial screening of electric potentials.

Escape energies for photon splittings \teq{\perp \to \parallel\parallel} have also been
plotted in Fig.~1 of Baring \& Harding (2001) {with numbers similar to those for pair creation:
when \teq{r_{\rm max}\lesssim 10}, photons above 10--15 MeV will not generally 
escape equatorial zones.  At lower altitudes} nearer the surface, 
{i.e., for \teq{r_{\rm max}\lesssim 4},} \teq{\perp} photons can be attenuated by splitting 
{at all energies} down to as low as 50 keV 
(e.g. Baring \& Harding 1998, 2001), so that this process can potentially aid in reducing 
the signal well above 150 keV, particularly since the \teq{\perp} mode is the 
dominant polarization in upscattered photons with 
\teq{0.1\, \erg_f^{\rm max}} --- see Fig.~\ref{fig:f9_spec4}.
This property affords the prospect of polarization diagnostics on the 
emission altitude and colatitude using future hard X-ray polarimeters.  Notwithstanding,
if \teq{\parallel} photons do not split, then there will be visible fluxes 
in this polarization mode extending up to the pair creation threshold somewhere 
above 1 MeV, and these signals should vary strongly with pulse 
phase.  This underlines the importance of developing 
hard X-ray/soft gamma-ray telescopes with substantially improved sensitivities 
and polarimetric capabilities that can probe this spectral cutoff domain.

\section{CONCLUSIONS}

In this paper, we have constructed an analytical framework of resonant
Compton upscattering and spectral generation in the context of magnetars 
and high-field pulsars {that experience transient magnetar-like activity}. We incorporate full QED cross sections and
kinematics appropriate for high fields and relativistic charges expected
in the inner magnetosphere of magnetars.   The collisional integrals for
spectral generation employ state-of-the-art spin-dependent ST cross
sections for treating the cyclotron resonance, surpassing other studies
that approximate scattering in the Thomson limit.
Specializing our formalism to monoenergetic electrons propagating along
individual dipole field lines, and ensembles thereof, we develop a
directed emission formalism for emergent spectra along arbitrary
observer line-of-sights.  These are sensitive to both the Lorentz factor
of an electron, and the values of the magnetic moment
inclination angle \teq{\alpha} and the viewing angle \teq{\zeta}
relative to the rotation axis.  Consequently, for magnetic axis
obliquities \teq{\alpha >0}, the emergent spectra vary as a function of
pulse phase, both in flux level at given energies, and the maximum energy 
resulting from resonant interactions.  Geometric shadowing by the star is also an important
consideration for certain portions of the phase space of viewing angles
and field loop radial extent.  Resonant interactions are almost always 
realized in the emergent spectra.  Yet they are generally exponentially 
suppressed for lower Lorentz factors, and also substantially reduced 
for fields that permit resonant interactions only with soft X-ray photons 
deep in the Wien portion of the Planck spectrum.

The kinematics for resonant interactions yields a one-to-one
correspondence between the final scattering angle and observer-frame
scattered energy for a given electron Lorentz factor.  Consequently, the
resonant Compton scattering spectrum is highly anisotropic, 
beamed within a relatively narrow solid angle of angular
extent \teq{\propto 1/\gamma_e} surrounding field line tangents from 
select locales that
point towards an observer. Most magnetospheric emission is beamed in
other directions, so that for most pulse phases, the observer only
detects a much softer spectrum, nominally below \teq{\sim 1}MeV if the
Lorentz factor is not too high.  Such a narrow beaming of emission may
also be important for inner-magnetospheric models of
high-field rotation-powered pulsars such as PSR 1846-0258 that show
magnetar-like activity but no high-energy emission following outbursts. For pulse phases 
that correspond to viewing angles
co-planar with a field loop (meridional viewing geometry), it is found
that Lorentz factors must be limited to \teq{\gamma_e \lesssim 30}, 
in order that the hardest inverse Compton emission will accommodate the
constraining \teq{200-500} keV COMPTEL bounds on magnetar emission. 
This constraint applies for a wide range of viewing angles.  Such low \teq{\gamma_e} may be the product 
of intense radiation reaction in the resonant scattering process.  Yet,
if such low Lorentz factors are not realized, super-MeV emission generated at
low altitudes is likely to be reprocessed by the QED processes of
magnetic photon splitting and pair creation, particularly because the
upscattered photons are produced remotely from polar locales. 
{The constraint on \teq{\gamma_e \lesssim 30} is also generally consistent with the 
simplified numerical experiments of the pair corona model in Beloborodov (2013a, 2013b). 
Therein, particle cooling limits Lorentz factors to below \teq{\gamma_e \lesssim 25} in 
equatorial zones with lower fields \teq{B \lesssim 1/4}. This general consistency between 
our work here and those two expositions is simply a result of resonant scattering kinematics. 
Angular and polarization dependencies of resonant Compton scattering are 
highlighted in this presentation, but not in the simplified approach of Beloborodov (2013a, 2013b),
with the potential utility of our work to considerations of hard X-ray attenuation being apparent.} 
We also remark that our illustration of spectra obtained when using the magnetic
Thomson approximation portrays hardening of the emission and
substantial violation of energy conservation for low altitude scattering
locales \teq{r_{\rm max}\lesssim 4}.

For meridional field lines, the spectral index of emergent spectra for
electrons of fixed \teq{\gamma_e} transiting single field lines is found
to be harder than is observed for the magnetar tails.  This softens somewhat to
approximately \teq{\propto (\erg_f)^0} forms when integrating over field
line azimuths, essentially assessing cumulative emission from toroidal
field surfaces of fixed radial extent \teq{r_{\rm max}}.  The need for
further steepening to match the source data may be satisfied by
introducing summations over \teq{r_{\rm max}} to model complete emission
volumes: the steepening apparent from the
envelope of the array of \teq{\rmax}-dependent spectra 
illustrated in Fig.~\ref{fig:f10_spec5} supports such a contention.
Yet it is also evident that a self-consistent simulation of acceleration and
cooling is necessary, treating the resulting distribution of
Lorentz factors.  This is perhaps best done using a 3D Monte Carlo
photon transport code that includes other QED processes such as photon
splitting and magnetic photon pair production that may operate in the
inner magnetosphere of magnetars.  Such an advance should also promote the possibility 
of using pulse profiles at different energies to constrain the values of 
\teq{\alpha} and \teq{\zeta} --- the sky maps in Fig~\ref{fig:toroid_phase_alpha_var} 
suggest that the double-peaked profiles seen in some of the
NuSTAR data for 1E 1841-045 are best matched by \teq{\alpha \sim 15^{\circ}}
in the resonant Compton upscattering model.

Finally, an important deliverable of our calculations consists of examples of
polarization-dependent spectra, primarily in Fig.~\ref{fig:f9_spec4}.
Therein, the \teq{\perp} polarization mode exceeds the \teq{\parallel}
state, reflecting the character of the magnetic Compton differential
cross section. The associated polarization degrees can range as high as
100\%. This feature appears mostly for energies \teq{\gtrsim 0.1
\erg_f^{\rm max}}, corresponding to small angles \teq{\lesssim
15^{\circ}} of emission relative to the direction of the field line
local to the resonant interaction. Such high polarization levels should 
emerge at infinity after propagation through the birefringent magnetosphere. 
The polarized spectra suggest that
future phase-resolved, energy-dependent X-ray polarization observations
will afford discrimination of magnetar geometry and upscattering model
parameters, as well as probing the attenuation action of magnetic photon
splitting and pair creation. Magnetars and high-field pulsars thus form
a key science goal for any future hard X-ray or Compton polarimetry
mission.  This identification looks forward to an era for which the IXPE initiative at
lower X-ray energies will help pave the way.

\vskip 10pt
\acknowledgments 
The authors thank the referee for a number of suggestions helpful to 
improving the communicability and clarity of the paper, Lucien Kuiper 
for comments germane to observational contexts, and Anna Watts for insightful 
perspectives. We also thank Sandro Mereghetti for an update concerning the spectroscopy of SGR 1900+14.
Z.W. is supported by the South African National
Research Foundation. Any opinion, finding and conclusion or
recommendation expressed in this material is that of the authors and
the NRF does not accept any liability in this regard.
M.G.B. acknowledges the generous support of the National Science Foundation
through grant AST-1009725 during the early phase of this work, and the 
NASA Astrophysics Theory and {\it Fermi} Guest Investigator Programs
through grants NNX13AQ82G and NNX13AP08G. 
P.L.G. thanks the National Science Foundation through grant AST-1009731, 
the NASA Astrophysics Theory Program through grant NNX13AO12G,
and the Michigan Space Grant Consortium for their generous support.
A.K.H. also acknowledges support through the NASA Astrophysics Theory
Program.

\def\mn{M.N.R.A.S.}
\def\aassupp{{Astron. Astrophys. Supp.}}
\def\apss{{Astr. Space Sci.}}
\def\apj{ApJ}
\def\nat{Nature}
\def\aaps{{Astron. \& Astr. Supp.}}
\def\aap{{A\&A}}
\def\apjs{{ApJS}}
\def\sp{{Solar Phys.}}
\def\jgr{{J. Geophys. Res.}}
\def\jphysb{{J. Phys. B}}
\def\ssr{{Space Science Rev.}}
\def\araa{{Ann. Rev. Astron. Astrophys.}}
\def\nature{{Nature}}
\def\asr{{Adv. Space. Res.}}
\def\rmp{{Rev. Mod. Phys.}}
\def\prc{{Phys. Rev. C}}
\def\prd{{Phys. Rev. D}}
\def\pr{{Phys. Rev.}}

\bibliographystyle{aasjournal}

\begin{thebibliography}{}


\bibitem[Abdo et al.(2010)]{2010ApJ...714..927A} 
Abdo, A.~A., Ackermann, M., Ajello, M., et al.\ 2010, \apj, \vol{714}{927} 

\bibitem[Adler(1971)]{1971AnPhy..67..599A} 
Adler, S.~L.\ 1971, Ann. Physics, \vol{67}{599} 

\bibitem[An et al.(2015)]{2015ApJ...807...93A} 
An, H., Archibald, R.~F., Hasco{\"e}t, R., et al.\ 2015, \apj,\vol{807}{93} 

\bibitem[An et al.(2013)]{2013ApJ...779..163A} 
An, H., Hasco{\"e}t, R., Kaspi, V.~M., et al.\ 2013, \apj,\vol{779}{163}

\bibitem[An et al.(2014)]{2014ApJ...790...60A} 
An, H., Kaspi, V.~M., Beloborodov, A.~M., et al.\ 2014, \apj,\vol{790}{60} 

\bibitem[Archibald et al.(2016)]{2016ApJ...829L..21A} Archibald, R.~F., Kaspi, V.~M., Tendulkar, S.~P., \& Scholz, P.\ 2016, \apjl, 829, L21 

\bibitem[Archibald et al.(2017)]{2017ApJ...849L..20A} Archibald, R.~F., Burgay, M., Lyutikov, M., et al.\ 2017, \apjl, 849, L20 

\bibitem[Baring(1994)]{baring94}
Baring, M.~G. 1994, in {\it Gamma-Ray Bursts}, eds. Fishman, G.,
	Hurley, K. \& Brainerd, J.~J., (AIP Conf. Proc. 307, New York) p. 572.

\bibitem[Baring et al.(2005)]{bgh05}
Baring, M.~G., Gonthier, P.~L., Harding A.~K. 2005, \apj,\vol{630}{430}

\bibitem[Baring \& Harding(1998)]{1998ApJ...507L..55B} 
Baring, M.~G., \& Harding, A.~K.\ 1998, \apjl, \vol{507}{L55} 

\bibitem[Baring \& Harding(2001)]{bh01}
Baring, M.~G. \& Harding A.~K. 2001, \apj,\vol{547}{929}

\bibitem[Baring \& Harding(2007)]{bh07}
Baring, M.~G. \& Harding A.~K. 2007, \apss,\vol{308}{109}

\bibitem[Baring et al.(2011)]{bwg11}
Baring, M.~G., Wadiasingh, Z., \& Gonthier, P.~L.\ 2011, \apj, \vol{733}{61} [BWG11]

\bibitem[terBeek(2012)]{terBeek12}
ter Beek, F. 2012, Masters thesis, University of Amsterdam.

\bibitem[Beloborodov(2009)]{Beloborodov09}
Beloborodov, A.~M.\ 2009, \apj,\vol{703}{1044}

\bibitem[Beloborodov(2013)]{Beloborodov13} 
Beloborodov, A.~M.\ 2013a, \apj, \vol{762}{13} 

\bibitem[Beloborodov(2013)]{2013ApJ...777..114B} Beloborodov, A.~M.\ 2013b, \apj, \vol{777}{114} 

\bibitem[Beloborodov \& Thompson(2007)]{2007ApJ...657..967B} 
Beloborodov, A.~M., \& Thompson, C.\ 2007, \apj, 657, 967

\bibitem[Bussard et al.(1986)]{bam86}
Bussard, R.~W., Alexander, S.~B. \& M\'esz\'aros, P. 1986, \prd,\vol{34}{440}

\bibitem[Canuto et al.(1971)]{clr71}
Canuto, V., Lodenquai, J. \& Ruderman, M., 1971, \prd,\vol{3}{2303}

\bibitem[Chen \& Beloborodov(2017)]{cb17} 
Chen, A.~Y., \& Beloborodov, A.~M.\ 2017, \apj,\vol{844}{133}

\bibitem[Daugherty \& Harding(1986)]{dh86}
Daugherty, J.~K., \& Harding, A.~K. 1986, \apj,\vol{309}{362}

\bibitem[Daugherty \& Harding(1996)]{dh96}
Daugherty, J.~K. \& Harding A.~K. 1996, \apj,\vol{458}{278}

\bibitem[De Angelis et al.(2017)]{2017arXiv171101265D} De Angelis, A., Tatischeff, V., Grenier, I.~A., et al.\ 2017, arXiv:1711.01265 

\bibitem[Dermer(1989)]{1989ApJ...347L..13D} Dermer, C.~D.\ 1989, \apjl, 347, L13 

\bibitem[Dermer(1990)]{derm90}
Dermer, C. D. 1990, \apj,\vol{360}{197}

\bibitem[Dermer \& Schlickeiser(1993)]{ds93}
Dermer, C. D. \& Schlickeiser, R. 1993, \apj,\vol{416}{458}

\bibitem[Duncan \& Thompson(1992)]{dt92}
Duncan, R. C. \& Thompson, C. 1992, \apj,\vol{392}{L9}

\bibitem[Enoto et al.(2010)]{Enoto10}
Enoto, T., Nakazawa, K., Makishima, K., et al.\ 2010, \apjl, \vol{722}{L162}

\bibitem[Enoto et al.(2017)]{Enoto17} 
Enoto, T., Shibata, S., Kitaguchi, T., et al.\ 2017, \apjs, 231, 8

\bibitem[Fern\'andez \& Thompson(2007)]{FT07}
Fern\'andez, R. \& Thompson, C. 2007, \apj,\vol{660}{615}

\bibitem[Gavriil et al.(2008)]{2008Sci...319.1802G} 
Gavriil, F.~P., Gonzalez, M.~E., Gotthelf, E.~V., et al.\ 2008, Science, \vol{319}{1802}

\bibitem[G{\"o}{\u g}{\"u}{\c s} et al.(2016)]{Gogus16}
G{\"o}{\u g}{\"u}{\c s}, E., Lin, L., Kaneko, Y., et al.\ 2016, \apjl, 829, L25

\bibitem[Goldreich \& Julian(1969)]{gj69}
Goldreich, P. \& Julian, W. H. 1969, \apj,\vol{157}{869}

\bibitem[G\"otz et al.(2006)]{goetz06}
G\"otz, D., Mereghetti, S., Tiengo, A., et al. 2006, \aap,\vol{449}{L31}

\bibitem[Gonthier et al.(2000)]{gonthier00}
Gonthier, P.~L., Harding A.~K., Baring, M. G., et al.
   2000, \apj,\vol{540}{907}

\bibitem[Gonthier et al.(2014)]{2014PhRvD..90d3014G} 
Gonthier, P.~L., Baring, M.~G., Eiles, M.~T., et al.\ 2014, \prd, \vol{90}{043014} 

\bibitem[Gradshteyn \& Ryzhik(1980)]{GR80}
Gradshteyn, I.~S. and Ryzhik, I.~M. 1980, {\it Table of Integrals, Series
   and Products}, (Academic Press, New York).

\bibitem[Greenstein \& Hartke(1983)]{Greenstein1983} 
Greenstein, G., \& Hartke, G.~J.\ 1983, \apj, \vol{271}{283} 

\bibitem[Guo et al.(2013)]{2013APh....41...63G} 
Guo, Q., Beilicke, M., Garson, A., et al.\ 2013, Astroparticle Physics, \vol{41}{63} 

\bibitem[G{\"u}ver et al.(2015)]{2015ApJ...801...48G} 
G{\"u}ver, T., G{\"o}{\u g}{\"u}{\c s}, E., \& {\"O}zel, F.\ 2015, \apj, \vol{801}{48}

\bibitem[Harding et al.(1997)]{1997ApJ...476..246H} 
Harding, A.~K., Baring, M.~G., \& Gonthier, P.~L.\ 1997, \apj, \vol{476}{246} 

\bibitem[Harding et al.(1999)]{hck99}
Harding, A.~K., Contopoulos, I., \& Kazanas, D. 1999, \apjl, \vol{525}{L125}

\bibitem[Harding et al.(2017)]{HK17}
Harding, A. K. \& Kalapotharakos, C. 2017, Proc. of Science, IFS2017, in press. 

\bibitem[Harding et al.(2008)]{hsdf08}
Harding, A.~K., Stern, J.~V., Dyks, J., \& Frackowiak, M. 2008, \apj,\vol{680}{1378}

\bibitem[den Hartog et al.(2008)]{hartog08a}
den Hartog, P.~R., Kuiper, L., Hermsen, W., et al. 2008a, \aap, \vol{489}{245}

\bibitem[den Hartog et al.(2008)]{hartog08b}
den Hartog, P.~R., Kuiper, L., \& Hermsen, W. 2008b, \aap, \vol{489}{263}

\bibitem[Hasco{\"e}t et al.(2014)]{hbdH14} 
Hasco{\"e}t, R., Beloborodov, A.~M., \& den Hartog, P.~R.\ 2014, \apjl,\vol{786}{L1}

\bibitem[Herold(1979)]{herold79}
Herold, H. 1979, \prd,\vol{19}{2868}

\bibitem[Ho \& Epstein(1989)]{he89}
Ho, C, \& Epstein, R.~I. 1989, \apj,\vol{343}{227}

\bibitem[Johnson et al.(1949)]{JL49}
Johnson, M.~H., \& Lippmann, B.~A. 1949, Phys. Rev.,\vol{76}{828}

\bibitem[Johnson et al.(2014)]{johnson14}
Johnson, T.~J., Venter, C., Harding, A.~K., et al. 2014, \apjs,\vol{213}{6}

\bibitem[Kalapotharakos et al.(2014)]{KHK14}
Kalapotharakos, C., Harding, A.~K., \& Kazanas, D.\ 2014, \apj, \vol{793}{97}

\bibitem[Kalapotharakos et al.(2012)]{KKHC12}
Kalapotharakos, C., Kazanas, D., Harding, A. K. \& Contopoulos, I. 2012, \apj, \vol{749}{2}

\bibitem[Kaspi \& Boydstun(2010)]{kb10} 
Kaspi, V.~M., \& Boydstun, K.\ 2010, \apj, \vol{710}{L115}

\bibitem[Kaspi et al.(2003)]{kaspi03}
Kaspi, V.~M., Gavriil, F.~P., Woods, P.~M., et al. 2003, \apj,\vol{588}{L93}

\bibitem[Kuiper \& Hermsen(2009)]{2009A&A...501.1031K} 
Kuiper, L., \& Hermsen, W.\ 2009, \aap, 501, 1031 

\bibitem[Kuiper \& Hermsen(2015)]{2015MNRAS.449.3827K} 
Kuiper, L., \& Hermsen, W.\ 2015, \mnras, 449, 3827 

\bibitem[Kuiper et al.(2018)]{2018MNRAS.475.1238K} Kuiper, L., Hermsen, W., \& Dekker, A.\ 2018, \mnras, 475, 1238 

\bibitem[Kuiper et al.(2006)]{kuip06}
Kuiper, L., Hermsen, W., den Hartog, P.~R., et al. 2006, \apj,\vol{645}{556}

\bibitem[Kuiper et al.(2004)]{kuip04}
Kuiper, L., Hermsen, W. \& Mende\'z, M. 2004, \apj,\vol{613}{1173}

\bibitem[Latal(1986)]{latal86}
Latal, H.~G. 1986, \apj,\vol{309}{372}

\bibitem[Li et al.(2017)]{2017ApJ...835...30L} Li, J., Rea, N., Torres, D.~F., \& de O{\~n}a-Wilhelmi, E.\ 2017, \apj, 835, 30 

\bibitem[Lin et al.(2012)]{2012ApJ...761..132L} Lin, L., G{\"o}{\v g}{\"u}{\c s}, E., G{\"u}ver, T., \& Kouveliotou, C.\ 2012, \apj, 761, 132 

\bibitem[Lyutikov \& Gavriil(2006)]{lg06}
Lyutikov, M. \& Gavriil, F.~P. 2006, \mnras,\vol{368}{690}

\bibitem[Manchester et al.(2005)]{2005AJ....129.1993M} 
Manchester, R.~N., Hobbs, G.~B., Teoh, A., \& Hobbs, M.\ 2005, \aj, \vol{129}{1993} 

\bibitem[Mereghetti et al.(2005)]{mereg05}
Mereghetti, S., G\"otz, D., Mirabel, I.~F., et al. 2005, \aap\  Lett., \vol{433}{L9}

\bibitem[Molkov et al.(2005)]{molk05}
Molkov, S., Hurley, K., Sunyaev, R., et al. 2005, \aap\  Lett., \vol{433}{L13}

\bibitem[Mushtukov et al.(2016)]{Mushtukov16} 
Mushtukov, A.~A., Nagirner, D.~I., \& Poutanen, J.\ 2016, \prd, \vol{93}{105003}

\bibitem[Nobili et al.(2008)]{ntz08a}
Nobili, L., Turolla, R. \& Zane, S. 2008a, \mnras,\vol{389}{989}

\bibitem[Nobili et al.(2008)]{ntz08b}
Nobili, L., Turolla, R. \& Zane, S. 2008b, \mnras,\vol{386}{1527}

\bibitem[Nobili et al.(2011)]{ntz11}
Nobili, L., Turolla, R., \& Zane, S. 2011, Adv. Space Res., \vol{47}{1305}

\bibitem[Olausen \& Kaspi(2014)]{2014ApJS..212....6O} 
Olausen, S.~A., \& Kaspi, V.~M.\ 2014, \apjs, \vol{212}{6} 	
	
\bibitem[Ozel(2002)]{ozel02} 
\"Ozel, F.\ 2002, Proc. Ninth Marcel Grossmann Meeting, 2321. arXiv:astro-ph/0106071

\bibitem[Parfrey \& Beloborodov(2013)]{Parfrey2013} 
Parfrey, K., Beloborodov, A.~M., \& Hui, L.\ 2013, \apj,\vol{774}{92}

\bibitem[Pennucci et al.(2015)]{Pennucci15}
Pennucci, T.~T., Possenti, A., Esposito, P., et al.\ 2015, \apj, \vol{808}{81} 

\bibitem[Perna et al.(2001)]{perna01}
Perna, R., Heyl, J.~S., Hernquist, L.~E., et al. 2001, \apj,\vol{557}{18}

\bibitem[Philippov \& Spitkovsky(2014)]{Philippov-2014-ApJ} 
Philippov, A.~A., \& Spitkovsky, A.\ 2014, \apjl, \vol{785}{L33}

\bibitem[Pierbattista et al.(2015)]{Pierbattista15}
Pierbattista, M., Harding, A.~K., Grenier, I.~A., et al.\ 2015, \aap, \vol{575}{3}

\bibitem[Rea \& Esposito(2011)]{2011ASSP...21..247R} Rea, N., \& Esposito, P.\ 2011, Astrophysics and Space Science Proceedings, 21, 247 

\bibitem[Rea et al.(2013)]{rea13} 
Rea, N., Israel, G.-L., Pons, J.~A., et al.\ 2013, \apj, \vol{770}{65} 

\bibitem[Rea et al.(2012)]{rea12} 
Rea, N., Pons, J.~A., Torres, D.~F., \& Turolla, R.\ 2012, \apjl, \vol{748}{L12}

\bibitem[Rea et al.(2008)]{rea08}
Rea, N., Zane, S., Turolla, R., et al. 2008, \apj,\vol{686}{1245}

\bibitem[Shapiro \& Teukolsky(1983)]{1983bhwd.book.....S} 
Shapiro, S.~L., \& Teukolsky, S.~A.\ 1983, Research supported by the National Science Foundation.~New York, Wiley-Interscience, 1983, 663 p.,  

\bibitem[Sokolov \& Ternov(1968)]{ST68}
Sokolov, A.~A. \& Ternov, I.~M. 1968, Synchrotron Radiation, 
   (Pergamon Press, Oxford).

\bibitem[Spitkovsky(2006)]{2006ApJ...648L..51S} 
Spitkovsky, A.\ 2006, \apjl, \vol{648}{L51} 

\bibitem[Story \& Baring(2014)]{sb14}
Story, S.~A., \& Baring, M.~G.\ 2014, \apj, \vol{790}{61}

\bibitem[Sturner(1995)]{sturn95}
Sturner, S.~J. 1995, \apj, \vol{446}{292}

\bibitem[Thompson \& Duncan(1996)]{td96}
Thompson, C. \& Duncan, R. C. 1996, \apj,\vol{473}{332}

\bibitem[Thompson et al.(2002)]{tlk02}
Thompson, C., Lyutikov, M. \& Kulkarni, S.~R. 2002, \apj\vol{574}{332}

\bibitem[Tiengo et al.(2002)]{tiengo02}
Tiengo, A., G{\"o}hler, E., Staubert, R., \& Mereghetti, S.  2002, \aap,\vol{383}{182}

\bibitem[Tiengo et al.(2013)]{2013Natur.500..312T} 
Tiengo, A., Esposito, P., Mereghetti, S., et al.\ 2013, \nat, \vol{500}{312} 

\bibitem[Timokhin \& Harding(2015)]{th15}
Timokhin, A.~N., \& Harding, A.~K.\ 2015, \apj,\vol{810}{144}

\bibitem[Vasisht \& Gotthelf(1997)]{vg97}
Vasisht, G. \& Gotthelf, E. V. 1997, \apj,\vol{486}{L129}

\bibitem[van Putten et al.(2016)]{vpwbw16}
van Putten, T., Watts, A.~L., Baring, M.~G. \& Wijers, R.~A.~M.~J. 2016, \mnras, \vol{461}{877}

\bibitem[Vigan{\`o} et al.(2013)]{vigano13}
Vigan{\`o}, D., Rea, N., Pons, J.~A., et al.\ 2013, \mnras, \vol{434}{123}

\bibitem[Vogel et al.(2014)]{Vogel2014} 
Vogel, J.~K., Hasco{\"e}t, R., Kaspi, V.~M., et al.\ 2014, \apj, \vol{789}{75}

\bibitem[Wang et al.(2014)]{Wang2014} 
Wang, W., Tong, H., \& Guo, Y.-J.\ 2014, Research in Astronomy and Astrophysics, \vol{14}{673} 

\bibitem[Watters et al.(2009)]{Watters09}
Watters, K.~P., Romani, R.~W., Weltevrede, P., \& Johnston, S.\ 2009, \apj, \vol{695}{1289}

\bibitem[Wu et al.(2013)]{Wu} 
Wu, J.~H.~K., Hui, C.~Y., Huang, R.~H.~H., et al.\ 2013, Journal of Astronomy and Space Sciences, \vol{30}{83}

\bibitem[Zavlin et al.(1996)]{zps96} 
Zavlin, V.~E., Pavlov, G.~G., \& Shibanov, Y.~A.\ 1996, \aap, \vol{315}{141}

\end{thebibliography}

\nopagebreak

\appendix

\section{The Integration over the Soft Photon Angular Distribution} 
  \label{sec:AppendixA}

In this Appendix, analytic development of the soft photon integral
\begin{equation}
     \int_{\erg_-}^{\erg_+} \dover{d \erg_s}{\erg_s^2} \, n_\gamma(\erg_s) \, f(\mu_i)
 \label{eq:soft_photon_int}
\end{equation}
that appears in the photon production rate in Eq.~(\ref{eq:scatt_spec_fin}) 
is detailed.  This integral also appears in the calculation of electron cooling rates at arbitrary interaction 
points in BWG11.  Such reductions are expedient for numerical computations, particularly 
because there is considerable sensitivity to the exponential in the integrand.
In addition, it should be noted that developments here do not involve the cross section physics, and 
so are applicable to a variety of inverse Compton scattering problems in proximity to stars, for example, 
gamma-ray binaries.
For uniform hemispherical thermal soft photons as described by Eq. (\ref{eq:Planck_spec}), 
with dimensionless temperature \teq{\Theta = kT/m_ec^2}, this integral specializes to
\begin{equation}
   \dover{\Omega_s}{\pi^2 \lambar^3} \int_{\erg_-}^{\erg_+}  \dover{f[\mu_i (\erg_s) ]}{e^{\erg_s/\Theta} -1}  d\erg_s
   \; =\; \dover{\Omega_s}{\pi^2 \lambar^3}  \dover{\beta_e \omega_i}{\gamma_e}
          \int^{\mu_+}_{\mu_-} \dover{f(\mu_i)}{(1+\beta_e \mu_i)^2}     \dover{d\mu_i}{e^{\erg_s (\mu_i)/\Theta}-1}
 \label{eq:soft_photon_integ}
\end{equation}
with 
\begin{equation}
   \erg_s (\mu_i) \;\equiv\; \dover{\omega_i}{\gamma_e(1+\beta_e \mu_i)}
   \quad \hbox{and} \quad
   \erg_{\pm} \; =\; \erg_s (\mu_{\mp})\quad .
 \label{eq:erg_s_pm_def}
\end{equation}
Here \teq{\mu_{\pm}} represent the bounds of the angle cosines of the 
soft photons relative to the field direction.  The analytical approximations and series developed in this Appendix 
for this integral speed up numerical computations of the spectral rates considerably.
Central to these developments is the recognition that the Planck spectrum is a 
perfect derivative of a logarithmic form:
\begin{equation}
   \dover{1}{e^{\Psi} -1} \; =\; \dover{d\Upsilon}{d\Psi} 
   \quad ,\quad
   \Upsilon (\Psi) \; =\;  \log_e \bigl( 1 - e^{-\Psi} \bigr) \quad .
 \label{eq:Planck_deriv_form}
\end{equation}
For applications here, \teq{\Psi > 0} and so \teq{\Upsilon (\Psi ) < 0}.
One can then write
\begin{equation}
   {\cal I} (\Theta ,\, \gamma_e,\, \omega_i) \;\equiv\; \int_{\erg_-}^{\erg_+}  \dover{f[\mu_i (\erg_s) ]}{e^{\erg_s/\Theta} -1}  d\erg_s
   \; =\; - \Theta  \int^{\mu_+}_{\mu_-}f(\mu_i) \, 
          \dover{\partial}{\partial \mu_i} \Bigl(  \Upsilon \bigl[ \erg_s (\mu_i)/\Theta \bigr] \Bigr)  \, d\mu_i \quad .
 \label{eq:soft_photon_integ_alt}
\end{equation}
Observe that the minus sign out the front appears because the change of the 
differentiation from the \teq{\erg_s /\Theta} variable to \teq{\mu_i}.
This form suggests an integration by parts, 
a protocol that proves expedient because the forms for the 
derivatives of \teq{f(\mu_i)} are relatively compact.  Numerical 
evaluations of the ensuing integrals are then more stable.

\subsection{Altitudinal and Colatitudinal Dependence of  $f(\mu_i)$}

The functional form of \teq{\mu_\pm} and \teq{f(\mu_i)} on location 
in the magnetosphere also depend on the sense of motion of the charge carriers 
relative to the magnetic field at an interaction point. Throughout this paper we adopt 
dipole field geometry \teq{r = \rmax \sin^2 \vartheta} for simplicity, and 
assume that electrons are propagating along field loops 
{\it from a southern to a northern magnetic footpoint} located at colatitudes 
\teq{\thetamin} and \teq{\thetamax}, respectively.  The definitions of \teq{\mu_\pm} 
also depend on the hemisphere, and for the present case of downward electrons 
for \teq{\thetamin \leq \vartheta \leq \thetamax} are defined as
\begin{equation}
   \mu_+ \; =\;  \left \{ \begin{array}{cl}
    1 & \mbox{if } \quad \thetamin \le \vartheta \le \pi/2 \, \, \, \mbox{and} \, \, \, \thetaBr \le \thetaC \\[4pt]
     \cos (\thetaBr - \thetaC) & \mbox{if } \quad 
            \Bigl( \thetamin \le \vartheta \le \pi/2  \,\,\, \mbox{and} \,\,\,\thetaBr > \thetaC \Bigr) 
            \;\; \mbox{or} \; \; \Bigl(\pi/2 \leq \vartheta \leq \thetamax \Bigr)
  \end{array} 
 \right.  
 \label{eq:mup_def}
\end{equation}
and
\begin{equation}
   \mu_- \; =\;  \left \{ \begin{array}{cl}
    \cos (\thetaBr + \thetaC) & \mbox{if } \quad 
     \Bigl( \thetamin \le \vartheta \le \pi/2 \Bigr) \;\; \mbox{or} \;\;  
          \Bigl( \pi/2 \le \vartheta \leq \thetamax \, \, \,\mbox{and} \, \, \, \pi - \thetaBr > \thetaC \Bigr) \\[4pt]
           -1 & \mbox{if } \quad \pi/2 \leq \vartheta \leq \thetamax \, \, \, \mbox{and} \, \, \,  \pi - \thetaBr \leq \thetaC \quad .
  \end{array} 
 \right. 
 \label{eq:mum_def}
\end{equation}
Here, for the branch \teq{0 \leq \arccos x \leq \pi}, we have defined 
\begin{equation}
   \thetaC \; =\; \arccos \sqrt{1 - \Bigl( \dover{\rns}{R} \Bigr)^2 }
   \quad \hbox{and}\quad
   \thetaBr \; =\; \arccos \left[ \dover{2 \cos \vartheta}{\sqrt{1+3 \cos^2 \vartheta}} \right] \quad ;
 \label{eq:theta_B_Br_def}
\end{equation}
see also Eq.~(\ref{eq:fmui_norm}).
These represent the opening angle \teq{\thetaC} of the cone of collimation of 
the soft photons at altitude \teq{R}, and the polar angle \teq{\thetaBr} of the local 
field vector with respect to the radial direction.
For polar locales, the definition Eq.~(\ref{eq:theta_B_Br_def}) can also be interpreted 
as defining a new \teq{\thetaBr^{\prime} \equiv \pi - \thetaBr} where 
\teq{0 \le \thetaBr^{\prime} \leq \pi/2} while \teq{\thetaC} remains unchanged,
since along a field line \teq{r = \rmax \sin^2 \vartheta = \rmax \sin^2 (\pi-\vartheta)}. 
This is equivalent to defining \teq{\thetaBr^{\prime}} in the first quadrant via
\begin{equation}
    \thetaBr^{\prime} \; =\; \arccos \left | \frac{2 \cos \vartheta}{\sqrt{1+3 \cos^2 \vartheta}}   \right|
    \quad ,\quad
    0 \le \thetaBr' \le \pi/2 \quad .
 \label{eq:thetaBr_prime}
\end{equation}
This form is convenient at polar locales when 
\teq{\thetaBr' \leq \thetaC} for \teq{\vartheta \ge \pi/2} and  \teq{\thetaBr \leq \thetaC} for  
\teq{\vartheta \le \pi/2}. The definitions of \teq{\mu_{\pm}} above then transform to those in BWG11.

Within the angle cosine range \teq{[\mu_- ,\, \mu_+]}, the angular distribution along a field line 
for electrons moving from the south pole outward is given by modifying the definition of
\teq{f(\mu_i)} given in Eq.~(71) of BWG11:
 \begin{equation} 
     f(\mu_i) \; =\; \left \{ \begin{array}{cl}
             f[\mu_i (\vartheta)]_{\rm down} & \mbox{if } \quad \thetamin \leq \vartheta \leq \pi/2 \\[4pt]
             f[\mu_i (\pi - \vartheta)]_{\rm up}  & \mbox{if} \quad \pi/2 \le \vartheta \leq \thetamax
 \end{array} 
 \right. \quad ,
 \end{equation}
where ``downward" electrons correspond to \teq{\thetamin \leq \vartheta \leq \pi/2} and 
``upward" electrons are for \teq{\pi/2 \le \vartheta \le \thetamax}.
 At high altitudes and \underline{equatorial locales} along a field line, when \teq{\thetaBr > \thetaC}, 
 the form of \teq{f(\mu_i)} in BWG11 is
\begin{equation}
   f_{\hbox{\sixrm E}}(\mu_i) \; =\; \dover{1}{\pi} \arcsin \biggl( \dover{{\cal D}}{\sin\thetaBr} \biggr)
   \quad ,\quad
   {\cal D} \; =\;  \sqrt{\frac{(\mu_+ - \mu_i)(\mu_i-\mu_-)}{1-\mu_i^2}} \quad ,
 \label{eq:fmui_equator}
\end{equation}
for \teq{ \mu_{\pm} \to \cos (\thetaBr \mp \thetaC )}.
This form is also valid when \teq{\vartheta \geq \pi/2},  i.e. when 
\teq{\thetaBr \geq \pi/2} as long as \teq{\pi - \thetaBr \geq \thetaC}.
The form of \teq{f(\mu_i)} at \underline{polar locales} along the field line has 
three pieces, adapting the branches of the \teq{\arcsin} function 
defined in BWG11 to our geometry.  For the case \teq{\vartheta \leq \pi/2}, 
the electrons are moving downward, and we can write the \teq{f(\mu_i)} distribution as
\begin{equation}
       f_{\hbox{\sixrm P}}(\mu_i) \; =\; \left \{ \begin{array}{cl}
            1 & \mbox{if } \quad   \cos (\thetaBr - \thetaC )\leq \mu_i \leq 1 \\[4pt]
            1 - \dover{1}{\pi} \arcsin \biggl( \dover{{\cal D}}{\sin\thetaBr} \biggr) 
                 & \mbox{if }\quad   \cos \thetaC \sec \thetaBr \leq \mu_i < \cos (\thetaBr - \thetaC) \\[4pt]
            \dover{1}{\pi} \arcsin \biggl( \dover{{\cal D}}{\sin\thetaBr} \biggr) 
                 & \mbox{if }\quad   \cos (\thetaBr + \thetaC ) \leq \mu_i < \cos \thetaC \sec \thetaBr  \\[6pt]
            0 & \mbox{otherwise\, .}
  \end{array} 
 \right.  
 \label{eq:fmui_pole}
\end{equation}
Note the \teq{\mu_\pm} limits are readily identifiable in this 
result and also in Eq.~(\ref{eq:fmui_equator}).
Similarly, for \teq{\vartheta > \pi/2}, the prescription for `upward electrons' applies 
using the substitutions \teq{\mu_i \to \mu_i'  = - \mu_i}  and \teq{\thetaBr \to \thetaBr' = \pi - \thetaBr}
in Eq.~(\ref{eq:fmui_pole}).  Graphical depictions of these angular distributions 
for colatitudes \teq{30^{\circ}} and \teq{60^{\circ}} are given in Fig.~9 of BWG11.

\subsection{Manipulating the Soft Photon Energy/Angle Integration}

As indicated above, integration by parts is a logical path to expediting the evaluation 
of Eq.~(\ref{eq:soft_photon_integ_alt}).  The exact details of this step depend on 
the mathematical nature of the soft photon angular distribution, and as has just 
been expounded, there are two general forms for \teq{f (\mu_i)}, 
appropriate for equatorial and polar zones; we treat these sequentially.

\subsubsection{Equatorial Locales}
 \label{sec:app_equator}
 
For the case of equatorial locales at moderate altitudes, the magnetic field vector 
lies outside the soft photon collimation cone, and so \teq{\pi \geq \thetaBr > \thetaC}.
Then Eq.~(\ref{eq:fmui_equator}) presents a simple form for 
\teq{f(\mu_i) \to f_{\hbox{\sixrm E}}(\mu_i)}.  The bounds \teq{\mu_- \leq \mu_i \leq \mu_+} 
to the angle integration suggest a change of variables to \teq{\phi} defined by
\begin{equation}
   \mu_i \; =\; \dover{\mu_+ + \mu_-}{2} + \dover{\mu_+ - \mu_-}{2} \, \cos\phi
   \quad ,\quad
   0 \;\leq\; \phi \;\leq\; \pi \quad .
 \label{eq:mui_phi_reln}
\end{equation}
Given the forms \teq{\mu_{\pm} = \cos (\thetaBr \mp \thetaC )} apparent in 
Eqs.~(\ref{eq:mup_def}) and~(\ref{eq:mum_def}), it follows that
\begin{equation}
   \dover{\mu_+ + \mu_-}{2} \; =\; \cos\thetaBr \cos\thetaC \;\equiv\; {\cal C}
   \quad ,\quad
   \dover{\mu_+ - \mu_-}{2} \; =\; \sin\thetaBr \sin\thetaC  \;\equiv\; {\cal S} \quad .
 \label{eq:mup_mum_ident}
\end{equation}
These define the angular quantities \teq{{\cal C}} and \teq{{\cal S}} that will appear 
in the algebra below, with \teq{\mu_{\pm} = {\cal C} \pm {\cal S}}.
If one then transforms the \teq{\arcsin} to an \teq{\arctan} function with the aid the 
standard trigonometric identity \teq{\arcsin \chi = \arctan (\chi/\sqrt{1-\chi^2})}, one 
arrives at an alternative form for the angular distribution:
\begin{equation}
   f_{\hbox{\sixrm E}}(\mu_i) \; \to\;  f_{\hbox{\sixrm E}}(\phi ) \;\equiv\;\dover{1}{\pi} 
   \arctan \biggl( \dover{\sin\thetaC \sin\phi}{\cos \thetaC \sin \thetaBr - \cos\phi \cos \thetaBr \sin \thetaC}  \biggr)
   \quad ,\quad
   \thetaBr \; >\; \thetaC \quad .
 \label{eq:fmui_equator_alt}
\end{equation}
At this juncture, since the relationship between \teq{\mu_i} and \teq{\phi} 
defines a cosine rule for a spherical triangle,
it is evident that \teq{\phi} represents an azimuthal angle about the 
radial vector that serves as the axis of the spherical cap defined by the soft photon 
propagation collimation cone.  From the form in Eq.~(\ref{eq:fmui_equator_alt}), 
using the identity \teq{d\mu_i = {\cal S}\, d(\cos \phi )}, one quickly obtains 
the derivative (valid for \teq{\thetaBr > \thetaC})
\begin{equation}
    2\pi {\cal S} \, \sin\phi \, \dover{\partial f_{\hbox{\sixrm E}}(\mu_i)}{\partial \mu_i} \; =\; 
     \dover{\cos\thetaC + \cos\thetaBr}{{\cal C} +1 + {\cal S} \cos\phi } 
    + \dover{\cos\thetaC - \cos\thetaBr}{{\cal C} - 1 + {\cal S} \cos\phi }
    \; <\; 0 \quad ,
 \label{eq:dfmui_eq_dmui}
\end{equation}
after some algebraic simplification.   Since \teq{\thetaBr > \thetaC}, the numerators of 
both terms on the right are positive, and it can be quickly demonstrated that
the \teq{\partial f_{\hbox{\sixrm E}}(\mu_i)/\partial \mu_i < 0}.   Clearly \teq{{\cal S}>0} and \teq{(1 + {\cal C})/{\cal S} > 1}.
Also, as \teq{1 - {\cal C} - {\cal S} = 1 - \cos (\thetaBr - \thetaC) >0} for \teq{\thetaBr > \thetaC},
it follows that \teq{({\cal C}-1)/{\cal S} < -1} applies to the denominator of the second term on the 
right.  Hence, no singularities arise in the integrand resulting from the integration by parts step, 
which introduces no residual boundary terms because these are zero at \teq{\mu_i = \mu_{\pm}}.
Thus, for equatorial locales,  since \teq{d\mu_i = {\cal S}\,\sin\phi\, d\phi},
\begin{equation}
   {\cal I}_{\hbox{\sixrm E}} (\Theta ,\, \gamma_e,\, \omega_i) 
   \; =\; \Theta  \int^{\pi}_0 \dover{d\phi}{2\pi}
    \left(  \dover{\cos\thetaC + \cos\thetaBr}{{\cal C} +1 + {\cal S} \cos\phi } 
    + \dover{\cos\thetaC - \cos\thetaBr}{{\cal C} - 1 + {\cal S} \cos\phi } \right) 
          \Upsilon \biggl[ \dover{\erg_s ({\cal C} + {\cal S}\cos\phi )}{\Theta} \biggr] \; .
 \label{eq:soft_photon_integ_equator}
\end{equation}
Both factors in the integrand are negative when  \teq{\thetaBr > \thetaC}.
The pieces of this integral take the form of that in Eq.~(\ref{eq:integ_master}), and
protocols for computing such integrations are outlined in Section~\ref{sec:app_ang_integ_approx}.

\subsubsection{Polar Locales}
 \label{sec:app_polar}
 
Without loss of generality, we specialize to \teq{\vartheta \le \pi/2}. The operational definition 
of polar locales is for \teq{\thetaBr} small enough for the magnetic field vector to lie inside 
the soft photon collimation cone.  This amounts to \teq{\thetaBr < \thetaC}, with Eq.~(\ref{eq:fmui_pole}) being the 
appropriate angular distribution.  The ensuing integration by parts proceeds much 
as for the equatorial case, except for the fact that there are three pieces to the integration,
as inferred from Eq.~(\ref{eq:fmui_pole}), 
and the boundary evaluations are different.  The total integration is
\begin{eqnarray}
    \int_{\erg_-}^{\erg_+}  \dover{f[\mu_i (\erg_s) ]}{e^{\erg_s/\Theta} -1}  d\erg_s 
    & = & -\, \Theta \int_{\cos ( \thetaBr - \thetaC )}^1 \dover{\partial}{\partial \mu_i} \Bigl(  \Upsilon \bigl[ \erg_s (\mu_i)/\Theta \bigr] \Bigr)  \, d\mu_i  \nonumber\\[2.0pt]
    && - \, \Theta \int^{\cos( \thetaBr-\thetaC )}_{\cos \thetaC /\cos \thetaBr} \Bigl[ 1 - f_{\hbox{\sixrm E}}(\mu_i) \Bigr] \,
       \dover{\partial}{\partial \mu_i} \Bigl(  \Upsilon \bigl[ \erg_s (\mu_i)/\Theta \bigr] \Bigr)  \, d\mu_i 
 \label{eq:polar_integ_setup}\\[2.0pt]
   && - \, \Theta \int_{\cos(\thetaBr+\thetaC)}^{\cos \thetaC /\cos \thetaBr} f_{\hbox{\sixrm E}}(\mu_i) \,
       \dover{\partial}{\partial \mu_i} \Bigl(  \Upsilon \bigl[ \erg_s (\mu_i)/\Theta \bigr] \Bigr)  \, d\mu_i \quad .\nonumber
\end{eqnarray}
The equatorial angular distribution \teq{f_{\hbox{\sixrm E}}(\mu_i)} of 
Eq.~(\ref{eq:fmui_equator}) is introduced to both render the 
algebra more compact, but also to identify mathematical similarities with the developments in 
Section~\ref{sec:app_equator}.  Adapting the equivalent form in 
Eq.~(\ref{eq:fmui_equator_alt}), we recast it 
and analytically continue it here via
\begin{equation}
   f_{\hbox{\sixrm E}}(\phi )  \; \to\;  f_{\hbox{\sixrm P}}(\phi ) \;\equiv\;
   \dover{1}{\pi} \, \hbox{Sign}\bigl( {\cal T} - \cos\phi \bigr) \;
   \arctan \biggl( \dover{\sin\phi}{\cos \thetaBr \vert {\cal T} - \cos\phi \vert} \biggr)
   \quad \hbox{for}\quad
   {\cal T} \; =\; \dover{\tan \thetaBr}{\tan \thetaC} \quad .
 \label{eq:fmui_equator_alt2}
\end{equation}
Since \teq{0 < \thetaBr < \thetaC}, it is evident that \teq{0 < {\cal T} < 1}, so that two different 
branches to this \teq{\arctan} function must be sampled on the interval \teq{0 \leq \phi\leq\pi}.
These are separated by the singular point \teq{\cos\phi = {\cal T}}, which is equivalent 
to \teq{\mu_i = \cos\thetaC/\cos\thetaBr \equiv \mu_m}, 
the ratio of the two pertinent cosines.
This discontinuity imposes itself on the integration by parts, and the explicit 
appearance of the Sign function factor in Eq.~(\ref{eq:fmui_equator_alt2}) 
compactly accommodates the two branches.
The integral evaluation progresses by isolating the terms not dependent on 
\teq{f_{\hbox{\sixrm E}}(\mu_i)}, which contain perfect derivatives in the integrand.
For the two other terms, we integrate by parts to yield
\begin{eqnarray}
    \dover{1}{\Theta} \int_{\erg_-}^{\erg_+}  \dover{f[\mu_i (\erg_s) ]}{e^{\erg_s/\Theta} -1}  d\erg_s 
    & = & - \Bigl[  \Upsilon \bigl[ \erg_s (\mu_i)/\Theta \bigr] \Bigr]^1_{\mu_m}  \nonumber\\
    &&  + \biggl[ f_{\hbox{\sixrm E}}(\mu_i)  \,
             \Upsilon \bigl[ \erg_s (\mu_i)/\Theta \bigr] \biggr]^{\mu_+}_{\mu_m} 
             - \int^{\mu_+}_{\mu_m}  \dover{\partial}{\partial \mu_i} 
       \Bigl[ f_{\hbox{\sixrm E}}(\mu_i) \Bigr] \, \Upsilon \bigl[ \erg_s (\mu_i)/\Theta \bigr]  \, d\mu_i  \quad
 \label{eq:polar_integ_dev1} \\
   &&  +  \int_{\mu_-}^{\mu_m} \dover{\partial}{\partial \mu_i} 
          \Bigl[  f_{\hbox{\sixrm E}}(\mu_i) \Bigr] \, \Upsilon \bigl[ \erg_s (\mu_i)/\Theta \bigr] \, d\mu_i 
          - \biggl[ f_{\hbox{\sixrm E}}(\mu_i)  \,
             \Upsilon \bigl[ \erg_s (\mu_i)/\Theta \bigr] \biggr]_{\mu_-}^{\mu_m}\quad ,\nonumber
\end{eqnarray}
for \teq{\mu_m = \cos\thetaC/\cos\thetaBr}.  For the non-integral terms, 
we observe that for \teq{\mu_m < \mu_i \leq \mu_+}, one has
\teq{{\cal T} < \cos\phi \leq 1} so that the negative branch of Eq.~(\ref{eq:fmui_equator_alt2})
is applicable.  Then \teq{f_{\hbox{\sixrm E}}(\mu_+)=0} since then \teq{\phi =0}, and also 
\teq{f_{\hbox{\sixrm E}}(\mu_m) \to - 1/2} at the singular point.  Next, 
when \teq{\mu_- < \mu_i < \mu_m}, the positive branch of Eq.~(\ref{eq:fmui_equator_alt2})
applies as \teq{0 \leq \cos\phi < {\cal T}}, so \teq{f_{\hbox{\sixrm E}}(\mu_-)=0} since then \teq{\phi =\pi}, and
\teq{f_{\hbox{\sixrm E}}(\mu_m) \to 1/2} for the other singular point.  Thus the
constant terms contribute a total of  \teq{\Upsilon \bigl[ \erg_s (1)/\Theta \bigr]}, 
with substantial cancellation arising in the associated sum.
For the remaining integrals, observe that the derivatives of 
\teq{f_{\hbox{\sixrm E}}(\mu_i)} select the negative branch on \teq{[\mu_m,\,\mu_+]}
and the positive one on \teq{[\mu_-,\, \mu_m]}, so that they can be blended 
into a single integral on the interval \teq{[\mu_-,\, \mu_+]}.  Then 
using the change of variables in Eq.~(\ref{eq:mui_phi_reln}), this integral can be 
cast in a form similar to that in Eq.~(\ref{eq:soft_photon_integ_equator}), 
remembering the concatenation of minus signs.  Thus the final 
polar result for \teq{\vartheta < \pi /2} is
\begin{equation}
   {\cal I}_{\hbox{\sixrm P}} (\Theta ,\, \gamma_e,\, \omega_i) 
   \; = \; -\, \Theta \,\Upsilon \biggl[ \dover{\erg_s (1)}{\Theta} \biggr]
     +  \Theta   \int^{\pi}_0 \dover{d\phi}{2\pi}
    \left(  \dover{\cos\thetaC + \cos\thetaBr}{{\cal C} +1 + {\cal S} \cos\phi } 
    + \dover{\cos\thetaC - \cos\thetaBr}{{\cal C} - 1 + {\cal S} \cos\phi } \right) 
          \Upsilon \biggl[ \dover{\erg_s ({\cal C} + {\cal S}\cos\phi )}{\Theta} \biggr] \;\;  . 
 \label{eq:soft_photon_integ_polar}
\end{equation}
Observe that now the integral is not always positive, as can be 
discerned by selecting the \teq{\thetaBr=0} polar axis case, so this 
serves to reduce the value of the positive constant term.
The corresponding \teq{\pi /2 < \vartheta < \pi} result can be obtained 
in similar fashion, generating the same integral contribution, but with the 
constant residual term being instead \teq{ + \Theta \Upsilon \bigl[ \erg_s (-1)/\Theta \bigr]}.

\subsection{Analytic Approximations to the Angular Integral}
 \label{sec:app_ang_integ_approx}

In the previous two sections of this Appendix, the end point of the integration by parts 
involves integrals of the form 
\begin{equation}
  {\cal J}(\alpha ,\, s,\, A) \; =\;  
  \int_0^{\pi} \dover{d\phi}{\alpha + \cos \phi}\, \log_e \left[ 1 - \exp \left\{ - \dover{A}{s + \cos \phi} \right\} \right] \quad ,
 \label{eq:integ_master}
\end{equation}
for the domains \teq{\vert\alpha\vert >1} and \teq{\vert s\vert > 1}.  
While \teq{\alpha} takes on various forms, the other two variables 
are fixed for all the integrals:
\begin{equation}
   A \; =\; \dover{2\hat{\omega}_i}{\gamma_e \Theta \beta_e (\mu_+ - \mu_-)}
   \; \equiv\; \dover{\hat{\omega}_i}{\gamma_e \Theta \beta_e {\cal S}}
   \quad \hbox{and} \quad
   s \; =\; \dover{2 + \beta_e (\mu_+ + \mu_-)}{\beta_e (\mu_+ - \mu_-)} 
   \;\equiv\; \dover{1 + \beta_e {\cal C}}{\beta_e {\cal S}} \quad .
 \label{eq:A_s_def}
\end{equation}
For all magnetospheric locales, \teq{s > 1}.  To facilitate numerical calculations of spectra, 
we now develop some analytic approximations to \teq{{\cal{J}}} in the limits of \teq{A} 
small and large. We note that the integral requires numerical computation only in a 
small gap of the parameter space where \teq{A} is neither 
small nor large, where the accuracy of the two protocols described hereafter rises above \teq{0.1\%} precision.
Taylor series expansion of the logarithm is an obvious path, and this is 
preferable when \teq{A\gg 1} and exponential contributions come into play.

\subsubsection{Small \teq{A}}

The series expansion of the logarithm for \teq{A \ll1} is not particularly practical, since 
the radius of convergence for \teq{A/(s+\cos\phi)} is only \teq{2 \pi}.  Term-by-term integration 
would then yield integrals expressible in terms of Legendre polynomials.  
We seek a more expedient path, and it starts by 
employing Euler's infinite product representation of the \teq{\sinh} function. 
Using identity 1.431.2 of Gradshteyn \& Ryzhik (1980), the infinite product identity 
for the \teq{\sinh x} function, then for all complex \teq{\chi}
a quick manipulation of the logarithmic form yields
\begin{equation}
   \log_e \Bigl(1 - e^{-\chi} \Bigr) \; =\; - \dover{\chi}{2} +  \log_e \chi
   + \sum_{k=1}^{\infty} \log_e \Bigl[ 1 + \dover{\chi^2}{4 k^2\pi^2}  \Bigr] \quad .
 \label{eq:log_ident2}
\end{equation}
For \teq{\chi = A/(s + \cos \phi)} this series must be carried to \teq{k \gg A/(2 \pi(s-1))}. 
Defining \teq{x_k \equiv A/(2 \pi k)}, the logarithm portion of the integrand 
in Eqs.~(\ref{eq:soft_photon_integ_equator}) and~(\ref{eq:soft_photon_integ_polar})
can be expressed as
\begin{equation}
   \log_e \Bigl(1 - e^{-\chi} \Bigr) \; =\; \log_e A - \dover{A}{2( s + \cos \phi )} 
   - \log_e [s + \cos \phi ] 
   + \sum_{k=1}^{\infty} \Bigl\{   \log_e \left[ (s + \cos \phi )^2 + x_k^2 \right]
   - 2 \log_e  [s + \cos \phi ] \Bigr\} \;\; .
 \label{eq:calJ_integrand_series}
\end{equation}
Term-by-term integration of the series requires several integral identities; 
these we shall not prove in detail here. The first is
\begin{equation}
   {\cal I}_1(\alpha ) \;\equiv\; \int_0^{\pi} \dover{d\phi }{\alpha + \cos\phi} 
   \; =\; \dover{\pi\, \mbox{Sign}(\alpha)}{\sqrt{\alpha^2-1}}\quad ,
 \label{eq:integ1}
\end{equation}
which is easily established by contour integration on the unit circle with \teq{z + 1/z = 2 \cos\phi} 
for poles that lie outside, i.e. \teq{\vert \alpha \vert  >1}. Using a partial fraction decomposition, 
one can then easily establish the second integral identity,
\begin{equation}
   {\cal I}_2(\alpha , s) \;\equiv\; \int_0^{\pi} \dover{dt}{\alpha + \cos\phi} 
   \, \dover{1}{s+\cos\phi}
   \; =\; \dover{\pi}{s-\alpha} \left\{ \dover{\mbox{ Sign}(\alpha)}{\sqrt{\alpha^2-1}} - \dover{1}{\sqrt{s^2-1}} \right\}\quad .
 \label{eq:integ2}
\end{equation}
The third integral identity can be established by recasting \teq{{\cal I}_2} as 
the perfect derivative of an integral with integrand proportional to \teq{\log_e (s + \cos\phi)}, 
or using identity 4.397.16 of Gradshteyn \& Ryzhik (1980).  If one defines a parameter 
\teq{\sigma} such that \teq{\alpha = (\sigma + 1/\sigma)/2},
\begin{equation}
   \sigma \; =\; \left \{ 
   \begin{array}{cl}
       \alpha - \sqrt{\alpha^2-1} & \mbox{if } \quad  \alpha >1 \quad ,\\[3pt]
       \alpha + \sqrt{\alpha^2-1}& \mbox{if } \quad \alpha < -1 \quad ,
   \end{array} 
   \right.  
 \label{eq:sigma_def}
\end{equation}
such that for \teq{\mu = s - \sqrt{s^2-1}}, 
\begin{equation}
    {\cal {I}}_3 (\alpha, s) \; \equiv\; \int_0^\pi \dover{\log_e \left[ s+\cos\phi \right] }{\alpha + \cos\phi} \, d\phi 
    \; =\; \dover{\pi \mbox{ Sign}(\alpha) }{\sqrt{\alpha^2-1}} \, 
    \log_e \left\{ \dover{[1- \mu \sigma ]^2}{2\mu } \right\} \quad .
 \label{eq:integ3}
\end{equation}
The fourth integral identity is harder to establish, but can be found by analytically continuing 
\teq{s} to the complex plane via \teq{s\to s_\pm = s \pm i x_k}.  Then summing over the 
factorization of the logarithm yields a sum integrals in the form of \teq{{\cal I}_3}. 
Defining \teq{2 \tau = \sqrt{(s+1)^2+x_k^2} + \sqrt{(s-1)^2+x_k^2}} and 
\teq{\kappa = \tau - \sqrt{\tau^2-1}}, the fourth identity is found to be
 \begin{equation}
     {\cal I}_4 (\alpha,s,x_k) \;\equiv\; \int_0^\pi 
         \dover{\log_e \left[ (s+\cos\phi)^2 +x_k^2 \right] }{\alpha + \cos\phi} d\phi
     \; = \;  \dover{2 \pi \mbox{ Sign}(\alpha) }{\sqrt{\alpha^2-1}} 
         \log_e \left\{ \dover{[1-\kappa \sigma ]^2}{2 \kappa} 
         - \dover{[\kappa - \mu ][1 - \kappa \mu] \sigma }{\mu (1+\kappa^2)}  \right\} \quad .
 \label{eq:integ4}   
\end{equation}
Note that in this form it is easily apparent that  as \teq{x_k \rightarrow 0},
\teq{\tau \rightarrow s} so that \teq{\kappa\to\mu}, and
\teq{ {\cal I}_4 (\alpha,s,x_k) \to 2 {\cal {I}}_3 (\alpha,s)}.
The original integral is now cast in terms of the four integrals,
\begin{equation}
    {\cal J}(\alpha, s, A) = \left( \log_e A \right) {\cal I}_1 (\alpha) - \dover{A}{2}  {\cal I}_2 (\alpha,s) 
    -  {\cal I}_3 (\alpha,s) + \sum_{k=1}^\infty \Bigl[  {\cal I}_4 (\alpha,s,x_k) -  2{\cal I}_3 (\alpha,s) \Bigr] \quad .
 \label{eq:calJ_eval}
\end{equation}
Numerical evaluation of the sum is carried to some \teq{k_{\rm max} \gg  A/(2 \pi (s-1))} 
bounded above such that the computation speed is superior to numerical integration, 
generally by factors up to \teq{10^1-10^2} for a modest amount of terms. 
If \teq{k_{\rm max}} becomes very large, it is possible to replace the sum by an 
integration over \teq{k} for contributions in the \teq{x_k\ll 1} range.
Alternatively, a remainder term to the summation can be identified and 
approximated by considering the leading order term in expansion of the sum's argument for \teq{x_k \ll 1}, 
which is quadratic in \teq{A x_k} due to the form of \teq{\tau}, and expressible in terms 
of the Polygamma function.
This protocol was adopted for the summation whenever large numbers of terms 
were required, for which small values of \teq{x_k} would eventually be encountered.

\subsubsection{Large \teq{A}}

The analytics developed for small \teq{A} in the previous section, although valid for 
arbitrarily large values of \teq{A}, quickly become numerically inefficient and 
prohibitive for a given accuracy when \teq{A} is large due to the high number of terms 
required for convergence. As an alternative protocol, here we develop asymptotic 
approximations for large \teq{A}, by first forming the series of logarithm,
\begin{equation}
   {\cal{J}}(\alpha, s, A)\;  =\; - \sum_{n=1}^{\infty} \frac{1}{n} \int_0^\pi \frac{e^{-A n /(s+\cos\phi)}}{\alpha + \cos\phi} d\phi \quad .
\end{equation}
Changing variables to \teq{ y=(s^2-1) \left[ u- 1/(s+1) \right]/2} with \teq{ u= (s+ \cos\phi)^{-1}} transforms the integral in the sum to,
\begin{equation}
   \int_0^\pi \frac{e^{-A n /(s+\cos\phi)}}{\alpha + \cos\phi} \, d\phi 
   \; =\; \dover{e^{-A n/(s+1)} \sqrt{s^2-1}}{(1+\alpha)(s-1)} 
   \int_0^1 \dover{dy}{\sqrt{y (1-y)}} \, (1- {\cal{B}} y)^{-1} 
   \exp \biggl\{ - \dover{2An y}{s^2-1} \biggr\} 
 \label{eq:largeA_integ_rep}
\end{equation} 
where 
\begin{equation}
   {\cal B} \; =\; \dover{2(s-\alpha)}{(1+\alpha)(s-1)} \quad ,
 \label{eq:calB_def}
\end{equation}
with \teq{  0<{\cal{B}}<1} when \teq{\alpha = a, s > a >1 }. This integral is given by a double sum in
identity 3.385 of Gradshteyn \& Rhyzik (1980), but it proves not to be an efficient path for computation.

For large \teq{A n}, the exponential dominates the integrand suppressing all contributions to the 
integral except those for small \teq{y}. Thus the method of steepest descents is appropriate, 
concomitant with the approximation of extending the upper limit to infinity. 
For the case \teq{\alpha >1} where \teq{  0<{\cal{B}}<1}, we form a Maclaurin series 
of the non-singular algebraic part in the integrand about \teq{y=0},
\begin{eqnarray}
     (1-y)^{-1/2} (1-{\cal{B}} y)^{-1} &=& \sum_{k = 0}^{\infty} \left[ \frac{{\cal{B}}^{k + 1/2}}{\sqrt{{\cal{B}}-1}} 
          +\frac{(-1)^k \sqrt{\pi} \, _2F_1(1, 3/2 + k, 2+ k, 1/{\cal{B}})}{{\cal{B}}     \Gamma (k+2) \Gamma[-(k + 1/2)] } \right] y^k  \nonumber  \\
    & \approx & 1 + \left(\frac{1}{2} + {\cal{B}} \right) y + \left( \frac{3}{8} + \frac{{\cal{B}}}{2} + {\cal{B}}^2 \right) y^2.
\end{eqnarray}
For each term in the \teq{k} series, the upper limit of the integral in 
Eq.~(\ref{eq:largeA_integ_rep}) can be set to infinity, one can then analytically express 
the integral using the definition of the Gamma function
\begin{equation}
    \int_0^\infty y^{k -\frac{1}{2}} e^{\frac{- 2 A n y}{s^2-1}} dy 
    \; =\; \left(\frac{s^2-1}{2 A n} \right)^{k + \frac{1}{2}} \Gamma \left[k + \frac{1}{2} \right] \quad .
\end{equation}
From the duplication formula for  the Gamma function, one can reduce 
\teq{\Gamma (k+ 1/2) = 2^{1- 2k} \sqrt{\pi} (2 k-1)!/(k-1)!} in the usual manner.
Thus to second order in \teq{k} for large \teq{A} and \teq{\alpha >1}, we obtain 
the asymptotic approximation
\begin{equation}
   \int_0^\pi \frac{e^{-A n /(s+\cos\phi)}}{\alpha + \cos\phi} d\phi  
   \; \approx \;  \sqrt{\dover{\pi}{2 A n}} \,  \dover{ s+1}{\alpha+1} 
    \left[ 1  + \frac{1}{2} \left(\frac{1}{2} +{\cal{B}} \right) \left(\frac{s^2-1}{2 A n} \right)  \right]  e^{-A n/(s+1)} \quad .
 \label{eq:integ_zw1}
\end{equation}

For the regime \teq{\alpha < -1} a better approximation is found by keeping the \teq{(1- {\cal{B}} y)} 
denominator intact, since \teq{{\cal{B}} <0} and \teq{|{\cal{B}}| \rightarrow \infty} as \teq{\alpha \rightarrow -1}. 
We note again that the dominant contribution to the integral comes from \teq{y \approx 0}. 
As such, we employ the first few terms in Taylor series of the \teq{(1-y)^{-1/2}} portion about \teq{y=0},
\begin{equation}
    (1-y)^{-1/2}  \; =\;  \sqrt{\pi} \sum_{k=0}^\infty \frac{(-1)^k}{k! \Gamma(1/2- k ) } y^k \approx  1 + (1/2)y + (3/8)y^2 \quad .
 \label{eq:1-y_binomial}
\end{equation}
As for the \teq{\alpha > 1} case, we send the upper limit of the integral to 
infinity for expediency, and note the identity
\begin{equation}
   \int_0^\infty y^{-1/2} (1-{\cal{B}} y)^{-1} y^m e^{-c y} dy 
   \; =\; (-{\cal{B}})^{-(m + 1/2)} \, e^{-c/{\cal{B}}} \, \Gamma \Bigl[ m+ 1/2\Bigr] \Gamma\Bigl[ 1/2-m,-c/{\cal{B}} \Bigr] 
\end{equation}
for \teq{ c>0} and \teq{{\cal{B}}<0}. Here the incomplete Gamma function 
simplifies to linear combinations complementary error functions by the identity 
\teq{\Gamma[ 1/2, x] = \sqrt{\pi}\, \mbox{erfc} (\sqrt{x})} when combined with 
recurrence relations of the gamma function. Thus, for \teq{\alpha <-1 } to second order
\begin{equation}
    \int_0^\pi \dover{e^{-A n /(s+\cos\phi)}}{\alpha + \cos\phi} d\phi
    \;\approx\; -\sqrt{ \dover{\pi (s+1)}{ 2 (\alpha + 1) (\alpha -s) }}  \, 
    \exp \biggl\{ - \dover{An}{s-\alpha} \biggr\}  
    \left( \Gamma\left[ \dover{1}{2}, \, - \dover{2 A n}{ {\cal B}(s^2-1)} \right] 
       - \dover{1}{4 {\cal B}} \, \Gamma \left[ - \dover{1}{2}, \, - \dover{2 A n}{ {\cal B}(s^2-1)} \right] \right) \; ,
 \label{eq:integ_zw2} \\[0.5pt]
\end{equation}
where Eq.~(\ref{eq:calB_def}) has been used to simplify the algebra for 
the argument of the exponential and other factors.  Numerically, for 
large \teq{A n}, only two or three terms of the binomial series
in Eq.~(\ref{eq:1-y_binomial}) are required for either case outlined 
above when summing over \teq{n}. For a minimal the gap region, these 
asymptotic approximations are accurate when \teq{2A/(s^2-1) \gg 1} and 
\teq{\exp[-A/(s+1)]/\sqrt{A} \lesssim 5}. The sum over \teq{n} is taken to 
\teq{n_{max} \sim n_c \mbox{Max}[(s^2-1)/(2A), (s+1)/A, (s-\alpha)/A] }, 
where \teq{n_c \gg 1} with the value chosen such that it is no slower than 
numerical integration as one approaches the gap where the accuracy threshold is not satisfied. 

\end{document}